\documentclass[12pt]{article}
\usepackage[normalem]{ulem}
\usepackage{tocloft}
\usepackage{amsfonts}
\usepackage{amsmath}
\usepackage{amssymb}
\usepackage{float,tikz, extarrows, tikz-cd}
\usetikzlibrary{decorations.pathmorphing,calc}
\usepackage{array}
\usepackage{bigints}
\usepackage{textcomp}
\usepackage{bm}
\usepackage{booktabs}
\usepackage[nosort]{cite}
\usepackage{color}
\usepackage{dsfont}
\usepackage{float}
\usepackage{framed}
\usepackage{graphicx}
\usepackage{tcolorbox}
\usepackage{indentfirst}
\usepackage{mathrsfs}
\usepackage{multirow}
\usepackage{pdflscape}
\usepackage{setspace}
\usepackage{titlesec}
\usepackage{wrapfig}
\usepackage{mathtools}
\usepackage[all]{xy}
\usepackage{young}
\usepackage[vcentermath]{youngtab}
\usepackage{relsize}
\usepackage{stackengine}
\usepackage{verbatim}
\usepackage{slashed}
\usepackage[compat=1.0.0]{tikz-feynman}
\usepackage{adjustbox}
\usepackage{subcaption}
\usepackage{caption}

\usepackage{mathtools}
\DeclarePairedDelimiter\ceil{\lceil}{\rceil}
\DeclarePairedDelimiter\floor{\lfloor}{\rfloor}

\usepackage{hyperref}
\hypersetup{colorlinks=true}
\hypersetup{linkcolor=black}
\hypersetup{citecolor=black}
\hypersetup{urlcolor=black}

\numberwithin{equation}{section}

\usepackage{arydshln}
\usepackage{blkarray}
\usepackage{physics}

\usepackage[left=2.5cm,right=2.5cm,top=2.5cm,bottom=3cm]{geometry}
\fontdimen2\font=1.2\fontdimen2{\jot}{5pt}

\titleformat{\section}{\large\bfseries}{\thesection.}{4pt}{}
\titlespacing{\section}{0pt}{20pt}{6pt}

\titleformat{\subsection}{\normalfont\bfseries}{\thesubsection.}{4pt}{}
\titlespacing{\subsection}{0pt}{15pt}{6pt}

\titleformat{\subsubsection}{\normalfont\itshape}{\thesubsubsection.}{4pt}{}
\titlespacing{\subsubsection}{0pt}{15pt}{6pt}

\titleformat{\paragraph}{\normalfont\itshape}{\theparagraph.}{4pt}{}
\titlespacing{\paragraph}{0pt}{15pt}{6pt}

\def\tilde{\widetilde}

\def\hat{\widehat}

\def\bar{\overline}

\def\pa{\partial}

\def\ep{\varepsilon}
\def\1{{\mathds 1}}
\def\Re{\mathop{\rm Re}}

\DeclareMathAlphabet{\mathbfsf}{OT1}{cmss}{bx}{n}

\def\CA{{\mathcal A}}

\def\CG{{\mathcal G}}

\def\CN{{\mathcal N}}
\def\CO{{\mathcal O}}
\def\CP{{\mathcal P}}

\def\CR{{\mathcal R}}

\def\CW{{\mathcal W}}

\newcommand{\beq}{\begin{equation}\begin{aligned}}
\newcommand{\eeq}{\end{aligned}\end{equation}}

\newcommand{\lam}{\lambda}

\newcommand{\al}{\alpha}
\newcommand{\be}{\beta}
\newcommand{\del}{\delta}

\newcommand{\ov}{\over}
\newcommand{\ex}[1]{\langle #1 \rangle}
\def\vp{{\varphi}}
\def\ve{{\varepsilon}}

\newcommand{\bea}{\begin{eqnarray}}
\newcommand{\eea}{\end{eqnarray}}

\newcommand{\beqa}{\begin{eqnarray}}
\newcommand{\eeqa}{\end{eqnarray}}
\newcommand{\beqar}{\begin{eqnarray*}}
\newcommand{\eeqar}{\end{eqnarray*}}
\newcommand{\e}{\epsilon}

\def\({\left(} \def\){\right)}
\def\[{\left[} \def\]{\right]}

\DeclareFontShape{OT1}{cmr}{mx}{n}%
{<->cmr10}{}
\newcommand{\mytitlefont}{\fontseries{mx}\selectfont}
\DeclareMathAlphabet{\titlemath}{OT1}{cmr}{mx}{n}

\newcommand{\Oo}{\mathcal{O}}

\begin{document}

% TITLEPAGE FOR PAPERS

\begin{titlepage}
%\begin{flushright} \small
%UUITP-17/19
 %\end{flushright}

\begin{center}
			
~\\[0.9cm]
			
{\fontsize{26pt}{0pt} \mytitlefont Surface Operators 
   and Exact Holography}
			
~\\[1cm]

Changha Choi,$^{a}$\,\footnote{\href{mailto:cchoi@perimeterinstitute.ca}{\tt cchoi@perimeterinstitute.ca}}
Jaume Gomis,$^{a}$\,\footnote{\href{mailto:jgomis@perimeterinstitute.ca}{\tt jgomis@perimeterinstitute.ca}}
Raquel Izquierdo Garc\'ia,$^{a}$\,\footnote{\href{mailto:rizquierdogarcia@perimeterinstitute.ca}{\tt rizquierdogarcia@perimeterinstitute.ca}}

~\\[0.5cm]

{\it $^{a}$Perimeter Institute for Theoretical Physics,\\ 
Waterloo, Ontario, N2L 2Y5, Canada}\

\end{center}

\vskip0.5cm
			
\noindent 

\makeatletter
\renewenvironment{abstract}{%
    \if@twocolumn
      \section*{\abstractname}%
    \else %% <- here I've removed \small
      \begin{center}%
        {\bfseries \normalsize\abstractname\vspace{\z@}}%  %% <- here I've added \Large
      \end{center}%
      \quotation
    \fi}
    {\if@twocolumn\else\endquotation\fi}
\makeatother

  \begin{abstract}  
  \normalsize

Surface operators are nonlocal probes of gauge theories capable of  distinguishing   phases  that are  not discernible by the classic Wilson-'t Hooft criterion.
We prove that the correlation function  of a surface operator with a chiral primary operator in ${\cal N}=4$ super Yang-Mills is a finite polynomial in the Yang-Mills coupling constant.  Surprisingly, in spite of these observables receiving nontrivial quantum corrections, we find that these correlation functions  are exactly captured in the 't Hooft limit by supergravity in asymptotically $AdS_5\times S^5$ \cite{Drukker:2008wr}\,!
We also calculate exactly  the surface operator vacuum expectation value and the correlator of a surface operator with 1/8-BPS Wilson loops using supersymmetric localization. We demonstrate that these correlation functions in ${\cal N}=4$ SYM  realize in a nontrivial fashion the conjectured action of $S$-duality. Finally, we perturbatively quantize ${\cal N}=4$ SYM around the surface operator singularity and identify the Feynman diagrams that when summed over reproduce the exact result obtained by localization.
\end{abstract}

\vfill 
%\begin{flushleft} 
%May 2020
 %\end{flushleft}

\end{titlepage}
	
		\setcounter{table}{1}
% TABLE OF CONTENTS
	
\setcounter{tocdepth}{3}
\renewcommand{\cfttoctitlefont}{\large\bfseries}
\renewcommand{\cftsecaftersnum}{.}
\renewcommand{\cftsubsecaftersnum}{.}
\renewcommand{\cftsubsubsecaftersnum}{.}
\renewcommand{\cftdotsep}{6}
\renewcommand\contentsname{\centerline{Contents}}
	
\tableofcontents

% MAIN TEXT

\vfill\eject

\section{Introduction}\label{sec:intro}

  In the  AdS/CFT   correspondence \cite{Maldacena:1997re,Gubser:1998bc,Witten:1998qj}    bulk supergravity    describes  the physics 
 of the boundary field theory in the strong coupling regime. Therefore, 
 there is  a priori no reason to expect    
 supergravity     to capture the
    weak coupling  expansion of   boundary field theory   observables that receive quantum corrections. 
    
    In this paper we exactly compute   correlation functions in ${\cal N}=4$ super-Yang-Mills (SYM) -- that in spite of receiving    quantum corrections --  are   exactly captured in the 't Hooft limit by supergravity in asymptotically $AdS_5\times S^5$\,! 
  Using our exact results, we demonstrate that these correlation functions in ${\cal N}=4$ SYM  realize in a nontrivial fashion the conjectured action of $S$-duality.

The main character  in our story are surface operators \cite{Gukov:2006jk}. These are nonlocal operators in gauge theories  that  are able to discern between phases of gauge theories that are not otherwise distinguishable by the classic Wilson-'t Hooft criterion \cite{Gukov:2013zka}. 

Surface operators   in ${\cal N}=4$ SYM are supported on a surface $\Sigma$ in spacetime, which henceforth we take to  be the maximally symmetric spaces $\Sigma=\mathbb R^2$ or $S^2$. Surface operators, which we denote by ${\cal O}_\Sigma$,\footnote{A   surface operator  depends on   $\mathbb L$ and   $(\alpha_I,\beta_I,\gamma_I,\eta_I)$, but  for readability we   write ${\cal O}_\Sigma\equiv {\cal O}^{\mathbb L}_\Sigma(\alpha,\beta,\gamma,\eta)$.}   are labeled by the choice of a Levi group \cite{Gukov:2006jk}
\beq
\mathbb L=\prod_{I=1}^M U(N_I)\subset U(N)
\eeq
that specifies how gauge symmetry is broken as the surface $\Sigma$  is approached,  and 
a collection of $4M$ 
continuous parameters \cite{Gukov:2006jk}
\beq
(\alpha_I,\beta_I,\gamma_I,\eta_I)\,,
\eeq
 which   are exactly marginal couplings for local operators on the surface defect (surface operators are described in more detail in  section 2).
 
\begin{figure}
\centering    \includegraphics[width=0.35\textwidth]{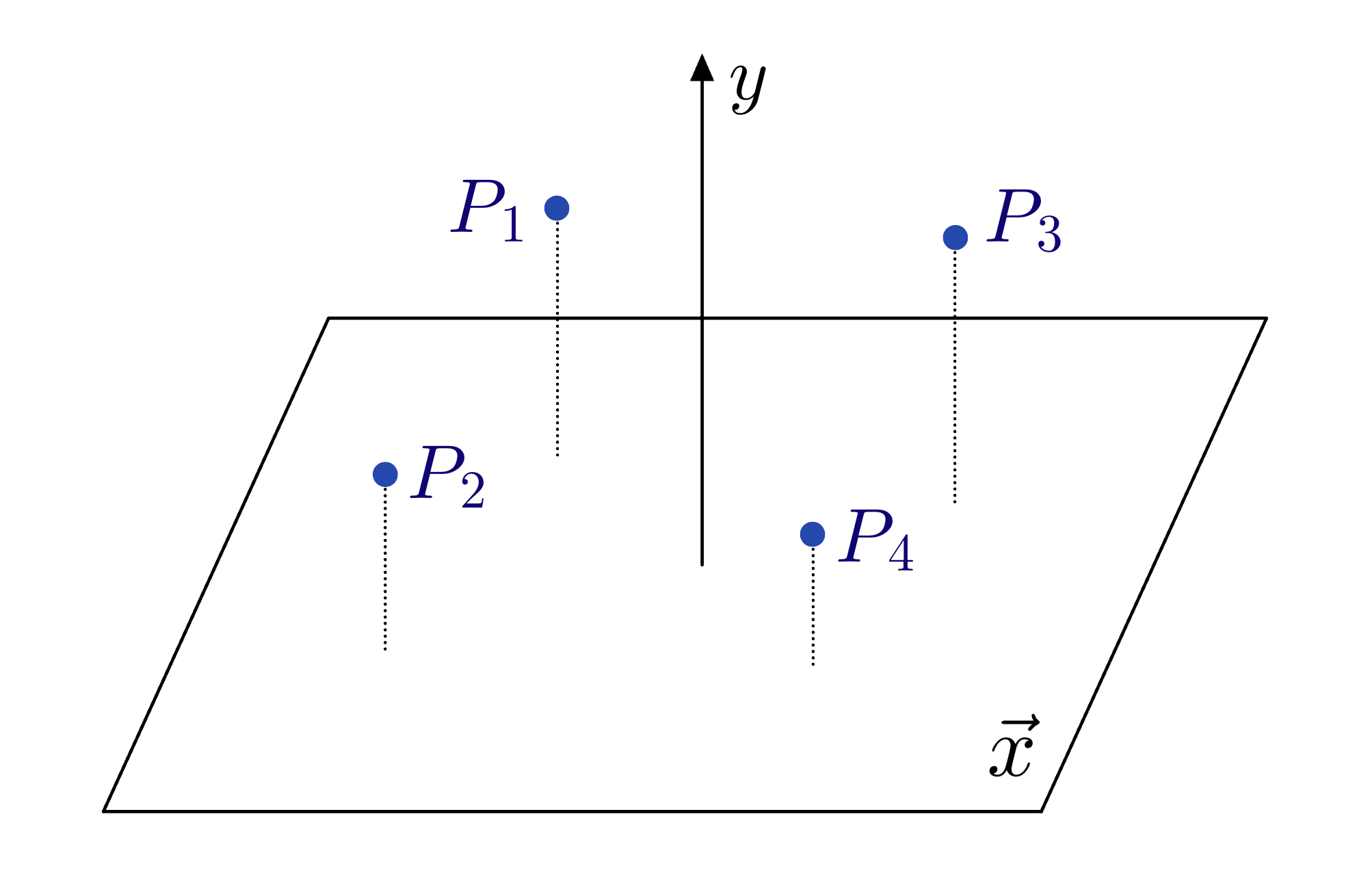}
\caption[.]{
Asymptotically $AdS_5\times S^5$ solution dual to ${\cal O}_\Sigma$ with Levi group $\mathbb L=\prod_{I=1}^M U(N_I)$ is determined by $M$ point sources $P_I$ that live in a $3d$ half-space $(\vec x, y\geq 0)$ over which $AdS_3\times S^3\times S^1$ is nontrivially fibered. The mapping of parameters is:
\medskip

\begin{minipage}{\linewidth}
\beq
{y_I^2\over R_{\text{AdS}}^4}&=N_I \qquad\qquad~~~
{\vec x_I\over 2\pi l_s^2}=(\beta_I,\gamma_I)\\[+2pt]
\int_{D_I}B_{\text{NS}}&= \alpha_I \qquad\qquad
\int_{D_I}B_{\text{R}}= \eta_I\nonumber
\eeq
where $D_I$ are  two-cycles in the asymptotically  $AdS_5\times S^5$ geometry (see \cite{Gomis:2007fi} for details).
\end{minipage}
}
\label{figure:bubbling}
\end{figure}

 The topologically nontrivial, smooth,  asymptotically $AdS_5\times S^5$ ``bubbling" supergravity dual description of surface operators in ${\cal N}=4$ SYM was put forward  in \cite{Gomis:2007fi}. This gave a physical interpretation to the  Type IIB supergravity solutions found earlier by Lin-Lunin-Maldacena \cite{Lin:2004nb,Lin:2005nh}. The choice of Levi group $\mathbb L$  and   parameters $(\alpha_I,\beta_I,\gamma_I,\eta_I)$  that characterize a surface operator in ${\cal N}=4$ SYM  are   encoded in a nontrivial fashion in   the asymptotically $AdS_5\times S^5$ supergravity solutions \cite{Gomis:2007fi} (see figure \ref{figure:bubbling} for the dictionary between surface operator data    
 and supergravity solution data).\footnote{Surface operators are described in the probe approximation  by D3-branes in $AdS_5\times S^5$ \cite{Gukov:2006jk} (see also \cite{Drukker:2008wr}). The  supergravity  backgrounds \cite{Gomis:2007fi}
 capture  the  nonlinear backreaction due to the D3-branes.}

%\vfill\eject

 The computation of the correlation functions of a surface operator ${\cal O}_\Sigma$ with  a chiral primary operator\footnote{The         $SO(4)_R\subset SU(4)_R$ invariant chiral operators ${\cal O}_{\Delta,k}$ have a nontrivial  correlator due  to the $SO(4)_R$ symmetry of ${\cal O}_\Sigma$.  $\Delta$ and $k=-\Delta, -\Delta+2,\ldots, \Delta-2, \Delta$ are   the scaling dimension and $U(1)_R$ charge of  ${\cal O}_{\Delta,k}$.}     
    ${\cal O}_{\Delta,k}$ was performed 
  using the dual asymptotically $AdS_5\times S^5$ supergravity solutions  \cite{Gomis:2007fi} as well as probe D3-branes in \cite{Drukker:2008wr}.   
  The correlators computed in the supergravity regime, where $\lambda\equiv g^2N\gg 1$,  take the form \cite{Drukker:2008wr}\footnote{The position dependence is completely fixed by confomal invariance and is omitted in the formula.}
\beq
\begin{aligned}
{\langle {\cal O}_\Sigma \cdot  {\cal O}_{\Delta,k}\rangle_{\text{sugra}}\over {\langle {\cal O}_\Sigma \rangle}_{\text{sugra}}}={1\over \lambda^{|k|\over 2}}\left(A_0+{A_1\over \lambda}+{A_2\over \lambda^2}  +\ldots {A_{{\Delta-|k|\over 2}}\over \lambda^{{\Delta-|k|\over 2}} } \right)\,,
\end{aligned}
\label{strongres}
\eeq
where $A_a$ are explicit functions of $N_I$ and of the surface operator parameters $(\beta_I,\gamma_I)$. 

Instead, in the planar    perturbative gauge theory regime, where $\lambda <<1$,  the correlator takes the form  
\beq
\begin{aligned}
{\langle {\cal O}_\Sigma \cdot {\cal O}_{\Delta,k}\rangle_{\text{YM}}\over
\langle {\cal O}_\Sigma\rangle_{\text{YM}}}
={1\over \lambda^{\Delta\over 2}}\left( B_0 +\sum_{n=1}^\infty B_n \lambda^n \right) \end{aligned}\,.
\label{weakexpan}
\eeq

$B_0$ was computed in \cite{Drukker:2008wr} and the following surprise was found: the leading perturbative gauge theory result (\ref{weakexpan}) exactly matched the most subleading term in the strong coupling expansion of the supergravity computation (\ref{strongres}), that is 
\beq
B_0=A_{{\Delta-|k|\over 2}}\,. 
\eeq

In this paper we compute these correlators  using supersymmetric localization, valid  at any value of the gauge coupling $g^2$,  and   prove  that the perturbative series   truncates at a finite order $\left(g^2\right)^{{\Delta-|k|\over 2}}$, that is $n_{\text{max}}={\Delta-|k|\over 2}$ in (\ref{weakexpan}).\footnote{This was conjectured in \cite{Drukker:2008wr}.     ${\Delta-|k|\over 2}$ is   an integer.} This implies that the planar correlator (\ref{weakexpan})  terminates and is a polynomial in the coupling 
\beq
\begin{aligned}
{\langle {\cal O}_\Sigma \cdot {\cal O}_{\Delta,k}\rangle_{\text{YM}}\over
\langle {\cal O}_\Sigma\rangle_{\text{YM}}}={1\over \lambda^{\Delta\over 2}}\left( B_0 + B_1 \lambda+  B_2 \lambda^2+ \ldots B_{\Delta-|k|\over 2}  \lambda^{\Delta-|k|\over 2}  \right) \end{aligned}\,.
\label{weakexpanf}
\eeq
We furthermore show that 
\beq
 B_n=A_{{\Delta-|k|\over 2}-n}\,,
\eeq
and therefore supergravity computes\footnote{The supergravity computations were performed  in \cite{Drukker:2008wr}.} the planar correlators exactly, even though they acquire quantum corrections!

We also calculate other correlation functions involving surface operators in ${\cal N}=4$ SYM. We  exactly determine the  vacuum expectation value of a  surface operator    $\langle {\cal O}_{S^2}\rangle$ supported on a two-sphere.\footnote{The planar surface defect is trivial $\langle {\cal O}_{\mathbb R^2}\rangle=1$.}
The surface operator is also probed  with the general  ${1/8}$-BPS Wilson-Maldacena loop operator \cite{Maldacena:1998im,Rey:1998ik,Drukker:2007dw,Drukker:2007yx,Drukker:2007qr} through the correlation function $\langle {\cal O}_{S^2}\cdot W^{1/8}_{\CR}(C)\rangle$, where $C$ is an arbitrary closed curve  on a two-sphere that links the surface operator $ {\cal O}_{S^2}$.  Summarizing, we exactly evaluate the following observables:
\begin{itemize}
    \item $\langle {\cal O}_{S^2}\rangle$
    \item $\langle {\cal O}_{S^2} \cdot {\cal O}_{\Delta,k}\rangle$
    \item $\langle {\cal O}_{S^2}\cdot W^{1/8}_{\CR}(C)\rangle$
\end{itemize}
Furthermore, we identify the Feynman diagrams that contribute to these correlation functions upon quantizing ${\cal N}=4$ SYM around the surface operator singularity.

A remarkable feature of ${\cal N}=4$ SYM is its   conjectured action of $S$-duality.
Under the action of an element of the $S$-duality group $PSL(2,\mathbb Z)$ 
\beq
{\cal M}=\begin{pmatrix}
a&b\\
c&d
\end{pmatrix}\,  
\eeq
 a surface operator  ${\cal O}_\Sigma$ is mapped to itself  up to a nontrivial action on the parameters  \cite{Gukov:2006jk} (see also \cite{Gomis:2007fi})
\beq 
\begin{aligned}
    (\beta_I,\gamma_I)\rightarrow |c\tau+d|(\beta_I,\gamma_I)\\
    (\alpha_I,\eta_I)\rightarrow (\alpha_I,\eta_I){\cal M}^{-1}\,,
    \label{actionSS}
\end{aligned}
\eeq
where 
\beq
\tau={\theta\over 2\pi}+{4\pi i \over g^2}
\label{complexco}
\eeq
is the complexified coupling constant, on which  $PSL(2,\mathbb Z)$ acts as
\beq \label{eq:s-dual}
\tau\rightarrow {a\tau+b\over c\tau+d}\,.
\eeq
We demonstrate that $\langle {\cal O}_{S^2}\rangle$
and $\langle {\cal O}_{S^2}\cdot {\cal O}_{\Delta,k}\rangle$ admit -- in a nontrivial fashion -- the 
 action of    $S$-duality and of the periodic  identifications of the surface operator parameters $\alpha_I\simeq \alpha_I+1$ and $\eta_I\simeq \eta_I+1$ (see section 2).\footnote{$\langle {\cal O}_{S^2}\rangle$ enjoys the desired duality and periodicity properties by virtue of $\langle {\cal O}_{S^2}\rangle$ being well-defined   up to a K\"ahler transformation $\langle {\cal O}_{S^2}\rangle\rightarrow \langle {\cal O}_{S^2}\rangle e^{f(t,\tau)+\bar f(\bar t,\bar \tau)}$,    just as the $S^2/S^4$ partition functions of $2d$ ${\cal N}=(2,2)$ SCFTs and of $4d$ ${\cal N}=2$ SCFTs are
 only meaningful up to a K\"ahler transformation \cite{Gomis:2012wy, Gerchkovitz:2014gta,Gomis:2014woa,Gomis:2015yaa} (see section \ref{sec:surfacevev}  for details).}\textsuperscript{,}\footnote{The    normalized chiral primary operators such that their two-point function   is independent of $g^2$ \newline -- such as  ${1\over g^{\Delta}}\hbox{Tr}\, \Phi^{\Delta}$ -- are       $S$-duality invariant \cite{Intriligator:1998ig,Argyres:2006qr,Gomis:2009xg}.}
 
 The plan of the rest of the paper is as follows. In section \ref{sec:surface operator}  we     discuss    surface operators in $\CN=4$ SYM. In section \ref{sec:surfacevev} we explain the implications that  conformal anomalies of surface operators have on the expectation value of     surface operators. We then compute   the expectation value of spherical surface operators  using supersymmetric localization in the description of surface operators 
 in terms of coupling $4d$ ${\cal N}=4$ SYM to $2d$ $\CN=(4,4)$ quiver gauge theories. 
In section \ref{sec:2dYMloc}  we compute the correlation functions of a surface operator with a    chiral primary operator and with a  1/8-BPS Wilson loop   by   supersymmetric localization on $B^3\times S^1$ using the disorder definition and suitable supersymmetric boundary conditions and boundary terms. In Appendix \ref{appsec:B3S1}  we explain the    $B^3\times S^1$ geometry, and in Appendix \ref{appsec:N=4} we spell out our conventions on spinors and $\CN=4$ SYM. In Appendix \ref{app:disorder}  we   compute $\ex{\CO_{S^2}}_{S^4}$ using the disorder definition and by localization on  $S^4$, reproducing the   result of section \ref{sec:surfacevev}. In Appendix \ref{appsec:bdry}  we derive the boundary conditions and the boundary terms describing the surface operators which were used in section \ref{sec:2dYMloc}. In Appendix \ref{appsec:CPO} and \ref{appsec:Wilson}  we perform the perturbative quantization of surface operators in  ${\cal N}=4$ SYM using the disorder point of view.  We use this to identify the Feynman diagrams      that reproduce   the exact computations using   localization. In Appendix \ref{appsec:higherCPO}, we present the   explicit  computation  of the correlation function of a surface operator with a CPO  with $\Delta=4,5$.

\section{Surface Operators in $\CN=4$ SYM} \label{sec:surface operator}

A surface operator ${\cal O}_\Sigma$ in \cite{Gukov:2006jk} induces an Aharonov-Bohm phase for a charged particle linking the surface $\Sigma$ (see \cite{Gukov:2014gja} for a review). This is created by a singular gauge field configuration that breaks the $U(N)$ gauge group to a Levi subgroup $\mathbb L= \prod_{I=1}^M U(N_I)\subset U(N)$ as $\Sigma$ is approached. The singularity in the gauge field is given by
\beq
A=
\begin{pmatrix}
\alpha_1\otimes   \mathds{1}_{N_1}& 0 &\ldots &0\\
0 & \alpha_2\otimes   \mathds{1}_{N_2}&  \ldots & 0\\
\vdots& \vdots & \ddots & \vdots\\
0&0&\ldots& \alpha_M\otimes   \mathds{1}_{N_M}
\end{pmatrix} d\psi\,,
\label{alphamatrix}
\eeq
where $\psi$ is the (local) angle transverse to the surface $\Sigma$. The parameters $\alpha_I\simeq \alpha_I+1 $ are circle valued, taking values on the maximal torus of $U(N)$.  

Since ${\cal O}_\Sigma$ breaks $U(N)$ to $\mathbb L= \prod_{I=1}^M U(N_I)\subset U(N)$ on the surface, two-dimensional  $\theta$-angles can be added on the surface $\Sigma$ for each $U(1)$ factor. These can be combined in an $\mathbb L$-invariant matrix
\beq
\eta=
\begin{pmatrix}
\eta_1\otimes   \mathds{1}_{N_1}& 0 &\ldots &0\\
0 & \eta_2\otimes   \mathds{1}_{N_2}&  \ldots & 0\\
\vdots& \vdots & \ddots & \vdots\\
0&0&\ldots& \eta_M\otimes   \mathds{1}_{N_M}
\end{pmatrix}\,.
\eeq
The parameters $\eta_I\simeq \eta_I+1 $ are also circle valued, taking values on the maximal torus of the
$S$-dual or Langlands dual gauge group  $^LG$, where  $^LG=G=U(N)$.
   
For a maximally supersymmetric surface operator  ${\cal O}_\Sigma$ in ${\cal N}=4$ SYM,\footnote{A different class of maximally supersymmetric surface operators was  constructed in \cite{Buchbinder:2007ar}.} which requires  $\Sigma=\mathbb R^2$ or $S^2$,  ${\cal O}_\Sigma$ induces a codimension two, $\mathbb L$-invariant   singularity in a complex scalar field $\Phi$
\beq \label{eq:scalarbackground}
\Phi={1\over \sqrt{2}z}
\begin{pmatrix}
\beta_1+i \gamma_1\otimes   \mathds{1}_{N_1}& 0 &\ldots &0\\
0 & \beta_2+i\gamma_2\otimes   \mathds{1}_{N_2}&  \ldots & 0\\
\vdots& \vdots & \ddots & \vdots\\
0&0&\ldots& \beta_M+i \gamma_M\otimes   \mathds{1}_{N_M}
\end{pmatrix}\,,
\eeq
  where $z=re^{i\psi}$ is a complex coordinate in the (local) transverse plane to  $\Sigma$ and $\Phi={1\over \sqrt{2}}(\phi^5+i\phi^6)$ is a complex scalar made from a choice of 2 out of 6 scalars\footnote{In this section we follow the standard notation that   scalars in $\CN=4$ SYM are labeled by $\phi^1,\dots,\phi^6$.}. The parameters $\beta_I$ and $\gamma_I$  take values in the Cartan subalgebra of $U(N)$ and are noncompact. 

${\cal O}_\Sigma$  is thus labeled by a Levi group $\mathbb L= \prod_{I=1}^M U(N_I)$ and $4M$ continuous parameters 
\beq
(\alpha_I,\beta_I,\gamma_I,\eta_I)\,,
\eeq
whose physical interpretation we now elucidate. ${\cal O}_\Sigma$ preserves half of the supercharges  of ${\cal N}=4$ SYM. Specifically, it preserves $PSU(1,1|2)\times PSU(1,1|2)\times U(1)\subset PSU(2,2|4)$. The $4M$   parameters $(\alpha_I,\beta_I,\gamma_I,\eta_I)$  are   couplings for exactly marginal local operators on the surface defect that when integrated over $\Sigma$  preserve  $PSU(1,1|2)\times PSU(1,1|2)\times U(1)$.\footnote{In spite of the absence of   defect currents, one can borrow the argument in \cite{Gomis:2016sab} to show that the conformal manifold of $PSU(1,1|2)\times PSU(1,1|2)\times U(1)$ invariant defect deformations is locally of the form ${SO(4M)\over SO(4)\times SO(M)}$. }  Explicitly, the corresponding defect operators are:
\beq
\begin{aligned}
\alpha_I:&\qquad \text{Tr} F^I_{z\bar z}\\
\beta_I:&\qquad \text{Tr} \left( D_{z} \bar \Phi^I+D_{\bar z} \Phi^I\right)\\
\gamma_I:&\qquad i \, \text{Tr}\left(  D_{z} \bar \Phi^I-D_{\bar z} \Phi^I\right)\\
\eta_I:&\qquad \text{Tr} F^I_{w\bar w}\,,
\end{aligned}
\eeq
where $I=1,\ldots,M$ and
$w$ is a holomorphic coordinate in $\Sigma$. These operators are the top components of  two-dimensional  $PSU(1,1|2)\times PSU(1,1|2)\times U(1)$ short multiplets.

Alternatively,  a surface operator ${\cal O}_\Sigma$ can be defined by coupling a $2d$ ${\cal N}=(4,4)$   theory supported on $\Sigma$ to $4d$ ${\cal N}=4$  SYM \cite{Gukov:2006jk,Gukov:2008sn}. The coupling is canonical and is obtained by gauging -- in a supersymmetric fashion -- the global symmetry of the $2d$ ${\cal N}=(4,4)$     theory with the $4d$ ${\cal N}=4$  SYM gauge group. Integrating out the localized $2d$ fields induces the aforementioned singular field configurations on the ${\cal N}=4$  SYM fields. This presentation of surface operators will  be elaborated on    in the next section where we    compute the expectation value of the surface operator $\langle {\cal O}_{\Sigma} \rangle$.

In this paper we probe   ${\cal O}_\Sigma$ with local operators and Wilson loop operators. The position dependence of these correlation functions is fixed by conformal invariance. Our goal     is to determine    the  dependence of these correlators on the Levi group $\mathbb L$, the   parameters $(\alpha_I,\beta_I,\gamma_I,\eta_I)$ and the Yang-Mills coupling $g^2$.  

\vfill\eject

We   probe ${\cal O}_\Sigma$ with chiral primary operators ${\cal O}_\Delta^A$, which transform  in the $(0,\Delta,0)$ representation of the $SU(4)_R$ symmetry. The operators (in the planar normalization) are given by
\beq
{\cal O}_\Delta^A={(8\pi^2)^{\Delta/2}\over \lambda^{\Delta/2} \sqrt{\Delta}}C^{A}_{i_1\ldots i_\Delta}\text{Tr}\left(\phi^{i_1}\ldots \phi^{i_\Delta}\right)\,,
\eeq
where $Y^A= C^{A}_{i_1\ldots i_\Delta}x^{i_1}\ldots x^{i_\Delta}$ are $SO(6)$ spherical harmonics and $\lambda=g^2N$. Since ${\cal O}_\Sigma$ preserves $SO(4)_R\subset SU(4)_R$, the correlator with a chiral primary is nontrivial only if it is an $SO(4)_R$ singlet. For fixed scaling dimension  $\Delta$, there are $\Delta+1$ $SO(4)_R$ invariant operators ${\cal O}_{\Delta,k}$ labeled by their $U(1)_R$ charge $k=-\Delta,-\Delta+2,\ldots, \Delta-2,\Delta$ under the commutant of $SO(4)_R$ in $SU(4)_R$. The relevant operators that couple  to ${\cal O}_\Sigma$ are thus
\beq
{\cal O}_{\Delta,k}={(8\pi^2)^{\Delta/2}\over \lambda^{\Delta/2} \sqrt{\Delta}}C^{\Delta,k}_{i_1\ldots i_\Delta}\text{Tr}\left(\phi^{i_1}\ldots \phi^{i_\Delta}\right)\,,
\eeq
where $Y^A= C^{\Delta,k}_{i_1\ldots i_\Delta}x^{i_1}\ldots x^{i_\Delta}$ are $SO(4)$ invariant  spherical harmonics. For concreteness, we explicitly write the first few operators here \cite{Drukker:2008wr} (see Appendix \ref{appsec:CPO})
\beq \label{eq:O23}
\begin{aligned}
    {\cal O}_{2,0}&={4\pi^2\over \sqrt{6}\lambda}
\hbox{Tr}\left(4\Phi\bar{\Phi}-\sum_{I=1}^4\phi^I\phi^I\right);\hskip.65in
{\cal O}_{2,2}={8\pi^2\over \sqrt{2}\lambda}\hbox{Tr}\left(\Phi^2\right);
\cr
{\cal O}_{3,1}&={8\pi^3\over \lambda^{3/2}}
\hbox{Tr}\left(2\Phi^2\bar{\Phi}-\Phi\sum_{I=1}^4\phi^I\phi^I\right);\qquad
{\cal O}_{3,3}={32\pi^3\over\sqrt{6}\lambda^{3/2}}
\hbox{Tr}\left(\Phi^3\right)\,,
\end{aligned}
\eeq
with 
 ${\cal O}_{\Delta,-k}={\bar {\cal O}}_{\Delta,k}$.

The nontrivial correlators in $\mathbb R^4$ are therefore
\beq \label{eq:CPOR4}
\langle {\cal O}_\Sigma \cdot  {\cal O}_{\Delta,k}(x)\rangle= {C_{\Delta,k}\over z^{{\Delta+k}\over 2}
\bar z^{{\Delta-k}\over 2}}\,,
\eeq
with 
\beq
z=
\begin{cases}
x_3+ix_4~\qquad &\Sigma=\mathbb R^2\\[+4pt]
-{1\ov2R}({x_1^2+x_2^2+x_3^2+(x_4-iR)^2})~\qquad &\Sigma=S^2
\end{cases}\,,
\eeq
where   the planar defect is at $x_3=x_4=0$ and
$R$ is the radius of $S^2$, defined by $\sum_{i=1}^3 x_i^2=R^2$ and $x_4=0$. $U(1)_R$ symmetry implies that $C_{\Delta,-k}={\bar C}_{\Delta,k}$.

An important operator   to probe ${\cal O}_\Sigma$ with is the energy-momentum tensor $T_{mn}$. This defines the scaling weight $h$ of   surface operator \cite{Gomis:2007fi}, extending the notion of scaling dimension of a local operator to  defects \cite{Kapustin:2005py}.  The position dependence is fixed by conformal symmetry.  For $\Sigma=\mathbb R^2$ it takes the form    
\beq
{\langle T_{\mu\nu}\cdot{\cal O}_\Sigma\rangle\over\langle{\cal O}_\Sigma \rangle}
=h{\eta_{\mu\nu}\over |z|^4},
\qquad
{\langle T_{ij}\cdot{\cal O}_\Sigma\rangle\over \langle{\cal O}_\Sigma\rangle}
={h\over |z|^4}\left[{4n_in_j-3\delta_{ij}}\right],
\qquad
{\langle T_{\mu i} \cdot{\cal O}_\Sigma\rangle\over \langle{\cal O}_\Sigma\rangle}=0\,,
\label{hvalue}
\eeq
  where $x^\mu$ are coordinates along
$\Sigma=\mathbb R^2$ and $n^i=x^i/|z|$ is the unit normal vector to  $\Sigma$.    $h$ is related by a Ward identity\footnote{This follows by adapting the proof in \cite{Gomis:2008qa} for Wilson loops.} to the correlator with the chiral primary operator ${\cal O}_{2,0}$
\beq
h=-{N\over 2\pi^2\sqrt{6}}\, C_{2,0}\,.
\eeq
This relation is valid for arbitrary $g^2$, ${\mathbb L}$ and $(\alpha_I,\beta_I,\gamma_I,\eta_I)$.

\section{Surface Operator Expectation Value}
\label{sec:surfacevev}

In this section we compute the exact expectation value $\langle {\cal O}_{S^2}\rangle_{S^4}$ of  spherical surface operators ${\cal O}_{S^2}$  in ${\cal N}=4$ SYM on the four-sphere $S^4$. Before delving into  the detailed computation, we   discuss  expectations that stem solely from   conformal anomalies.

Conformally invariant surface operators  have intrinsic  Weyl anomalies \cite{Graham:1999pm}. A cohomological analysis  implies that there are three parity even Weyl anomalies: $b,c_1$ and $c_2$ \cite{Schwimmer:2008yh}. Given an arbitrary surface $\Sigma$ embedded on  a   manifold with metric $g_{mn}$, one finds that under a Weyl transformation with parameter $\delta\sigma$
\beq 
\delta_\sigma\log \langle {\cal O}_\Sigma \rangle= {1\over 24 \pi}\int d^2x \sqrt{h}\delta\sigma\left(b R_{\Sigma}+c_1g_{mn}  h^{\mu\sigma}h^{\nu\rho} \hat K^m_{\mu\nu} \hat K^n_{\rho\sigma}-c_2 W_{\mu\nu\rho\sigma}h^{\mu\rho}h^{\nu\sigma}\right)\,,
\label{Weylanom}
\eeq
where $h_{\mu\nu}$, $\hat K^m_{\mu\nu}$ and $W_{\mu\nu\rho\sigma}$ are the induced metric on $\Sigma$, the traceless part of extrinsic curvature and pullback of Weyl tensor respectively. The  scaling weight $h$ defined in  (\ref{hvalue}) can be written in terms of surface Weyl  anomalies, as mentioned in \cite{Drukker:2009sf}, and shown in \cite{Lewkowycz:2014jia,Bianchi:2015liz}
\beq
h=-{c_2\over 36\pi^2}\,,
\eeq
while $c_1$ is related to the two-point function of the displacement operator \cite{Lewkowycz:2014jia,Bianchi:2015liz}
\beq
c_1={3\pi^2\over 4} C_D\,.
\eeq
In   $\mathbb R^4$,   the geometric invariants (\ref{Weylanom}) vanish for $\Sigma=\mathbb R^2$. For  $\Sigma=S^2\subset \mathbb R^4$ or a maximal $S^2\subset S^4$,  
  the first invariant   is nontrivial while the other two vanish. This implies, upon integrating the anomaly, that the expectation value of ${\cal O}_{S^2}$   depends on the  radius $r$ of $S^2$ via\footnote{We discuss the vev of  surface operators on $S^2\subset S^4$ momentarily.}
\beq
\langle {\cal O}_{S^2} \rangle_{\mathbb R^4} \propto \left({r\over r_0}\right)^{b/3}\,,
\eeq
where $r_0$ is a scheme dependent scale. This means that, in the absence of   other symmetries,   the finite part of $\langle {\cal O}_{S^2} \rangle$  is scheme dependent.  $b$ is, however, unambiguous and physical.
Indeed, $b$ is monotically decreasing in  renormalization group flows triggered by  surface defect local operators  \cite{Jensen:2015swa,Wang:2020xkc,Casini:2022bsu}. 

Thus far, our discussion of surface operator vacuum expectation  values  (vev)  holds for a  generic CFT, that is, for CFTs with no additional spacetime symmetries.    We 
now discuss  the  salient new features obeyed by  the vev of  supersymmetric surface operators in ${\cal N}=4$ SYM. Actually, the discussion applies to the vev of any  superconformal surface defect in a $4d$ ${\cal N}=2$ SCFT. More precisely the surface defect must preserve  a $2d$ ${\cal N}=(2,2)$ superconformal algebra.

Our discussion builds on the  work on the $S^2$ partition function of $2d$ ${\cal N}=(2,2)$ SCFTs \cite{Doroud:2012xw,Benini:2012ui,Doroud:2013pka}.
The dependence of  the $S^2$ partition function of a  generic $2d$ SCFT on exactly marginal couplings $\lambda$ is scheme dependent, as it can be shifted by the local counterterm $\int d^2x \sqrt{g} f(\lambda) R$. If the CFT is an ${\cal N}=(2,2)$ SCFT,     the counterterms must be supersymmetrized, and the ambiguity of the $S^2$ partition function is drastically reduced from an arbitrary function of couplings to a    K\"ahler transformation \cite{Gomis:2012wy, Gerchkovitz:2014gta,Gomis:2015yaa}
\beq
Z_{S^2} \longrightarrow e^{f(t)+\bar f(\bar t)}Z_{S^2} \,,
\label{vevsurfa}
\eeq
where $t$ are exactly marginal couplings that are bottom components of  background  $2d$ ${\cal N}=(2,2)$ multiplets. Depending on the $SU(2|1)_A$ or $SU(2|1)_B$  ``massive subalgebra" \cite{Doroud:2012xw,Doroud:2013pka} preserved in computing the partition function, $t$ is be either a background twisted chiral multiplet or chiral multiplet. Furthermore, the $S^4$ partition function of a $4d$ ${\cal N}=2$ SCFT is also subject to K\"ahler transformation ambiguities of the exactly marginal couplings $\tau$ \cite{Gomis:2014woa,Gomis:2015yaa}
\beq
Z_{S^4} \longrightarrow e^{f(\tau)+\bar f(\bar \tau)}Z_{S^4} \,,
\label{vevsurfab}
\eeq
where,  for ${\cal N}=4$ SYM,  $\tau$ is the complexified coupling constant (\ref{complexco}).
  This implies that the expectation value of an spherical  surface operator in $S^4$ that preserves ${\cal N}=(2,2)$ superconformal symmetry is well-defined up to K\"ahler transformations
\beq
\langle {\cal O}_{S^2}\rangle_{S^4}\longrightarrow e^{f(t,\tau)+\bar f(\bar t,\bar \tau)}\langle {\cal O}_{S^2}\rangle_{S^4}\,. 
\label{vevsurf}
\eeq

Therefore the  vev of a surface operator supported on a maximal $S^2\subset S^4$  preserving $2d$ ${\cal N}=(2,2)$ superconformal symmetry takes the form
\beq
\langle {\cal O}_{S^2}\rangle_{S^4} = \left({r\over r_0}  \right)^{-4a+b/3} F(t,\tau,\bar t,\bar \tau)\,,
\eeq
and is subject to the K\"ahler transformation  (\ref{vevsurf}).
 $b$ is the  the Euler     surface anomaly we discussed earlier and $a$ is the celebrated   $4d$ Euler       anomaly, which also decreases along renormalization group flows \cite{Komargodski:2011vj}.

In the interest of readers mainly concerned with the  physics of $\langle {\cal O}_{S^2}\rangle_{S^4}$  we present the result now and show the detailed computation below. 
Explicit computation  using $SU(2|1)_A$ supersymmetric localization \cite{Doroud:2012xw,Benini:2012ui} finds\footnote{
Since the overall numerical coefficient is ambiguous we omit it for readability.}  
\beq
\langle {\cal O}_{S^2}\rangle_{S^4} =\left({r\over r_0}  \right)^{ -\text{dim} \,\mathbb L } \left({g^2\over 4\pi}\right)^{\text{dim} \,\mathbb L/2}e^{-8\pi^2 \alpha^2/g^2}\,,
\label{vevanswer}
\eeq
where $\alpha$ the matrix in (\ref{alphamatrix}). $\langle {\cal O}_{S^2}\rangle_{S^4}$ is one-loop exact, unlike  the correlators we study later.

Using that $a=\text{dim}(G)/4$ for $4d$ ${\cal N}=4$ SYM with gauge group $G$,  (\ref{vevanswer}) implies that the surface operator anomaly $b$ for Levi group $\mathbb L$ is \cite{Jensen:2018rxu,Chalabi:2020iie,Wang:2020xkc}
\beq
b=3\left(\text{dim}\, G -  \text{dim} \,\mathbb L  \right)\,.
\eeq
 See also recent work \cite{Jiang:2024wzs}.

 We now show that $\langle {\cal O}_{S^2}\rangle_{S^4}$    in (\ref{vevanswer}) exhibits  the action of $S$-duality as well as the periodicity of the angular variables $(\alpha,\eta)$ of the surface operator in a nontrivial fashion. Since $\langle {\cal O}_{S^2}\rangle_{S^4}$ is subject to an ambiguity (\ref{vevsurfa}),
 we must prove that $\langle {\cal O}_{S^2}\rangle_{S^4}$       realizes   the desired properties up to a K\"ahler transformation. Here,  $t$          are the  bottom components of   $2d$ ${\cal N}=(2,2)$ background twisted chiral multiplets  \cite{Gaiotto:2009fs} 
 \beq
t=\eta+\tau \alpha\,,
\eeq
 where $\tau$ is the complexified $4d$ coupling constant (\ref{complexco}).

We start by showing the  action of the $PSL(2,\mathbb Z)$ $S$-duality group by looking at its generators $S$ and $T$:\footnote{We thank D. Gaiotto for discussion.}  

\begin{itemize}

    \item $S$: $\tau\rightarrow -1/\tau $, which acts as
\beq
\begin{aligned}
&t\rightarrow  -{t\over \tau}\qquad \bar t\rightarrow  -{\bar t\over \tau}\,.
\end{aligned}
\eeq

$\langle {\cal O}_{S^2}\rangle_{S^4}$ in (\ref{vevanswer}) is invariant  up to the K\"ahler transformation $f=-{i\pi t^2/\tau}+{1\over 2}\,\text{dim}\,\mathbb L \log \tau$.
 
\item $T$: $\tau\rightarrow \tau+1 $, which acts as
\beq
\begin{aligned}
t\rightarrow t+\alpha\qquad \bar t\rightarrow \bar t+\alpha\,.
\end{aligned}
\eeq
$\langle {\cal O}_{S^2}\rangle_{S^4}$ in (\ref{vevanswer}) is invariant.  

\end{itemize}

Periodicity of the surface operator parameters $(\alpha,\eta)$ is also realized:

\begin{itemize}
\item $\alpha\rightarrow \alpha+1$, which acts as
\beq
t\rightarrow t+\tau\qquad \bar t\rightarrow \bar t+\bar \tau\,.
\eeq
$\langle {\cal O}_{S^2}\rangle_{S^4}$ in (\ref{vevanswer}) is   periodic  up to the K\"ahler transformation $f= i\pi (2t+ \tau)$.

\item   $\eta\rightarrow \eta+1$, which acts as
\beq
t\rightarrow t+1\qquad \bar t\rightarrow \bar t+1
\eeq
$\langle {\cal O}_{S^2}\rangle_{S^4}$ in (\ref{vevanswer}) is periodic.

\end{itemize}

\noindent
We now proceed to deriving (\ref{vevanswer}).

\subsection{$\langle {\cal O}_{S^2}\rangle$ from $4d/2d$} \label{sec:4d2d}

\begin{figure}
\centering    \includegraphics[width=1\textwidth]{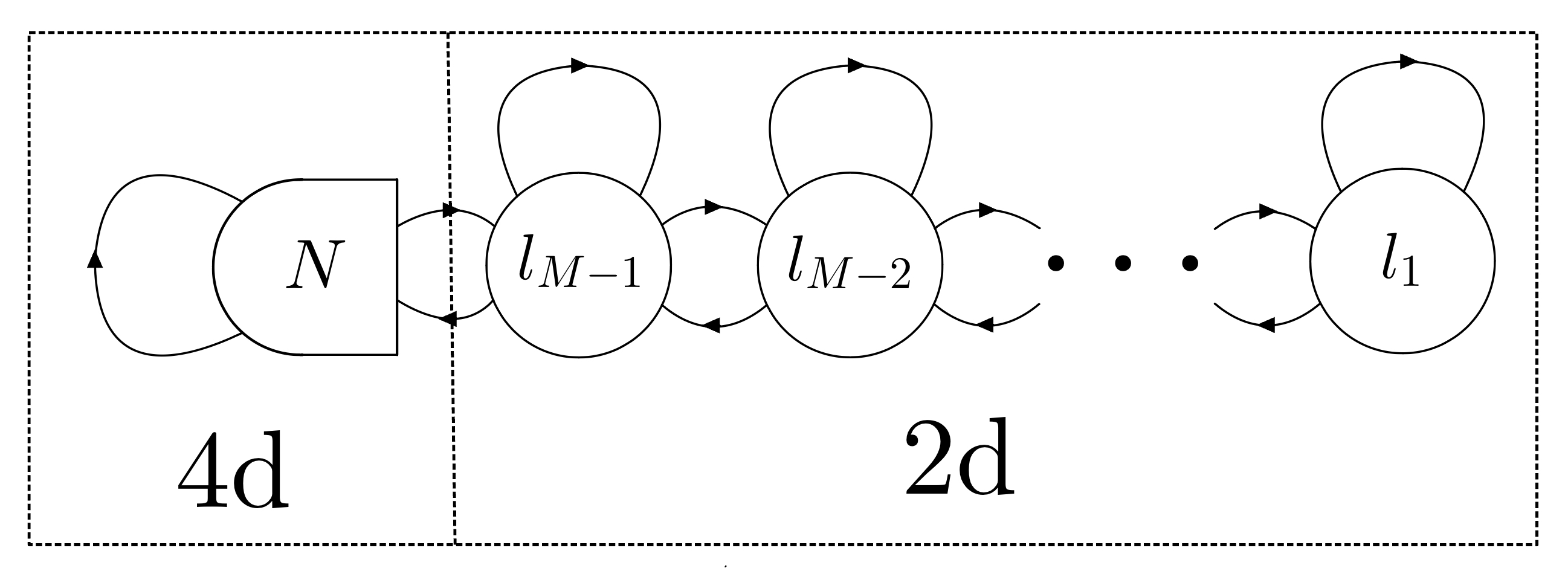}
\caption[]{$4d/2d$ quiver description of ${\cal O}_\Sigma$.\footnotemark}
\label{quiverfig}
\end{figure}

${\cal O}_{S^2}$ can also be defined by coupling a specific  $2d$ ${\cal N}=(4,4)$ quiver gauge theory on $S^2$ to ${\cal N}=4$ SYM  \cite{Gukov:2006jk,Gukov:2008sn}. The $2d$ gauge theory encodes the choice of Levi group $\mathbb L=\prod_{I=1}^M U(N_I)$, where $N=N_1+N_2+\ldots N_M$, in the ranks of the consecutive gauge groups  
\beq
l_i=\sum_{j=1}^iN_j\,.
\eeq
The $2d$ quiver gauge theory  can be found in Figure \ref{quiverfig}. It has ${\cal N}=(4,4)$ supersymmetry, but the quiver is written using ${\cal N}=(2,2)$ multiplets. The Higgs branch of this quiver gauge theory describes the the hyperk\"ahler manifold $T^*(U(N)/\mathbb L)$ in a choice of complex structure. ${\cal N}=4$ SYM is represented in the diagram as an ${\cal N}=2$ vector multiplet coupled to an adjoint hypermultiplet.

The $SU(N)$ flavor symmetry acting on the leftmost   $2d$ chiral multiplets is gauged with the ${\cal N}=4$ SYM gauge group in a way that preserves ${\cal N}=(4,4)$ supersymmetry. This induces a superpotential coupling between the leftmost chiral multiplets  and one of the chiral multiplets inside the $4d$ adjoint ${\cal N}=2$ hypermultiplet that makes up ${\cal N}=4$ SYM. 

\footnotetext{See \cite{Gadde:2013dda}. We use the mixed circle-square notation of $4d$ node in \cite{Gomis:2014eya}.}

The parameters $(\alpha_I,\beta_I,\gamma_I,\eta_I)$
defining ${\cal O}_\Sigma$ appear in this description as ${\cal N}=(4,4)$   FI parameters and $2d$ theta angles. In the  ${\cal N}=(2,2)$ language the parameters $(\alpha_I,\beta_I,\gamma_I,\eta_I)$ organize into background chiral and twisted chiral multiplets. They appear 
 in a  linear superpotential for each adjoint chiral multiplet and twisted superpotential for each $U(1)$  field strength multiplet. 

The computation of $\langle{\cal O}_{S^2}\rangle_{S^4}$ by supersymmetric localization is a very special  convolution of  the $S^4$ partition function of $4d$ ${\cal N}=2$ theories \cite{Pestun:2007rz,Hama:2012bg} and the $SU(2|1)_A$ invariant $S^2$ partition function of $2d$ ${\cal N}=(2,2)$   gauge theories \cite{Doroud:2012xw,Benini:2012ui}. It takes the form  (see also \cite{Lamy-Poirier:2014sea,Nawata:2014nca,Gomis:2016ljm,Pan:2016fbl})
 \beq
\langle{\cal O}_{S^2}\rangle_{S^4}=\int da \sum_B \int d\sigma Z^{4d}_{\text{cl}}(a)Z^{4d}_{\text{1-loop}}(a)Z^{2d}_{\text{cl}}(\sigma)(a)Z^{2d}_{\text{1-loop}}(\sigma,a) |Z_\text{inst,vortex}|^2\,.
\label{combo2d4d}
 \eeq
 $a$ and $(\sigma,B)$   parametrize $4d$ and $2d$ supersymmetric saddle points of the path integral respectively. $B$ is quantized flux on $S^2$, and must be summed over. Then we have the $4d$ and $2d$ classical action and one-loop determinants around the localization locus.
$Z_\text{inst,vortex}$ is the instanton-vortex partition function (see e.g. \cite{Gaiotto:2014ina,Pan:2016fbl,Gorsky:2017hro}). It is determined by the coupling of $4d$ ${\cal N}=4$ SYM to the leftmost chiral multiplets of the $2d$ gauge theory in Figure \ref{quiverfig}. 

We   introduce  an ${\cal N}=(2,2)$ preserving mass deformation of the $4d/2d$ system by turning on a mass deformation $m$ associated to the $U(1)$ flavor symmetry that is present when the theory is viewed as a $4d$ ${\cal N}=2$ gauge theory coupled to a ${\cal N}=(2,2)$ theory. This describes the coupling of  $4d$ ${\cal N}=2^*$ to a $2d$ ${\cal N}=(2,2)^*$ gauge theory. Setting $m=0$ recovers the half-supersymmetric surface operator in ${\cal N}=4$ SYM. 

Let us proceed with the computation by focusing on the $2d$ contribution first. The partition function of the $2d$ gauge theory on $S^2$ for general Levi group $\mathbb L=\prod_{I=1}^MU(N_I)$ is\footnote{We have set the radius of the spheres to $1$, but will reintroduce the scale below.}
\beq \label{eq:Z2d}
Z_{2d}&={1\ov l_1!\, l_2!\dots l_{M-1}!}\hskip-3pt\sum_{\substack{B^{(I)} \\ I=1,\dots, M-1}}
\int \prod_{I=1}^{M-1} \prod_{s=1}^{l_I} {d\sigma_s^{(I)}\ov 2\pi i}e^{-i4\pi \xi^{(I)}  \sigma_s^{(I)}-i\theta^{(I)}B_s^{(I)}}Z_{vec}Z_{adj}Z_{bf}Z_{f/af}\,.
\eeq
$(\xi^{(I)}, \theta^{(I)})$ are FI parameters and $2d$ theta angles that realize  the surface operator parameters $(\alpha^I,\eta^I$) and enter the $4d/2d$ theory in twisted superpotentials.\footnote{The relation is $\theta^{(I)}=\eta^I$ and $\xi^{(I)}={4\pi \alpha^I\over g^2}$.} The localization one-loop determinants are\footnote{The ${\cal N}=(2,2)$ flavor symmetry   that gives rise to   ${\cal N}=(2,2)^*$ is chosen to have charges $(1,0,-1)$ for adjoint, fundamental and antifundamental chirals respectively. It is a symmetry of the superpotential couplings.}
\beq
~&Z_{vec}=\prod_{I=1}^{M-1}\prod_{s<t}^{l_I}\left({(B_{st}^{(I)})^2 \ov 4} +(\sigma_{st}^{(I)})^2\right)
\\& Z_{adj}=\prod_{I=1}^{M-1} \prod_{s\neq t}^{l_I}{\Gamma(1-i\sigma_{st}^{(I)}-{B_{st}^{(I)}\ov 2}-i m)\ov \Gamma(i\sigma_{st}^{(I)}-{B_{st}^{(I)}\ov 2}+i  m)}
\\& Z_{bf}=\prod_{I=1}^{M-2}\prod_{s=1}^{l_I}\prod_{t=1}^{l_{I+1}}{\Gamma\left(-i\sigma_s^{(I)}+i\sigma_t^{(I+1)}-{B_s^{(I)}-B_t^{(I+1)}\ov 2} \right)
\Gamma\left(i\sigma_s^{(I)}-i\sigma_t^{(I+1)}+{B_s^{(I)}-B_t^{(I+1)}\ov 2}+im \right)
\ov
\Gamma\left(1+i\sigma_s^{(I)}-i\sigma_t^{(I+1)}-{B_s^{(I)}-B_t^{(I+1)}\ov 2} \right)
\Gamma\left(1-i\sigma_s^{(I)}+i\sigma_t^{(I+1)}+{B_s^{(I)}-B_t^{(I+1)}\ov 2}-im \right)
}
\\& Z_{f/af}=\prod_{s=1}^{l_{M-1}} \prod_{t=1}^{N} {\Gamma(-i\sigma_s^{M-1}-{B_s^{(N-1)\ov 2}+ia_t})
\Gamma(i\sigma_s^{M-1}+{B_s^{(N-1)\ov 2}-ia_t}+im)
\ov 
\Gamma(1+i\sigma_s^{M-1}-{B_s^{(N-1)\ov 2}-ia_t})
\Gamma(1-i\sigma_s^{M-1}+{B_s^{(N-1)\ov 2}+ia_t}-im)
}\,,
\eeq
where we use the notation $x_{st}\equiv x_s-x_t$. Upon coupling the $2d$ gauge theory to $4d$ ${\cal N}=4$ SYM, the $4d$ Coulomb branch parameters $a$ in the partition function (\ref{combo2d4d}) appears as twisted masses of the leftmost chiral multiplets.
The integration variables $\sigma_s^{(I)}$ run  just below the real axis and the contour is closed along $\text{Im}(\sigma_s^{(I)})<0$. 

 $Z_{2d}$ can be written as a sum over Higgs vacua \cite{Doroud:2012xw,Benini:2012ui}.  This  representation is obtained by summing over the poles
\beq
Z_{2d}=\sum_{v} Z^{2d}_{\text{cl}} (v)\text{res}_{\sigma=v}\left[Z^{2d}_{\text{1-loop}}(\sigma,a)\right] |Z_\text{vortex}|^2\,,
\label{Higgsbranch}
\eeq
where $Z_\text{vortex}$ is the vortex partition function, which is a nonpertubative contribution.

We first focus on 
the poles with $B_s^{(I)}=0$. The poles  with $B_s^{(I)}\neq 0$ contribute to $Z_\text{vortex}$ \cite{Doroud:2012xw,Benini:2012ui}, to which we turn below.
The strategy is to start with the poles of the  leftmost chiral multiples and ``tie-down" the poles of the other chiral multiplets as we move to the right of the quiver.

From $Z_{f/af}$  we have leading poles of $\{\sigma_s^{(M-1)}\}$ which are labeled by $l_{M-1}$-tuples of integers $\vec k^{(M-1)}=(k^{(M-1)}_1,\dots,k^{(M-1)}_{l_{M-1}})$, where $k^{(M-1)}_i=1,\dots,l_M(=N)$,    such that
\beq
\{\sigma_s^{M-1}=a_{k^{(M-1)}_s}\}|_{s=1\dots l_{M-1}}\,. 
\eeq
Only the mutually distinct poles of $\ex{\sigma_s^{(M-1)}}$ contribute to the residue because of the zeros from $Z_{vec}$. Since different tuples of $\vec k^{(M-1)}$ related by permutations give  the same residue, we can remove the  $1/l_{M-1}!$ factor in \eqref{eq:Z2d} and consider $_{l_M} C_{l_{M-1}}$ possibilities of poles. Therefore, we can treat tuples as a set so that $k^{(M-1)}=\{ k_1^{(M-1)},\dots, k_{l_{M-1}}^{(M-1)}\}$.

Next, we look at $I=M-2$ part of $Z_{bf}$. Similarly, from $I=M-2$ part of $Z_{bf}$ and $Z_{vec}$, the leading poles of $\{\sigma_s^{(M-2)}\}$ are labeled by $l_{M-2}$-tuples without repetitions $\vec k^{(M-2)}=(k^{(M-2)}_1,\dots,k^{(M-2)}_{l_{M-2}})$ such that $\vec k^{(M-2)}$ is a subset of $\vec k^{(M-1)}$. Similarly, we can remove the  $1/l_{M-2}!$ factor by identifying different $\vec k^{(M-2)}$ related by permutations and therefore poles are labeled by a set $k^{(M-2)}=\{ k_1^{(M-2)},\dots, k_{l_{M-2}}^{(M-2)}\}$.

Repeating  these steps to the last $I=1$ node, then we see that poles are labeled by a filtration of sets
\beq
 k^{(1)} \subset  k^{(2)}  \subset \cdots \subset  k^{(M-1)}\subset  \{1,\dots, N\}\,.
\eeq
Therefore total number of sets of poles are given by
\beq
\prod_{I=1}^{M-1}{_{l_{I+1}}}C_{l_{I}}={N!\ov N_1! \cdots N_M!}\,.
\eeq
It is convenient to realize  filtration such that 
\beq
\forall I<J:~ k_a^{(I)}=k_a^{(J)},\quad a=1\dots l_I, \quad I,J=1,\dots, M\,,
\eeq
where we defined $\{k_a^{(M)}\}=\{1,\dots,N\}$ with any assignments of the label $a$ where the above filtration is satisfied.

Now, for each set of poles,  the residues in (\ref{Higgsbranch}) for each multiplet are (up to an irrelevant numerical factor):
\beq
~&\text{res}\, Z^{1-loop}_{vec}=\prod_{I=1}^{M-1}\prod_{s<t}^{l_I}\left(a_{k^{(I)}_s}-a_{k^{(I)}_t}\right)^2=\prod_{I=1}^{M-1}\prod_{s \neq t}^{l_I} {1\ov  \gamma\left(ia_{k^{(I)}_s}-ia_{k^{(I)}_t}\right)}
\\&\text{res}\, Z^{1-loop}_{adj}=\prod_{I=1}^{M-1}\prod_{s\neq t}^{l_I}{1\ov \gamma\left(i a_{k^{(I)}_s}-ia_{k^{(I)}_t}+im\right)} 
\\& 
\text{res}\,Z_{f/af}^{1-loop}=\prod_{s=1}^{l_{M-1}} \prod_{\substack{t=1\\  \quad t\neq  k_s^{(M-1)}}}^{N} \gamma\left(-ia_{k^{(M-1)}_s}+ia_t\right) \gamma\left(ia_{k^{(M-1)}_s}-ia_t+im\right)
\\&
\text{res}\,Z_{bf}^{1-loop}=\prod_{I=1}^{M-2}\prod_{s=1}^{l_I}\prod_{\substack{t=1\\  ~~ k_t^{(I+1)}\neq  k_s^{(I)}}}^{l_{I+1}} \gamma\left(-ia_{k^{(I)}_s}+ia_{k^{(I+1)}_t}\right) \gamma\left(i a_{k^{(I)}_s}-ia_{k_t^{(I+1)}}+im\right)\,,
\eeq
so that $\text{res} \, Z^{2d}_{\text{1-loop}}$ in (\ref{Higgsbranch}) is 
\beq
\text{res} \, Z^{2d}_{\text{1-loop}}=\text{res} \,Z^{1-loop}_{vec}\cdot \text{res} \,Z^{1-loop}_{adj}\cdot \text{res} \,Z^{1-loop}_{f/af}Z^{1-loop}_{bf}\,,
\eeq
where 
\beq
\gamma(x)\equiv {\Gamma(x)\over \Gamma(1-x)}\,.
\eeq
 
This can be  simplified  by embedding $k_a^{(I)}$ into $M$-tuples $k_a$ such that  
\beq
k_a\equiv k_a^{(I)}, \quad  \forall a\leq l_I, \quad I=1\dots M\,,
\eeq
so the final answer is
\beq
\text{res} \, Z^{2d}_{\text{1-loop}}=\prod_{I=1}^{M-1} \prod_{s=1}^{l_M }\prod_{t=l_M+1}^{l_{M+1}} \gamma\left(-i(a_{k_s}-a_{k_t})\right)  \gamma\left(i(a_{k_s}-a_{k_t}+m)\right)\,.
\label{finansloop}
\eeq

Our main interest is in ${\cal N}=4$ SYM, obtained by setting the ${\cal N}=(2,2)$ mass deformation to vanish, that is $m=0$. For $m=0$ the formula (\ref{finansloop}) simplifies to a ratio of Vandermonde determinants. The numerator is the Vandermonde determinant  for the Levi group $\mathbb L=\prod_{I=1}^MU(N_I)$ and the denominator the  $U(N)$ Vandermonde determinant:
\beq
\text{res} \, Z^{2d}_{\text{1-loop}}=\prod_{I=1}^{M-1} \prod_{s=1}^{l_M }\prod_{t=l_M+1}^{l_{M+1}} {1\ov (a_{k_s}-a_{k_t})^2}\equiv {\Delta_{\mathbb L}(a)\ov \Delta_{U(N)}(a)}\,,
\label{ratiovander}
\eeq
where 
\beq
\Delta_{\mathbb L}(a)=\prod_{\bm\alpha>0, \bm\alpha\cdot \alpha=0} (\bm\alpha\cdot a)^2\,,
\eeq
with $\bm\alpha$ the roots of the ${\cal N}=4$ SYM gauge group $G$.

The contribution from the poles that arise from $B_s^{(I)}\neq 0$ build up the vortex partition function $Z_\text{vortex}$ in (\ref{Higgsbranch}). Remarkably, when $m=0$, explicit computation shows that  $Z_\text{vortex}=1$. This  signals the enhancement of supersymmetry and the vanishing contributions in the nontrivial vortex sector is due to   fermion zero modes. Likewise, also  $Z_\text{inst,vortex}=1$ in (\ref{combo2d4d})   due to the fermion zero modes when $m=0$, where supersymmetry is enhanced (see related discussion in \cite{Okuda:2010ke} for  unramified instantons). 

Putting everything together in (\ref{combo2d4d})  we have therefore shown that  
\beq \label{eq:2d4dvev}
\langle{\cal O}_{S^2}\rangle_{S^4}= \int da\,  \Delta_{\mathbb L}(a) e^{-{8\pi^2\over g^2} \left(a^2+2ia \alpha\right)}\,,  
\eeq
where the $U(N)$ Vandermonde in (\ref{ratiovander})   cancels the one-loop determinant of ${\cal N}=4$ SYM. 
The integral can be evaluated by shifting integration variables
\beq
a\rightarrow a+ i \alpha
\eeq
and using that 
\beq
\Delta_{\mathbb L}(a+ i \alpha) =\Delta_{\mathbb L}(a)\,,
\eeq
as a consequence of $\alpha$ being $\mathbb L$-invariant.

Reintroducing the sphere radius dependence by   dimensional analysis, sending $a\rightarrow ra$,  and performing the integral we get   (\ref{vevanswer})\footnote{See similarity with one-loop  vev of 't Hooft loop \cite{Gomis:2009ir}, which, however,  does receive higher loop corrections.}
\beq \label{eq:finalvevS2S4}
\langle {\cal O}_{S^2}\rangle_{S^4}  ={r}^{- \text{dim} \,\mathbb L } \left({g^2\over 4\pi}\right)^{\text{dim} \,\mathbb L/2}e^{-8\pi^2 \alpha^2/g^2}\,.
\eeq

\noindent
In Appendix \ref{app:disorder}, we compute $\langle {\cal O}_{S^2}\rangle_{S^4}$ using the disorder definition   of section \ref{sec:surface operator}.

\section{Surface Operator Correlation Functions Via Localization } \label{sec:2dYMloc}

In this section  we perform a localization computation of non-trivial correlation functions in the presence of surface operators in $\CN=4$ SYM. We initiate by establishing a geometric framework for localization, followed by the computation of correlation functions involving local chiral primary operators and 1/8-BPS Wilson loops. 

\subsection{Geometry of Surface Operator}

We perform supersymmetric localization, which we explain in a moment,
by placing   ${\cal N}=4$ SYM on the conformally flat warped geometry $B^3\times S^1$ using a Weyl transformation, where $B^3$  is a three-dimensional ball. 
This space is described by coordinates $(\tau,\tilde x_i)$ with $i=2,3,4$ where $\tau\simeq \tau +2\pi$ and $\tilde x_i$ is confined as $\tilde x^2\equiv\sum_{i=2}^4 \tilde x_i^2\leq  R$, and the metric is (this geometry is further explained in Appendix \ref{appsec:B3S1})
\beq \label{eq:B3S1 metric}
ds^2(B^3\times S^1)=d\tilde x_i^2+{R^2\ov 4}\left(1-{\tilde x^2\ov R^2} \right)^2d\tau^2 \,.
\eeq

From now on, we omit tilde in $\tilde x_i$ for clarity. This geometry was originally introduced in \cite{Pestun:2009nn} to evaluate correlation functions involving 1/8-BPS Wilson loops living on $S^2\subset \mathbb R^4$ \cite{Drukker:2007dw,Drukker:2007yx,Drukker:2007qr}. After the Weyl transformation,  this $S^2$ is mapped to the boundary of $B^3$, and the localization computation was performed using a   superconformal charge $Q\in PSU(4|4)$. This supercharge  generates the $SU(1|1)$ algebra
\beq \label{eq:Qsquare}
Q^2={2\ov R}( R_{05}-R_{\tau})\,,
\eeq
where $R_{05}$ is an $SO(2)$ R-symmetry generator which rotates the scalars $(\phi_0,\phi_5)$  and $R_\tau$ generates $SO(2)$ rotation along $S^1$ which shifts $\tau$ (see Appendix \ref{appsec:N=4} for our conventions for $\CN=4$ SYM).

The supersymmetry transformation generated by  $Q$ is parametrized by a   conformal Killing spinor. The conformal Killing spinors on $B^3\times S^1$ depend on  two $Spin(10)$ Weyl spinors $\e_s$, $\e_c$ of opposite chirality. The explicit expression of conformal Killing spinors on $B^3\times S^1$ is \cite{Gomis:2011pf}
\beq \label{eq:cksB3S1}
\e=\cos{\tau \ov 2}(\e_s+x_i \tilde{\Gamma}^i  \e_c) +\sin{\tau \ov 2} \tilde{\Gamma}^1(R \e_c+ {x_i \ov R}\Gamma^i \e_s)\,.
\eeq   The choice of  $\e_{s,c}$ corresponding to $Q$ is
\beq \label{eq:locQ}
\e^Q_s=\begin{pmatrix} 1\\ 0\\0\\0
\end{pmatrix} \otimes \begin{pmatrix} 1\\ 0\\0\\0
\end{pmatrix},\quad \e^Q_c={1\ov R}\begin{pmatrix} 0\\ 0\\0\\i
\end{pmatrix} \otimes \begin{pmatrix} 1\\ 0\\0\\0
\end{pmatrix}.
\eeq

Localization of the path integral is then performed by deforming the Euclidean action on $B^3\times S^1$ by $S\rightarrow S+ t Q V$ under the condition that $Q^2V=0$ and operators inserted are invariant under $Q$. A   feature of the choice of $Q$ and $V$ in \cite{Pestun:2009nn} is that upon reducing fields  to the localizing locus $QV=0$, the action becomes a total derivative and hence the path integral reduces to the effective $2d$ theory on $S^2$, the boundary of $B^3$. 

Our main point is that certain correlation functions (namely with local operators and Wilson loops) in the presence of the surface operator ${\cal O}_\Sigma$ can be incorporated into the above localization framework and hence can be computed exactly. We achieve this by utilizing the disorder description of the surface operator as described in section \ref{sec:surface operator}.  

In order to do this, we need to place the surface operator ${\cal O}_\Sigma$ on $B^3\times S^1$ in a way that is compatible with  the localizing supercharge $Q$. Namely, we need to insert the surface operator in such a way that  it is annihilated by   $Q$. 

\begin{figure}
\centering    \includegraphics[width=0.8\textwidth]{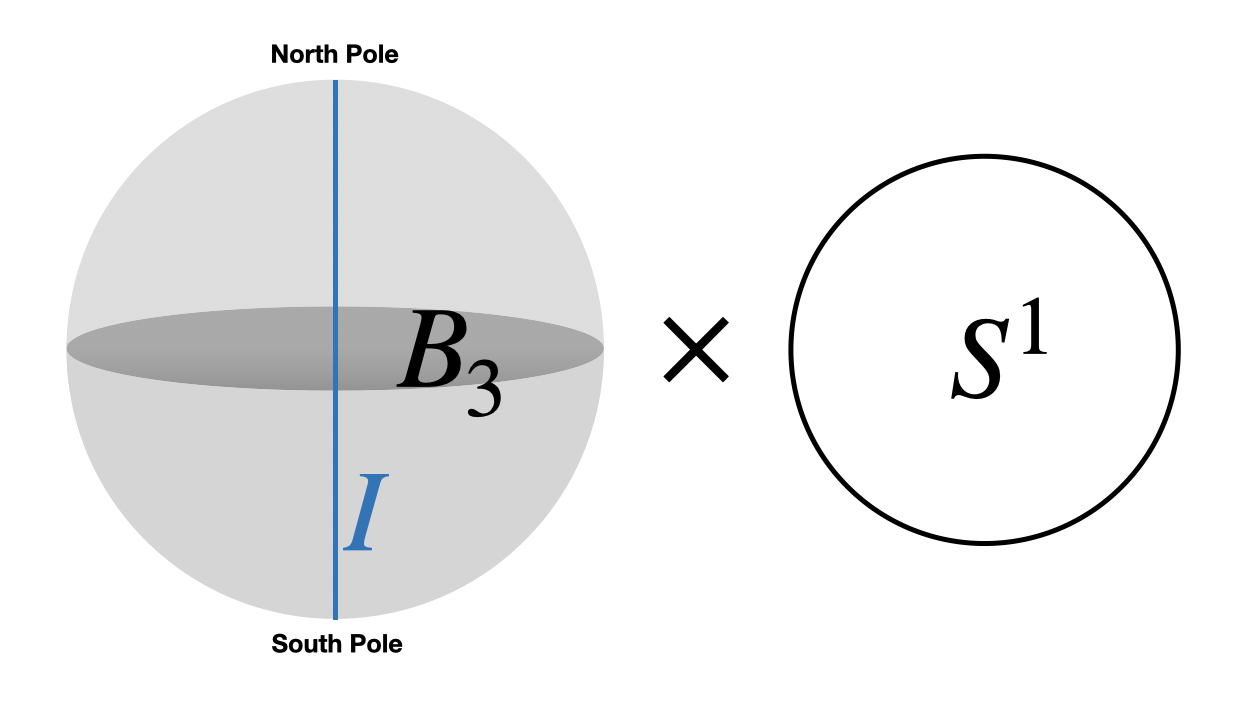}
\caption{The $B^3\times S^1$  geometry. The surface operator is supported on $I\times S^1$.} \label{figure:B3S1}
\end{figure}

We find that inserting the surface operator ${\cal O}_\Sigma$  on $I\times S^1\subset B^3\times S^1$ satisfies the above requirements, where $I$ is any axis of rotation of $B^3$ (see also \cite{Wang:2020seq}). Using the $SO(3)$ isometry of $B^3$, we  choose the surface $\Sigma$ to be
\beq
\Sigma=I \times S^1:~x_3=x_4=0, \quad -R\leq x_2\leq R, \quad 0\leq \tau\leq 2\pi,
\eeq
where   $I$ is an interval   passing through the North and South poles of $B^3$ (see figure \ref{figure:B3S1}). The surface is   topologically $\Sigma\simeq S^2$, due to the warp factor vanishing at N and S poles (see however discussion at end of section \ref{sec: localization}).

The surface operator ${\cal O}_\Sigma$ supported on this surface generates the following  Q-invariant singular field configurations on $B^3\times S^1$  
\beq \label{eq:B3Sqprofile}
\Phi= \phi_w={\be+i\gamma \ov \sqrt{2}(x_3+ix_4)},\quad \qquad A=\al d\psi ,\quad 
\eeq
where $\phi_w={\phi^7-i\phi^8\ov \sqrt{2}}$ and $\psi=\text{arg}(x_3+ix_4)$.\footnote{Here we   choose the opposite orientation (up to $SO(4)$ rotations) for $\Phi$ as compared to \eqref{eq:scalarbackground}. This is necessary for the background being $Q$-invariant, and does not produce any qualitative differences. \label{ft:orientation} } 
This is a  simple consequence of the Weyl transformation between $B^3\times S^1$ and $\mathbb R^4$ (see Appendix \ref{appsec:B3S1}).

The  conformal Killing spinors preserved by the background (\ref{eq:B3Sqprofile}),  which are solutions to the   BPS equations $\del_\e \psi=0$, are   characterized by
\beq \label{eq:B3S1halfBPS}
\Gamma^{zw}\e_s=0,\quad \tilde \Gamma^{zw}\e_c=0\,,
\eeq
where $z=x_3+ix_4$.
This includes the spinor \eqref{eq:locQ} which parameterizes the localizing supercharge, and thus the  insertion    ${\cal O}_\Sigma$  is $Q$-invariant.

\subsection{Localization Onto $2d$ Yang-Mills: Review} \label{eq:Pestun review}

In this section  we review how the localization of $\CN=4$ SYM on $B^3\times S^1$ reduces to $2d$ Yang-Mills on $S^2=\pa B^3$ in the zero-instanton sector \cite{Pestun:2009nn}, which was anticipated in \cite{Drukker:2007yx,Drukker:2007qr}. Throughout the analysis, we set the radius of $B^3$ to be $R=1$, and restore $R$ dependence when it is necessary.

The localization is achieved by deforming the Euclidean action $S\rightarrow S+t QV$ with some fermionic field $V$ (see \cite{Choi:2021yuz} and references therein for an introduction to localization). The choice of fermionic field is given by $V=(\lam,\bar{Q \lam})$ where we have the $Spin(10)$ invariant pairing $(\,,
):S^+\otimes S^-\rightarrow \mathbb C$ given by $(\chi,\psi)=\sum_{\al=1}^{16} \chi_\al \psi_\al$. Therefore the deformation term $Q V=(Q\lam,\bar {Q\lam})$ is positive semi-definite and the localization in the limit of large $t$ makes path integral localize  onto the solution of the BPS equations $Q\lam=0$.

The BPS equations consist  of 16 complex (32 real)
equations, and we can split $Q\lam$ into top 8 components $Q\lam^t$ and bottom 8 components $Q\lam^b$ according to the eigenvalues of  $-i\Gamma^0 \Gamma^1$. Seven auxiliary fields $K_I$ (see \cite{Berkovits:1993hx,Pestun:2007rz}) only appear in top equations and we can remove auxiliary fields and remain with 9 real equations from top components where only 8 equations are independent. Then the bottom 16 real equations together with one top equation can be simplified to a constraint that all the bosonic fields be constant along $S^1$ up to a gauge transformation, i.e.
\beq \label{eq:circleinv}
\pa_\tau A_\mu=0, \quad \pa_\tau \phi_I=0 \quad \text{(up to gauge transformation)},
\eeq
which is also natural from the fact that $Q^2$ generates translation along $\tau$ \eqref{eq:Qsquare}. Hence we can effectively focus on the $\tau=0$ slice by assuming all the fields are invariant along $S^1$.

The remaining seven real equations in $Q\lam^t$ after imposing \eqref{eq:circleinv} at $\tau=0$ and resolving the auxiliary fields are given by
\beq \label{eq:seveneqs}
\text{Re}Q\lam^t|_{e_k
}=&F_{9k}(1-x^2)-{1\ov 2}F_{ij}\e_{ijk}(1+x^2)+{1\ov 2}F_{i+4\, j+4}\e_{ijp}(\del_{pk}-x^2\del_{pk}+2x_px_k)
\\&-F_{5\, j+4}(\del_{jk}+x^2\del_{jk}-2x_jx_k)+2\phi_9x_k
\\ \text{Re}Q\lam^t|_{e_{k+4}
}=&F_{9\,i+4}(\del_{ik}+x_i x_k-x^2\del_{ik})-F_{i5}(\del_{ik}-x_ix_k +x^2\del_{ik})+2\phi_5(1-x^2)^{-1}x_k
\\& +F_{i\,j+4}(\ep_{ijk}-x_ix_p\e_{jpk}-x_jx_p\e_{ipk})-2\phi_{i+4}\e_{ijk}x_j\e_{k+4}
\\ \text{Re}Q\lam^t|_{e_5
}=&F_{9k}(1-x^2)-F_{i\,j+4}(\del_{ij}+\del_{ij}x^2-2x_ix_j)-2\phi_{j+4}x_j,
\eeq
where $|_{e_n}$ means n-th octonion component  and  $i,j,k,p=2,3,4$ are space indices on $B^3$ (see Appendix \ref{appsec:N=4}).

By looking at the linearized equations of \eqref{eq:seveneqs} around the trivial solution and considering the full equations, it was argued in \cite{Pestun:2009nn} that $\Phi_5=\Phi_9=0$ is a solution which is regular at the $S^2$ boundary. Together with the $S^1$ invariance on the Euclidean ${\cal N}=4$ SYM action, then the bosonic part of the action can be decomposed into a sum of squares of top components BPS equations and a total derivative as\footnote{Throughout the paper, we use a convention that the bilinear form $(\cdot,\cdot)$ on $\mathfrak g$ is implicit and to be $(a,b)=-\Tr(ab)$ where $\Tr$ is a trace in the fundamental representation. \label{ft:bilinear}}
\beq
S\,\stackrel{loc}{=}\,\,&{2\pi\ov 2g^2}\int_{B^3}  d^3 x \, \left[ a_n \text{Re}(Q\lam^t|_{e_n})^2+b_n \text{Im}(Q\lam^t|_{e_n}x)^2 \right]
\\&+{2\pi \ov 2g^2}\int_{B^3} d^3 x \Big[-\nabla_i ((1-x^2)(\phi_{i+4}\nabla_j \phi_{j+4}-\phi_{j+4}\nabla_j \phi_{i+4})) 
\\&\,\quad \quad\quad\quad \quad\quad+4\nabla_j(x_ix_k \phi_{k+4}\nabla_i \phi_{j+4}-x_ix_j\phi_i \nabla_k \phi_{k+4})
\\&\,\quad \quad\quad\quad \quad\quad+6\nabla_j (x_i \phi_{i+4}\phi_{j+4})\Big],
\eeq
where the  coefficients $a_n$, $b_n$ can be found in \cite{Pestun:2009nn}. Therefore, using Stokes theorem, the localized action becomes an effective two-dimensional action on $S^2=\pa B^3$
\beq \label{eq:S2d}
S_{2d}={\pi \ov g^2}\int_{S^2}d\Omega \left[4\phi_n(\nabla_n\phi_n-\nabla_i \phi_{i+4})+6\phi_n^2\right],
\eeq
where we defined 
\beq
\phi_n\equiv n_i \phi_{i+4},
\eeq
with $n_i=x_i/|x_i|$. Now the $e_5$ part of the localizing equations \eqref{eq:seveneqs} on $S^2$ can be simplified to
\beq \label{eq:e5}
\nabla_n\phi_n-\nabla_i \phi_{i+4}=-\phi_n,
\eeq
and hence the localized action is reduced to 
\beq
S_{2d}={2\pi \ov g^2}\int_{S^2}d\Omega \,\phi_n^2.
\eeq

Remarkably, it can be further massaged so that the localizing action can be written as a $2d$ YM theory in the zero-instanton sector. To explain this, let us first introduce a one-form and a connection on $B^3$
\beq
\bm{\phi}\equiv \phi_{i+4}dx^i, \quad  A=A_idx^i.
\eeq
Now from the natural inclusion map $\iota:S^2\rightarrow B^3$, we introduce the pullback of the one-form $\bm \phi_t=\iota^* \bm \phi\in \Omega^1(S^2)$ and similarly $ A_t$. Then there is an identity
\beq
d_{A}^{*2d} \bm \phi_t=\nabla_i \phi_{i+4}-\nabla_n \phi_n-2\phi_n,
\eeq
which together with \eqref{eq:e5} gives
\beq
S_{2d}={2\pi\ov g^2}\int_{S^2}d\Omega (d^{*2d}_{ A}\bm \phi_t)^2.
\eeq
Now we define a complex connection $\bm A_{\mathbb C}$ on $S^2$ such that
\beq
\bm A_{\mathbb C}= A_t-i*_{2d}\bm\phi_t\equiv  A-i*\bm \phi,
\eeq
where from now on we omit the subscript in $ A_t$, $\bm \phi_t$ and $*_{2d}$ by understanding that we are working strictly on $S^2$. Now the boundary fields are purely constrained by the tangential part of the localizing equations in \eqref{eq:seveneqs} which are
\beq \label{eq:2dconstraints}
~&   F_{ A}-\bm \phi\wedge \bm\phi=0
,\quad   d_{ A} \bm \phi =0.
\eeq
Then we see that 
\beq 
\bm F_{\mathbb C}= F_{ A}-\bm \phi \wedge \bm \phi-id_{ A}*\bm\phi.
\eeq
Therefore, we see that the $2d$ action becomes a complexified YM action (with an important overall minus sign )
\beq
S_{2d}=-{1 \ov g^2_{2d^2}}\int_{S^2}d\Omega \,(* \bm F_{\mathbb C})^2,
\eeq
which obeys the two constraints \eqref{eq:2dconstraints} and the two-dimensional gauge coupling is identified as 
\beq \label{eq:g2d}
g^2_{2d}=-{g^2\ov 2\pi R^2}.
\eeq
Notice that the two constraints \eqref{eq:2dconstraints} are two elements of Hitchin equations on $S^2$. In fact, it is possible to formulate $2d$ action in terms of constrained Hitchin/Higgs-Yang-Mills theory.

To do that, we briefly review the relevant aspects Hitchin/Higgs-Yang-Mills system following \cite{Kapustin:2006pk}. The field  content is given by a gauge field $A$ under compact Lie group $G$ and one form $\phi$ in adjoint representation on a Riemann surface $X$. Then the space of fields $\mathcal W=(A,\phi)$ can be regarded as an infinite-dimensional manifold with a natural flat metric induced from $X$
\beq
g(\del A_1,\del \phi_1;\del A_2, \del\phi_2)={1\ov 4\pi}\int \del A_1\wedge * \del A_2+\del \phi_1\wedge *\del \phi_2.
\eeq
Moreover, $\mathcal W$ is a hyperK\"ahler manifold with three canonical complex structures $I,J,K$ acting on $T^* \CW$ as 
\beq
~&I^t(\del A)=*\del A, \quad I^t(\del \phi)=-*\del \phi
\\& J^t(\del A)=-\del \phi, \quad J^t(\del \phi)=\del A
\\& K^t(\del A)=-*\del \phi, \quad K^t(\del \phi)=-*\del A,
\eeq
which obeys the quaternionic algebra $IJ=-JI=K$, $I^2=J^2=K^2=-1$. The corresponding three natural symplectic structures are
\beq \label{eq:symplectic form}
~&w_I(\del A_1,\del \phi_1;\del A_2, \del\phi_2)={1\ov 2\pi}\int_\Sigma \del A_1\wedge \del A_2-\del   \phi_1\wedge\del \phi_2
\\&w_J(\del A_1,\del \phi_1;\del A_2, \del\phi_2)={1\ov 2\pi}\int_\Sigma \del A_1\wedge *\del \phi_2-\del A_2\wedge *\del \phi_1
\\& w_K(\del A_1,\del \phi_1;\del A_2, \del\phi_2)={1\ov 2\pi}\int_\Sigma \del A_1\wedge \del \phi_2-\del A_2\wedge\del \phi_1.
\eeq
Now there is a natural algebra of vector fields $\mathfrak g_{gauge}$ in $\mathcal W$ generated by infinitesimal gauge transformations. Then are three natural moment maps $\mu: \mathcal W\rightarrow\mathfrak g_{gauge}$ corresponding to above symplectic structures given by
\beq
~&\mu_I(h)=\int (h,F-\phi\wedge \phi)
\\& \mu_J(h)=\int (h,d_A*\phi)
\\& \mu_K(h)=\int (h,d_A \phi),
\eeq
for any element $\phi\in \mathfrak g_{gauge}$.

After this digression, we can write the effective $2d$ path integral as a constrained $2d$ YM theory with gauge group $G_{\mathbb C}$ as follows. In terms of $\bm A_{\mathbb C}=A-i*\bm\phi$ and $\bm F_{\mathbb C}=\mu_I-i\mu_J$, our path integral is\footnote{We note that this form of the path integral is based on the main assumption in \cite{Pestun:2009nn} that the one-loop determinant of the localization is trivial. However, we would expect it to be non-trivial and at least produces a non-trivial powers of $g^2$. See discussions around \eqref{eq:vev2dYM}.} 
\beq \label{eq:cHYM}
\int_{\mu_I=\mu_K=0} D\bm A_{\mathbb C}D\bar{\bm A}_{\mathbb C}\, e^{{1\ov g^2_{2d}}\int_{S^2} d\Omega \bm (*\bm F_{\mathbb C})^2 }.
\eeq
Note that while the action itself has $G_{\mathbb C}$ gauge symmetry, there is only $G$ gauge symmetry because of the constraint $\mu_I=\mu_K=0$. The constraints in terms of $\bm A_{\mathbb C}$ are given by
\beq
\text{Re}(\bm F_{\mathbb C})=0,\quad d_{\text{Re} (\bm A_{\mathbb C})}*\text{Im}(\bm A_{\mathbb C})=0.
\eeq

Now under the assumption that we can analytically continue the integration contour so that $\text{Re}(\bm F_{\mathbb C})=\mu_I=0$ becomes $\text{Im}(\bm F_{\mathbb C})=\mu_J=0$, we get
\beq \label{eq:2dYMpathintegral}
\int_{\mu_J=\mu_K=0} D\bm A_{\mathbb C}D\bar{\bm A}_{\mathbb C}\, e^{-{1\ov g^2_{2d}}\int_{S^2} d\Omega \bm \mu_I^2 }.
\eeq
Of course, the analytic continuation of the contour should be accompanied by the analytic continuation of the coupling $g^2_{2d}$. 

Focusing on $G=U(N)$, we can plausibly assume that for generic sections of $A$ the solutions of the constraints $\mu_J=\mu_K=0$ (i.e. $d_A\phi=d_A^*\phi=0$) are trivial such that $\phi=0$. Then we see that the bosonic part of the localized theory is more or less identical to ordinary $2d$ Yang-Mills theory. The standard non-abelian localization argument \cite{Witten:1992xu} shows that the path integral of $2d$ Yang-Mills on Riemann surface localized on the instantons which are solutions of the equations of motion $d*F=0$. On $S^2$, such instantons are labeled by n-tuples of integers $\{n_1,\dots, n_N\}$ (modulo identifications by the Weyl group). The observation of \cite{Pestun:2009nn} is that on non-trivial instanton profiles, the kernels of $d_A$ and $d_A^*$ are non-trivial which suggests that the constrained $2d$ Yang-Mills integral \eqref{eq:2dYMpathintegral} is subtley different from the ordinary $2d$ Yang-Mills integral. If we additionally assume that the resulting localized path integral \eqref{eq:2dYMpathintegral} is equipped with a symplectic measure \eqref{eq:symplectic form} with respect to complex structure $I$ and associated with symplectic fermions, then there are non-trivial fermionic zero modes around the non-trivial instantons. This conceptually explains that only the zero-instanton sector contributes to the path integral, and effectively reduces the $2d$ path integral to the ordinary $2d$ YM with zero-instanton sector as
\beq \label{eq:2dYM0instanton}
\int DA\, e^{-{1\ov g^2_{2d}}\int_{S^2} d\Omega F^2 }\Big|_{\text{0-instanton}}.
\eeq

\subsection{Localization Onto $2d$ Yang-Mills With Surface Operator} \label{sec: localization}

The aforementioned localization framework has been successfully applied and tested in various generalizations, see e.g. \cite{Giombi:2009ds,Giombi:2009ek,Giombi:2012ep, Giombi:2017cqn,Giombi:2018qox,Wang:2020seq,Dedushenko:2020vgd}. Here we extend this localization framework to the case when a surface operator ${\cal O}_\Sigma$ is inserted. We take the disorder point of view such that in the path integral description we integrate over fields with desired singular behavior near the surface operator, as in section \ref{sec:surface operator}. 

\begin{figure}
\centering    \includegraphics[width=0.4\textwidth]{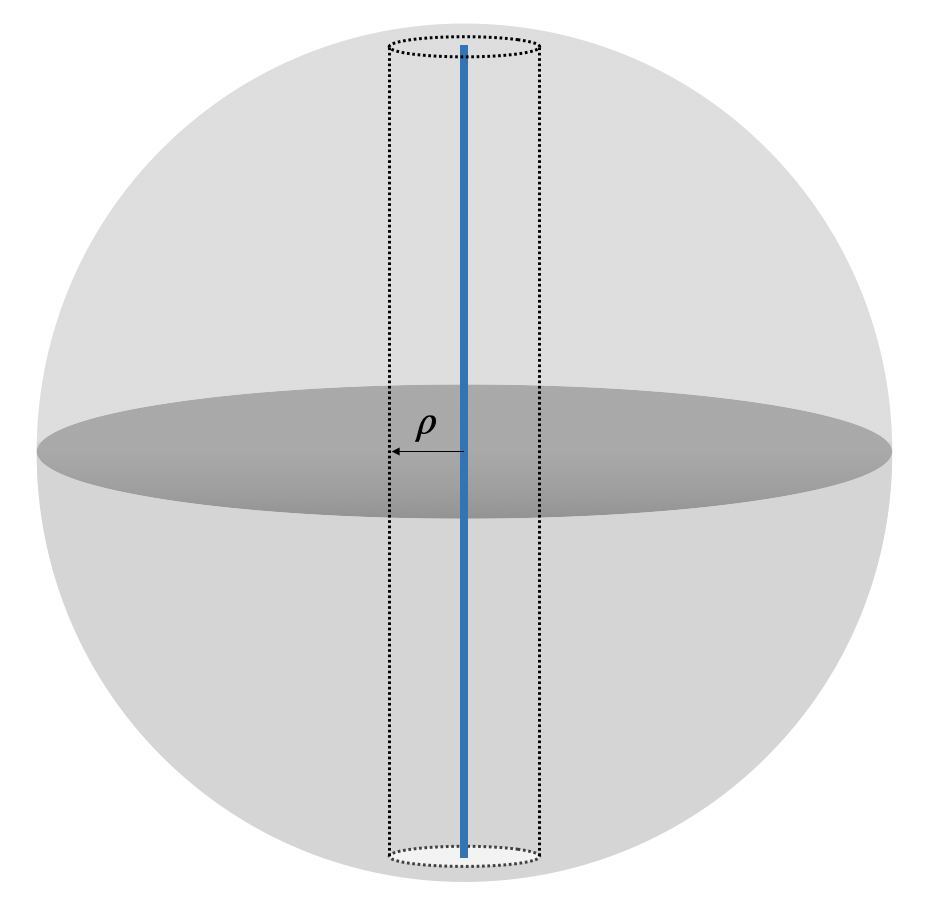}
\caption{The regularized surface operator ${\cal O}_\Sigma$ on $B^3$, where the boundary is an infinitesimal tube with radius $\rho$.} \label{figure:B3S1tube}
\end{figure}

In general, the singular behavior of fields produced by disorder operator generates divergences and therefore a proper regularization is needed. One natural choice of regularization is to excise the infinitesimal tubular neighborhood $T$ around the disorder operator. Therefore, the regulated spacetime $T \backslash M$ contains the boundary $\pa T$ and therefore we need   additional data at the boundary, which are boundary conditions and possibly   boundary terms. In our case of the surface operator in $B^3\times S^1$, we regularize it by introducing an infinitesimal tube-like cut-off at $\pa M: x_3^2+x_4^2=\rho^2$ with $\rho\ll 1$ as in figure \ref{figure:B3S1tube}.

To determine them, there are two fundamental guiding principles for the consistency of theory with a  boundary (see e.g. \cite{Lindstrom:2002mc,Belyaev:2008xk,Gaiotto:2008sa,Hori:2013ika,DiPietro:2015zia,Moore:2015szp,Gava:2016oep,Dedushenko:2018aox,Dedushenko:2018tgx,Hosomichi:2021gxe,Bason:2023bin}). Firstly, the total bulk+boundary action must have a well-defined Euler-Lagrange action principle. Namely, the total bulk+boundary action $S_{tot}$ must obey $\del S_{tot}=0$ under the bulk Euler-Lagrange equations of motions and the boundary conditions. Secondly, total bulk+boundary action and boundary conditions must be supersymmetric. 
In Appendix \ref{appsec:bdry}, we apply these principles to the planar and the spherical surface operator in $\mathbb R^4$.

Based on the successful story of the planar and the spherical surface operators in $\mathbb R^4$ described in the Appendix \ref{appsec:bdry}, we streamline the determination of boundary conditions and the supersymmetric boundary term in $B^3\times S^1$. We introduce labels $i,j,\dots=1,2$,~ $a,b,\dots=3,4$, ~$I,J,\dots=5,6,9,0$,~ $\hat A,\hat B,\dots=7,8$. 

Recalling \eqref{eq:cksB3S1}, the conformal Killing spinors on $B^3\times S^1$ satisfying $\nabla_\mu=\tilde \Gamma_\mu \tilde \e$ are given by
\beq 
\e&=(\cos{\tau \ov 2}+\sin{\tau \ov 2}{x_i\ov R}\tilde\Gamma^1 \Gamma^i )\e_s+(R \sin{\tau \ov 2} \tilde\Gamma^1 +\cos{\tau \ov 2} x_i \tilde\Gamma^i)\e_c
\\ \tilde \e&=-{1 \ov R} \sin{\tau \ov 2} \Gamma^1 \e_s+\cos{\tau \ov 2}\e_c.
\eeq
In the half-BPS sector of conformal Killing spinors satisfying \eqref{eq:B3S1halfBPS}, $\e$ satisfies $\Gamma^{wz}\e=o(\rho)$ at the boundary while $\Gamma^{wz}\tilde \e=0$ holds exactly in the bulk. Since we are going to eventually remove the cutoff by taking the limit $\rho\rightarrow 0$, we see that the situation is similar to the case of the planar surface operator in $\mathbb R^4$ analyzed in Appendix \ref{appsec:bdry}.

Therefore it is natural to impose the boundary condition for the fermions up to $o(\rho)$ as
\beq \label{eq:B3S1fbc}
\Gamma^{w\bar z}\lam=0 \quad  \text{or} \quad (1+\Gamma^{3478})\lam=0.
\eeq
The superpartner of the fermionic boundary condition gives ($P=\Gamma^{3478}$) up to $o(\rho)$ as
\beq
0=\del {P_+\lam}=&F_{12}\Gamma^{12}\e+F_{iI}\Gamma^{iI}
\e
+{1\ov 2}F_{IJ}\Gamma^{IJ}\e
+F_{z\bar z}\Gamma^{z\bar z}\e
+F_{w  \bar w}\Gamma^{w \bar w}\e
\\&+F_{z\bar  w}\Gamma^{z  \bar w}\e
+F_{\bar z  w}\Gamma^{\bar z w}\e
 -2\phi_I \tilde\Gamma^I \tilde \e,
\eeq
which determines the bosonic boundary condition up to $o(\rho)$ to be
\beq \label{eq:B3S1bc}
~&A_i=\phi_I=0
\\& D_z\phi_{\bar w}=D_{\bar z}\phi_w=F_{z\bar z}-{1\ov 2}F_{w\bar w}=0.
\eeq
Finally, the supersymmetric boundary term is (the integration contour for $dz$, $d\bar z$ is counter-clockwise)
\beq \label{eq:B3S1bt}
S_{bdry}=\int_{\pa M} -2i\phi_{\bar w}(D_z\phi_w dz +D_{\bar z}\phi_{w}d\bar z ) {1\ov 2}\left(1-{x^2}\right)dx_2 d\tau,
\eeq
In principle, one can recursively make the perturbative (in $\rho$) corrections to the boundary conditions and the boundary term to make the total action $S_{tot}=S+S_{bdry}$ to have a well-defined variational principle and   supersymmetry. However, only the leading contributions specified \eqref{eq:B3S1fbc}, \eqref{eq:B3S1bc}, \eqref{eq:B3S1bt} will effectively contribute in the $\rho\rightarrow 0$ limit.

Within this setup of supersymmetric boundary conditions and boundary term, it is legitimate to apply the localization with $QV=(Q\lam,\bar{Q\lam})$. After the localization which includes restricting field configuration covariantly constant along $S^1$, we get the bulk and the boundary action
\beq \label{eq:B3S1bulkloc}
g^2 S_{bulk}={2\pi \ov 2}\int_{B^3} d^3 x &\Big[-\nabla_i ((1-x^2)(\phi_{i+4}\nabla_j \phi_{j+4}-\phi_{j+4}\nabla_j \phi_{i+4})) 
\\&+4\nabla_j(x_ix_k \phi_{k+4}\nabla_i \phi_{j+4}-x_ix_j\phi_i \nabla_k \phi_{k+4})
\\&+6\nabla_j (x_i \phi_{i+4}\phi_{j+4})\Big],
\eeq
and
\beq
g^2 S_{bdry}={\pi}\int dx_2 d\vartheta(1-x^2)\Big[x_4(\phi_7 D_3\phi_8-\phi_8 D_3 \phi_7) -x_3 (\phi_7D_4\phi_8-\phi_8 D_4 \phi_7)\Big],
\eeq
where $\vartheta$ is a polar coordinate angle in $x_3-x_4$ plane.

We naturally separate bulk term into 3 pieces $g^2 S_{bulk}=S_{bulk}^1+S_{bulk}^2+S_{bulk}^3$ according to \eqref{eq:B3S1bulkloc}. We know the bulk term's contribution at the outer boundary $S^2$ which is \eqref{eq:S2d}. Therefore, we focus on what happens at the inner tube-like boundary. 

We can simplify the first term in $S^1_{bulk}$ at the inner tube-like boundary, which gives (by using the Dirichlet boundary condition $\phi_6=0$, which also implies $D_2\phi_6=0$) 
\beq
g^2 S^1_{bulk}|_{tube}&=\pi \int dx_2 d\vartheta\, {x_a}(1-x^2)(\phi_{a+4}D_b \phi_{b+4}-\phi_{b+4}D_b\phi_{a+4})
\\&=\pi\int dx_2 d\vartheta\, (1-x^2)\Big[x_3(\phi_7 D_4\phi_8-\phi_8D_4\phi_7) +x_4(\phi_8 D_3\phi_7-\phi_7 D_3 \phi_8)\Big],
\eeq
and hence leads to a cancellation
\beq
S_{bdry}+S^1_{bulk}|_{tube}=0.
\eeq
Therefore the effective tube action after the localization is $S_{tube}=S^{2}_{bulk}|_{tube}+S^{3}_{bulk}|_{tube}$, which gives
\beq \label{eq:Stube1}
S_{tube}&=\int{\pi}\Big[-4 (\vec x\cdot \vec \phi) x_i \hat \rho_j \nabla_i \phi_{j+4}+4 (\vec x\cdot \hat \rho) (\vec x\cdot \vec \phi) \nabla_k \phi_{k+4}-6 (\vec x\cdot \vec \phi) (\hat \rho \cdot \vec \phi) \Big]   \ve d\vartheta dx_2,
\eeq
where $\hat \rho=(0, x_3,x_4)/\rho$.

Let us try to express the tube action by expanding around the surface operator background as $\phi_{i+4}=\phi^{surf}_{i+4}+\del \phi_{i+4}$. We take the  $\del \phi_{i+4}$  fluctuation   around the background to be  regular (i.e. $o(1)$) near the defect. Now since $S_{tube}$ is quadratic in $\phi$, we have quadratic polynomial $\del \phi$. There is a finite 0-th order term which is 
\beq
S_{tube}^{surf}=-{8\pi^2\be^2 \ov g^2}+o(\rho^2).
\eeq 
On the other hand, the 2nd order term vanishes at $\rho\rightarrow 0$, because the integral goes like $\int o(\rho)d\vartheta d x_2$. Hence, the potential non-vanishing contributions would come from the linear terms in $\del \phi$. However, it is easy to see that the first and second terms in \eqref{eq:Stube1} which are linear terms in $\del\phi$ are at least $o(\rho)$. Therefore, only potentially finite contribution comes from  the linear term in $\del \Phi$ in the third term in \eqref{eq:Stube1} which gives (using $\vec x \cdot \vec \phi^{surf}=\be$)
\beq
S_{tube}-S_{tube}^{surf}=-6\pi \be \int_{tube} 
 z  \phi_6  d\vartheta dx_2+o(\rho).
\eeq
However, this also vanishes by the boundary condition \eqref{eq:B3S1bc}! Therefore, the localized $2d$ action in the presence of the surface operator is simply given by 
\beq \label{eq:surf2d1}
S_{2d}&={2\pi\ov g^2}\int_{S^2,|x|^2=1}\phi_n^2 d\Omega  - {8\pi^2 \be^2 \ov g^2}
\\&={2\pi\ov g^2}\int_{S^2,|x|^2=1} (\phi_n^2 -\phi_{surf,n}^2)d\Omega,
\eeq
where now the important new feature is that there are punctures located at the North and South poles of $S^2$.

The final step is to translate the localized action \eqref{eq:surf2d1} in terms of relevant $2d$ Yang-Mills theory. The procedure explained in section \ref{eq:Pestun review} is adaptable where the main difference is that there is a  non-trivial background for $\bm A_{\mathbb C}=A-i*\bm \phi$ produced by the surface operator  given by 
\beq
\ex{\bm A_{\mathbb C}}=\al d\psi-i\gamma \csc\theta d\theta -i\beta \cos\theta d\psi, \quad \ex{\bm F_{\bm A_{\mathbb C}}}=i\beta \text{vol}_{S^2}.
\eeq
Note that on the punctured sphere, the $\gamma$ dependent part of $\ex{\bm A_{\mathbb C}}$ is   pure gauge and hence can be removed.\footnote{There is a technicality: the $\gamma$ dependent part is   pure gauge in $G_{\mathbb C}$ but not   pure gauge in $G$ since for generic $(A,\phi)$ the constraints only preserves $G$ gauge symmetry as we discussed in \eqref{eq:cHYM}. However, profiles $ A_{\mathbb C}=\al d\psi-i\gamma \csc\theta d\theta -i\beta \cos\theta d\psi$ and $ A_{\mathbb C}=\al d\psi-i\gamma \csc\theta d\theta$ both satisfies the constraint and hence belong to the path integral domain. Therefore, we should think of removing $\gamma$ as a choice of physically equivalent background in the path integral. \label{ft:gamma}}

Therefore, $2d$ Yang-Mills theory with connection $A$ has a non-trivial background whose gauge invariant characterization is given by the holonomy around the North and South poles 
\beq \label{eq:2dbackground}
\text{hol}_N(\nabla_{\bm A}) = e^{2\pi(-\al+i\beta)},\quad \text{hol}_S(\nabla_{\bm A}) = e^{2\pi(\al+i\beta)},
\eeq
where we choose an orientation so that the path ordering is clockwise (counter-clockwise) at the north (south) pole.

Actually, there is an important subtle difference in our puncture when we compare to   standard $2d$ Yang-Mills on a Riemann surface $X$ with punctures with prescribed holonomies \cite{Witten:1991we,Nguyen:2021naa}. In the standard $2d$ Yang-Mills, the space of gauge transformations $\CG(X)$ is usually taken to be $G$-valued everywhere and consequently holonomies are defined only up to conjugacy classes in $G$. On the other hand, our punctures on the sphere are the outcome of the regularization of the surface operator. If we try to make sense of the gauge singularity $A=\al d\psi$ as a connection on a $G$-bundle $E$ in the case without UV cutoff, then the structure group of $E$ has to be restricted to the maximal torus $T_G$ \cite{Gukov:2006jk}. This naturally suggests that in the presence of the UV cutoff, we should restrict our gauge group to $T_G$ at the regularized boundary (c.f. \cite{Giombi:2009ek}). Therefore, we should understand that the boundary holonomies at the North and South poles \eqref{eq:2dbackground} are absolutely prescribed, since they are gauge invariant with $T_G$-valued gauge transformations at the boundary.

So the picture emerging is a $2d$ YM theory with a compact gauge group $G$ (with structure group $T_G$ at the punctures), with an analytically continued background generated by $G_{\mathbb C}$-valued holonomies at the boundary. This means that the space of connections $\bm A$ can be decomposed into $\bm A=\ex{\bm A}+\bm A^{qu}$ such that the background $\ex{\bm A}$ is $\mathfrak g_{\mathbb C}$-valued while $\bm A^{qu}$ is $\mathfrak g$-valued. 

The remaining ingredient is an analog of the restriction to the zero-instanton sector as in \eqref{eq:2dYM0instanton}.  The natural guess is that the effective contribution comes only from the fluctuations $\bm A^{qu}$ restricted to the zero-instanton sector, or equivalently,  the perturbative sector. In summary, our localization result  with surface operator ${\cal O}_\Sigma$ is that
\begin{tcolorbox}

$\CN=4$ SYM on $B^3\times S^1$ with surface operator ~=~ Perturbative sector of $2d$ YM on punctured $S^2$ with $G_{\mathbb C}$-valued holonomies \eqref{eq:2dbackground} around the North and South poles.

\end{tcolorbox}

Finally, we comment on the semiclassical expectation value of the surface operator on $B^3\times S^1$. From the $2d$ action \eqref{eq:surf2d1}, we see that the semiclassical expectation value is $1$. One might have anticipated a non-trivial expectation value since the surface operator on $B^3\times S^1$ is topologically $S^2$. However, we argue that the surface operator on $B^3\times S^1$ is more analogous to the planar surface operator on $\mathbb R^4$. First, one can observe that the background field profile, boundary conditions and boundary term resemble the planar case discussed in Appendix \ref{sec:R2R4} rather than the spherical case in Appendix \ref{sec:S2R4}.

Moreover, we claim that the surface Weyl anomalies on $B^3\times S^1$ are trivial and hence it is similar to the planar defect case. The reason is that the geometry $B^3\times S^1$ is singular on $S^2=\pa B^3$ since the scalar curvature blows up there. Therefore, we actually need an additional regularization on $B^3\times S^1$, such as excising an infinitesimal tubular neighborhood around $S^2=\pa B^3$. This makes the topology of the surface defect $I\times S^1$ to be a punctured sphere at the North and South poles. And due to the boundary,  the surface anomalies \eqref{Weylanom} have to be modified such that we add an additional boundary term contributing to the anomaly
\beq \label{eq:surfweylbdry}
{1\ov 12\pi} \int_{\pa \Sigma}  ds\, \del \sigma k_g,
\eeq
where $k_g$ is the geodesic curvature of the boundary. 

In the present case, the 2nd and the 3rd terms in \eqref{Weylanom} vanish and hence the only contributions are from the 1st term in \eqref{Weylanom} and the boundary term \eqref{eq:surfweylbdry}. These two combines so that the integrated anomaly becomes proportial to the Euler characteristic of the punctured sphere, which is zero and hence the surface anomaly is trivial.

\subsection{Correlation Function with Local Operators} \label{sec:CPOloc}

One of our main observables is a normalized correlator of chiral primaries $\CO_{\Delta,k}$ with the surface operator   ${\cal O}_\Sigma$. The Weyl transformation of the CPO correlator in $R^4$ \eqref{eq:CPOR4} from $\mathbb R^4$ to $B^3\times S^1$ described in Appendix \ref{appsec:B3S1} gives the structure of the CPO correlator on $B^3\times S^1$ as \footnote{Conformal invariance dictates that ${ \ex{\CO_{\Delta,k} \CO_\Sigma}\ov \ex{\CO_\Sigma}}\bigg|_{\mathbb R^4}={ \ex{\CO_{\Delta,k} \CO_\Sigma}\ov \ex{\CO_\Sigma}}\bigg|_{B^3\times S^1}$.}
\beq \label{eq:originalcorrelator}
{ \ex{\CO_{\Delta,k} \CO_\Sigma}\ov \ex{\CO_\Sigma}}={C_{\Delta,k}\ov z^{\Delta+k\ov 2}\bar z^{\Delta-k\ov 2}},
\eeq
where $|z|$ is the conformally invariant distance in $B^3\times S^1$. Hence our task is to compute the defect OPE coefficient $C_{\Delta,k}$.

We first notice that $ \ex{\CO_{\Delta,k} \CO_\Sigma}$ is not compatible with the localization on $B^3\times S^1$, since $\CO_{\Delta,k}$ is not invariant under the localizing supercharge $Q$. Instead, we can place a position-dependent local chiral primary operator on $S^2$ \cite{Drukker:2009sf} compatible with the localization \cite{Giombi:2009ds}, given by
\beq \label{eq:localizable CPO}
O_{\Delta}=\Tr\left(n_i \phi_{i+4}-i \phi_9\right)^\Delta,
\eeq
which is indeed $Q$-invariant. Therefore what we can compute from our localization is the following correlator 
\beq \label{eq:localizablecorrelator}
{ \ex{O_{\Delta} \CO_\Sigma}\ov \ex{\CO_\Sigma}},
\eeq
where position dependence is fixed by symmetry.

A natural question is what is the relation between two correlators \eqref{eq:originalcorrelator} and \eqref{eq:localizablecorrelator}. Surprisingly, we show that the localizable correlators \eqref{eq:localizablecorrelator} contain full information about the original correlator of   interest \eqref{eq:originalcorrelator}. The first step is to perform an $SO(4)$ R-symmetry average on $O_\Delta$ 
\beq
\mathcal P[O_\Delta]\equiv {\int dg\, O_{\Delta}\ov \int dg },\quad g\in SO(4),
\eeq
where $dg$ is a Haar measure. Then because of the $SO(4)$ symmetry of the surface defect, the correlation function is invariant under averaging
\beq
 \ex{O_{\Delta} \CO_\Sigma}= \ex{\CP[O_{\Delta}] \CO_\Sigma}.
\eeq
 $\CP[O_{\Delta}]$ is a nicer object because of the manifest $SO(4)$  symmetry. Together with the $U(1)$ invariance of $\CP[O_{\Delta}]$, it can be organized in terms of $O_{\Delta,k}$ as (recalling that we normalized the radius of ${S^2}$ to be 1)
\beq \label{eq:remainder}
\CP[O_{\Delta}]=\sum_{k=-\Delta}^\Delta z^{\Delta+k\ov 2} \bar z^{\Delta-k\ov 2} c_{k} \CO_{\Delta,k} + \mathcal R[O_{\Delta}],
\eeq
where we define the remainder $\mathcal R[O_{\Delta}]$ as terms involving at least some $SO(4)$ fields. It is uniquely determined from $c_k$, which is independent of $\be,\gamma$, and fixed by the terms only involving $\Phi$ and $\bar \Phi$. Therefore, the correlator becomes
\beq 
{ \ex{O_{\Delta} \CO_\Sigma}\ov \ex{\CO_\Sigma}}=\sum_{k=-\Delta}^\Delta  c_kC_{\Delta,k} + {\ex{\mathcal R[O_{\Delta}]\CO_\Sigma} \ov \ex{\CO_\Sigma}}.
\eeq
Note that the first term is   independent of position, while the second term has a non-trivial explicit dependence on $x_2$. Now we   shortly demonstrate that   localization makes the correlator $\ex{O_\Delta \CO_\Sigma}$     topological, i.e. independent of the position. This suggests that the contribution from $\CR[O_\Delta]$ vanishes  identically,\footnote{In Appendix \ref{appsec:higherCPO}, we show that for $\Delta=4,5$ this vanishing happens intrinsically at the level of $4d$, independent of the localization, by showing that $\CR[O_\Delta]$ is exact under $SO(4)$ global symmetry and hence vanishes by a related Ward identity. It would be nice to prove that   $\CR[O_\Delta]$ is $SO(4)$-exact for any $\Delta$.} and hence we are left with 
\beq \label{eq:CPOrelation}
{ \ex{O_{\Delta} \CO_\Sigma}\ov \ex{\CO_\Sigma}}=\sum_{k=-\Delta}^\Delta c_k C_{\Delta,k}.
\eeq

While this looks like a single equation, it is remarkable to observe that we can recover each OPE coefficient $C_{\Delta,k}$ separately once we know the left hand side of  \eqref{eq:CPOrelation}. 

In order to see this, let us  introduce a $U(1)_{\text{spur}}$ rotation which acts only on the exactly marginal couplings $\be$ and $\gamma$ as $(\be+i\gamma) \rightarrow (\be+i\gamma )e^{i\theta}$. Then  $C_{\Delta,k}$ has a definite charge $k$ under $U(1)_{\text{spur}}$ since ${\ex{\CO_{\Delta,k}\CO_\Sigma}\ov \ex{\CO_\Sigma}}$ has a charge $k$ under the $U(1)$ described in section \ref{sec:surface operator}. This implies that $C_{\Delta,k}$ can be extracted from the charge $k$ sector of the left hand side of \eqref{eq:CPOrelation}, namely
\beq
C_{\Delta,k}=c_k^{-1}{ \ex{O_{\Delta} \CO_\Sigma}\ov \ex{\CO_\Sigma}}\bigg|_{U(1)_{\text{spur}}~\text{charge } k}.
\eeq

From now on we consider the case of $G=SU(N)$ to compare with  the supergravity computation in \cite{Drukker:2008wr}. However, for   technical simplicity, it is convenient to take $G=U(N)$ and impose traceless constraints on marginal parameters $(\al,\be,\gamma,\eta)$. Let's consider examples of $\Delta=2,3$ explicitly (see Appendix \ref{appsec:higherCPO} for more cases). This case is quite simple since $\CR[O_\Delta]=0$ and using \eqref{eq:O23} we have
\beq  \label{eq:PO23}
~&\CP[O_2]={z^2\ov 2}{\sqrt{2}\lam\ov 8\pi^2}\CO_{2,2}+{\bar z^2\ov 2}{\sqrt{2}\lam\ov 8\pi^2}\CO_{2,-2}+{\sqrt{6}\lam \ov 16\pi^2}\CO_{2,0} 
\\&\CP[O_3]={\sqrt 3\lam^{3/2}\ov 64\pi^3} (z^3\CO_{3,3}+\bar z^3\CO_{3,-3})+{3\lam^{3/2}\ov 32\sqrt{2} \pi^3}(z^2\bar z\CO_{3,1}+z\bar z^2 \CO_{3,-1}).
\eeq
In fact, it is quite suggestive to recall   the correlators $\ex{O_\Delta O_\Sigma}$ using the supergravity results of the correlator $\ex{\CO_{\Delta,k}O_\Sigma}$ from \cite{Drukker:2008wr}\footnote{There are some relative sign differences in the formulas since we are using the anti-Hermitian convention as opposed to the Hermitian convention used in \cite{Drukker:2008wr}).}
\beq \label{eq:DGM}
~  &{ \ex{O_{2,2} \CO_\Sigma}\ov \ex{\CO_\Sigma}}\bigg|_{\text{sugra}}=
{1\ov z^2}{4\pi^2\ov \sqrt{2}\lam}\sum_{l=1}^MN_l(\be_l+i\gamma_l)^2
\\&{ \ex{O_{2,0} \CO_\Sigma}\ov \ex{\CO_\Sigma}}\bigg|_{\text{sugra}}
=
{1\ov |z|^2}{8\pi^2\ov \sqrt{6}\lam}\left(\sum_{l=1}^M N_l\left( (\be_l^2+\gamma_l^2)-{\lam\ov 4\pi^2}{N-N_l\ov 2N}\right) \right)
\\&{ \ex{O_{3,3} \CO_\Sigma}\ov \ex{\CO_\Sigma}}\bigg|_{\text{sugra}}
={1\ov z^3}{8\pi^3\ov \sqrt{3}\lam^{3/2}}\sum_{l=1}^MN_l(\be_l+i\gamma_l)^3
\\&{ \ex{O_{3,1} \CO_\Sigma}\ov \ex{\CO_\Sigma}}\bigg|_{\text{sugra}}
={1\ov z|z|^2}{8\pi^3\ov \sqrt{2}\lam^{3/2}}\left(\sum_{l=1}^MN_l\left( (\be_l^2+\gamma_l^2)-{\lam\ov 4\pi^2}{N-2N_l\ov N}\right)(\be_l+i\gamma_l) \right).
\eeq
Inserting \eqref{eq:DGM} to \eqref{eq:PO23}, we obtain 
\beq \label{eq:O23sugra}
~&{ \ex{O_{2} \CO_\Sigma}\ov \ex{\CO_\Sigma}}\bigg|_{\text{sugra}}=\sum_l N_l\left(\be_l^2+{\lam\ov 16\pi^2}{N-N_l\ov N}\right)
\\&{ \ex{O_{3} \CO_\Sigma}\ov \ex{\CO_\Sigma}}\bigg|_{\text{sugra}}=\sum_l N_l\left(\be_l^3+{3\lam\ov 32\pi^2}{N-2N_l\ov N}\be_l\right),
\eeq
where we see that the expected correlators of $O_\Delta$ only depend on $\be$ and are independent of $\gamma$\,! This is remarkably consistent with the localization picture on $B^3\times S^1$ with the surface operator insertion described in section \ref{sec: localization}, where the effective $2d$ theory is a zero-instanton sector of $2d$ Yang-Mills theory on a punctured 2-sphere with a background flux $\be$.

To compute the correlator with $O_\Delta$ from the localization, the first thing we need is the mapping of $O_\Delta$ to the $2d$ YM theory. Now under the assumption that the moduli space of solutions for the localizing equations forces $\phi_9=0$ just as in the case without surface operator inserted \cite{Pestun:2009nn}, we see that the mapping is given by (recall that we fixed $R=1$)
\beq
~&\phi_n=n_i\phi_{i+4}-i\phi_9 \quad \Longleftrightarrow \quad -i * \bm F
\\& O_\Delta \quad \Longleftrightarrow \quad   \Tr \left[ (-i * \bm F)^\Delta\right].
\eeq
Therefore we see that the the correlator becomes 
\beq
\ex{O_\Delta\CO_\Sigma}_{4d} \quad \Longleftrightarrow \quad  \ex{\Tr \left[ (-i *\bm F)^\Delta\right]}_{2d},
\eeq
where the $2d$ expectation value is evaluated in the singular background specified in \eqref{eq:2dbackground}.  
The first thing we need to know is the correlator of $-i*\bm F$ on the holonomy background. We introduce a natural complex coordinate on $S^2$ using stereographic projection which gives a metric
\beq
ds^2={4dzd\bar z\ov {(1+z\bar z)^2}}, \quad g_{z\bar z}={2\ov (1+z\bar z)^2},
\eeq
and we use a convention such that the volume form on $S^2$ is $
\sqrt{g}dz \wedge d\bar z=ig_{z\bar z}dz \wedge d\bar z$. 

Next we decompose $2d$ connection $\bm A$ in terms of background $\ex{A}$ (see footnote \ref{ft:gamma}) and quantum fluctuations $ A^{qu}$ as
\beq
\bm A=\ex{\bm A}+ \bm A^{qu}, \quad \ex{\bm A}=\al d\psi-i\be \cos\theta d\psi,
\eeq
where $A^{qu}$ is continuously connected to the trivial gauge so that it is purely perturbative. We can impose a background holomorphic gauge so that
\beq \label{eq:gauge fixing}
 \bm A^{qu}_{\bar z}=0.
\eeq
Then the field strength in this gauge is given by
\beq
 \bm F_{z\bar z}=\ex{\bm F_{z\bar z}}-D_{\bar z}^{cl} \bm A_z^{qu},\quad  \quad \ex{\bm F_{z\bar z}}=-\be g_{z\bar z},
\eeq
where $D_{\bar z}^{cl}\equiv \pa_{\bar z}+\text{ad}_{\bm A_{\bar z}^{cl}}$. Therefore the resulting gauge fixed $2d$ YM action becomes
\beq
S_{2d}^{hol}&={1\ov g^2_{2d}}\int \sqrt{g} g^{z\bar z}g^{\bar z z}(\bm F_{z\bar z}^2-\ex{\bm F_{z\bar z}}^2)dzd\bar z
\\&={1\ov g_{2d}^2}\int -i g^{z\bar z}(D_{\bar z}^{cl}  \bm A^{qu}_z)^2 dz d\bar z .
\eeq
Let's denote $\vp_n\equiv -i* \bm F$, then we have (henceforth we omit various superscripts for simplicity)
\beq \label{eq:vp}
\vp_n=\ex{\vp_n}+\vp,\quad \ex{\vp_n}=\be,\quad \vp=g^{z\bar z}D_{\bar z} \bm A_z,
\eeq
which simplifies the $2d$ action as
\beq
S_{2d}^{hol}=-{1\ov g^2_{2d}}\int \vp^2 \nu^2 d^2x,\quad \nu(z,\bar z)={2\ov 1+z\bar z},
\eeq
where we used $dzd\bar z=-2i d^2x$. Note that since the action is Gaussian in $\vp$, it is clear that the computation of the CPO correlator $\ex{O_\Delta \CO_\Sigma}$ reduces to the Wick contraction in the $2d$ theory.

To compute the propagator, we might naively change the path integral variable from $A_z$ to $\vp$, which gives the naive normalized correlator of $\vp$ as\footnote{Remarkably, there is no effect of non-trivial Jacobian from $A_z$ to $\vp$ since it is exactly canceled by the Faddeev-Popov determinant associated to the gauge fixing condition \eqref{eq:gauge fixing}.} 
\beq
\ex{\vp_{ij}(x) \vp_{ji}(x')}_{naive}={g^2_{2d}\ov 2} \nu^{-2}\del^2(x-x').
\eeq
However, there's an issue of zero modes. Namely, since $\ex{A_{\bar z}}\in \mathfrak l$, it is natural to use a decomposition $\mathfrak g=\mathfrak l \oplus \mathfrak p$, such that $A_z=A_z^{\mathfrak l}+A_z^{\mathfrak p}$ and $\vp=\vp^{\mathfrak l}+\vp^{\mathfrak p}$, which gives a relation 
\beq
\vp^{\mathfrak l}=g^{z\bar z}\pa_{\bar z}A_z^{\mathfrak l} ,\quad  \vp^{\mathfrak p}=g^{z\bar z}D_{\bar z}A_z^{\mathfrak p}.
\eeq
Accordingly, $\vp^{\mathfrak l}$ has to satisfy the constraint
\beq
\int g_{z\bar z}\vp^{\mathfrak l} d^2 x=\int {1\ov 2}\nu^2\vp^{\mathfrak l} d^2 x=0,
\eeq
which gives the correct normalized propagator
\beq \label{eq:llprop}
\ex{\vp_{ij}^{\mathfrak l}(x) \vp^{\mathfrak l}_{ji}(x')}&={g^2 _{2d} \ov 2}\left(  \nu^{-2}\del^2(x-x')- {1\ov \int \nu^{2} d^2 x} \right)
\\&={g^2_{2d} \ov 2} \left(  \nu^{-2}\del^2(x-x')- {1\ov 4\pi } \right).
\eeq
On the other hand, since $D_{\bar z}$ has a trivial kernel, the propagator of $\vp^{\mathfrak p}$ is simply given by
\beq \label{eq:ppprop}
\ex{\vp_{ij}^{\mathfrak p}(x) \vp^{\mathfrak p}_{ji}(x')}&={g^2_{2d} \ov 2}   \nu^{-2}\del^2(x-x').
\eeq

To compute the correlator of CPO from \eqref{eq:llprop}, \eqref{eq:ppprop}, we need to take a coincident limit $x\rightarrow x'$ of the propagator. Because of the divergence arising from the contact term, we need to renormalize the two-point function. The standard prescription is to subtract by the propagator with a trivial background (i.e. without a defect), whose normalized propagators are \cite{Giombi:2009ds}
\beq
\ex{\vp_{ij}^{\mathfrak l}(x) \vp^{\mathfrak l}_{ji}(x')}_{0}=\ex{\vp_{ij}^{\mathfrak p}(x) \vp^{\mathfrak p}_{ji}(x')}_{0}={g^2_{2d}\ov 2} \left(  \nu^{-2}\del^2(x-x')- {1\ov 4\pi } \right),
\eeq
since the zero modes are present in both $\mathfrak l$ and $\mathfrak p$ parts. Therefore, the renormalized two-point function is
\beq \label{eq:2dfinalprop}
~&\ex{\vp_{ij}^{\mathfrak l}(x) \vp^{\mathfrak l}_{ji}(x')}_{ren}=0
\\&\ex{\vp_{ij}^{\mathfrak p}(x) \vp^{\mathfrak p}_{ji}(x')}_{ren}={g^2_{2d}\ov 8\pi},
\eeq
where now we can take the coincident limit harmlessly.

Armed with \eqref{eq:vp} and \eqref{eq:2dfinalprop}, we can finally compute the exact correlator with $O_\Delta$ simply by Wick contractions! This implies  that $\ex{O_\Delta\CO_\Sigma}$ has corrections to order $\floor{\Delta/2}$. Alternatively, $\langle {\cal O}_\Sigma \cdot {\cal O}_{\Delta,k}\rangle\over
\langle {\cal O}_\Sigma\rangle$ is a polynomial in the gauge coupling $g^2$. It has corrections to order $(g^2)^{(\Delta-|k|)/2}$.

Actually, the correlator $\ex{O_\Delta\CO_\Sigma}$  is a polynomial in $g^2/\beta^2$, and since $g^2$  and $\beta^2$ transform the same way under the $PSL(2,\mathbb Z)$ $S$-duality group (see section 1), the correlators  
$\langle {\cal O}_\Sigma \cdot {\cal O}_{\Delta,k}\rangle\over
\langle {\cal O}_\Sigma\rangle$ are $S$-duality invariant!\footnote{Similarly, a properly normalized correlator $\ex{{1\ov g^{\Delta}}  O_\Delta\CO_\Sigma}$ is also invariant under the $S$-duality.}

As an illustration, we consider $\Delta=2,3$ and get (see Appendix \ref{appsec:higherCPO} for more examples)
\beq  \label{eq:2dYMO23}
~&{\ex{O_2\CO_\Sigma} \ov \ex{\CO_\Sigma}}=\ex{\Tr \vp_n^2}+\ex{\Tr \vp \vp}=\sum_l N_l \be_l^2+{g^2_{2d}\ov 8\pi}N_l(N-N_l)
\\&{\ex{O_3\CO_\Sigma} \ov \ex{\CO_\Sigma}}=\ex{\Tr \vp_n^3}+3 \ex{\Tr \vp_n \vp \vp}=\sum_l N_l \be_l^3+{3g^2_{2d}\ov 8\pi}N_l(N-N_l)\be_l,
\eeq
which matches with \eqref{eq:O23sugra} under the identification between gauge couplings \eqref{eq:g2d} and restricted to $G=SU(N)$ by imposing a traceless conditions on $\be$. It is striking that the supergravity computation with bubbling geometry, whose validity is for $\lam=g^2 N \gg 1$ and $o(N_l/N)\simeq 1$ turns out to be applicable at finite $g^2$ and for arbitrary $\mathbb L$.

Finally, the   CPO correlator being a finite polynomial in $g^2$ suggests that there are a finite number of Feynman diagrams that effectively contribute in the  perturbative $4d$ computation around the surface operator background. We   show  in Appendix \ref{appsec:CPO}    using the background field method and choosing the Feynman gauge, that the relevant   diagrams are precisely the quantum corrections obtained from   Wick contractions with the     propagator in the presence of surface operator background.

\subsection{Correlation Function with 1/8-BPS Wilson Loops} \label{sec:Wilsonloc}

In this section  we compute the correlator of the $1/8$-BPS Wilson with the surface operator using localization. On $\mathbb R^4$, the 1/8-BPS Wilson loop  restricted on $S^2$, defined by $\sum_{i=2,3,4}x_i^2=R^2$,  is given by \cite{Drukker:2007dw,Drukker:2007yx,Drukker:2007qr}
\beq
W^{1/8}_{\CR}(C)=\Tr_{\CR} P \exp \left [\oint_C \left( A_i -i\e_{ijk}\Phi_{j+4} {x^k\ov |x|}\right)dx^i \right],
\eeq
which preserves the 4 supercharges satisfying 
\beq
\Gamma_{jk} \e_c+i\ep_{ijk}\Gamma_{k+4}\e_s=0.
\eeq
The conformal mapping from $\mathbb R^4$ to $B^3\times S^1$ in Appendix \ref{appsec:B3S1} maps 1/8-BPS Wilson loop on $\mathbb R^4$ to 1/8-BPS Wilson loop on $B^3\times S^1$.

\begin{figure}
\centering    \includegraphics[width=0.4\textwidth]{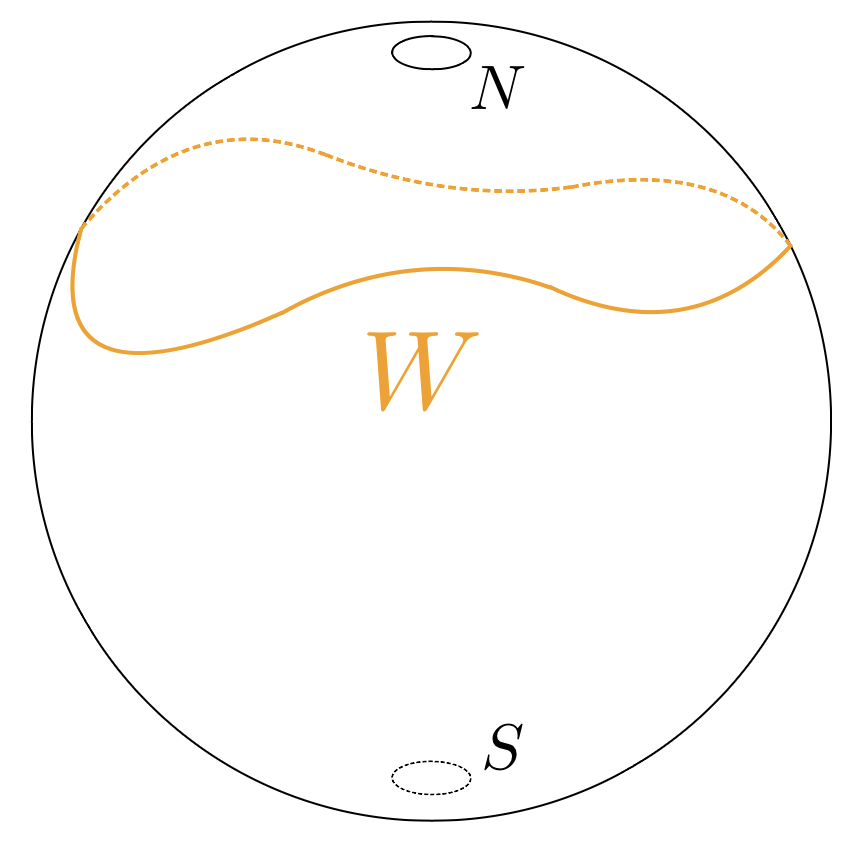}
\caption{1/8-BPS Wilson loop on $S^2$ wrapping around the surface operator.} \label{figure:wilson}
\end{figure}

In terms of the connection $\bm A_{\mathbb C}$, we have
\beq
W^{1/8}_\CR (C)=\Tr_\CR P \exp \left [\oint_C \bm A_{\mathbb C} \right].
\eeq
Therefore, we immediately recognize that the 1/8-BPS Wilson loop in $4d$ is simply mapped to the ordinary Wilson loop in $2d$ YM
\beq
~&W^{1/8}_\CR(C) \quad \Longleftrightarrow \quad \Tr_\CR P \exp \left [\oint_C \bm A \right].
\eeq
Hence we need to compute the correlator of the Wilson loop in the zero-instanton sector with the specified holonomies at the North and South poles. Our strategy is to use heat-kernel formulation \cite{Migdal:1975zg,Witten:1991we,Nguyen:2021naa} of $2d$ YM to compute the full non-perturbative correlator, and subsequently extract the zero-instanton sector \cite{Bassetto:1998sr}. For the comparison with supergravity \cite{Drukker:2008wr}, we consider the Wilson loop in the fundamental representation $\CR=F$ and non-trivially winding around the surface operator as in figure \ref{figure:wilson}. In terms of the $2d$ sector, this means that we consider a non-simply connected $2d$ Wilson loop on the punctured sphere. 

Since we are interested in the normalized correlator, we start with the computation of the expectation value of the surface operator. In the heat kernel lattice formulation, we cover the Riemann surface with arbitrary polygons and assign each link $l$ with a group variable $U_l\in G$ which is a parallel transport of the gauge connection along $l$. The partition function in this formulation reduces to   finite integrals over $U_l$ with appropriately normalized Haar measure and for the integrand, we assign for each plaquette $p$ a factor of
\beq
Z_p(U_p,\CA_p)=\sum_{\CR \in \text{Irrep} \,G}d_\CR  \chi_\CR (U_p)e^{-{1\ov 4}g^2_{2d} \CA_p C_2(\CR)},
\eeq
where $\CA_p$ is an area of plaquette $p$.

\begin{figure}
\centering    \includegraphics[width=0.4\textwidth]{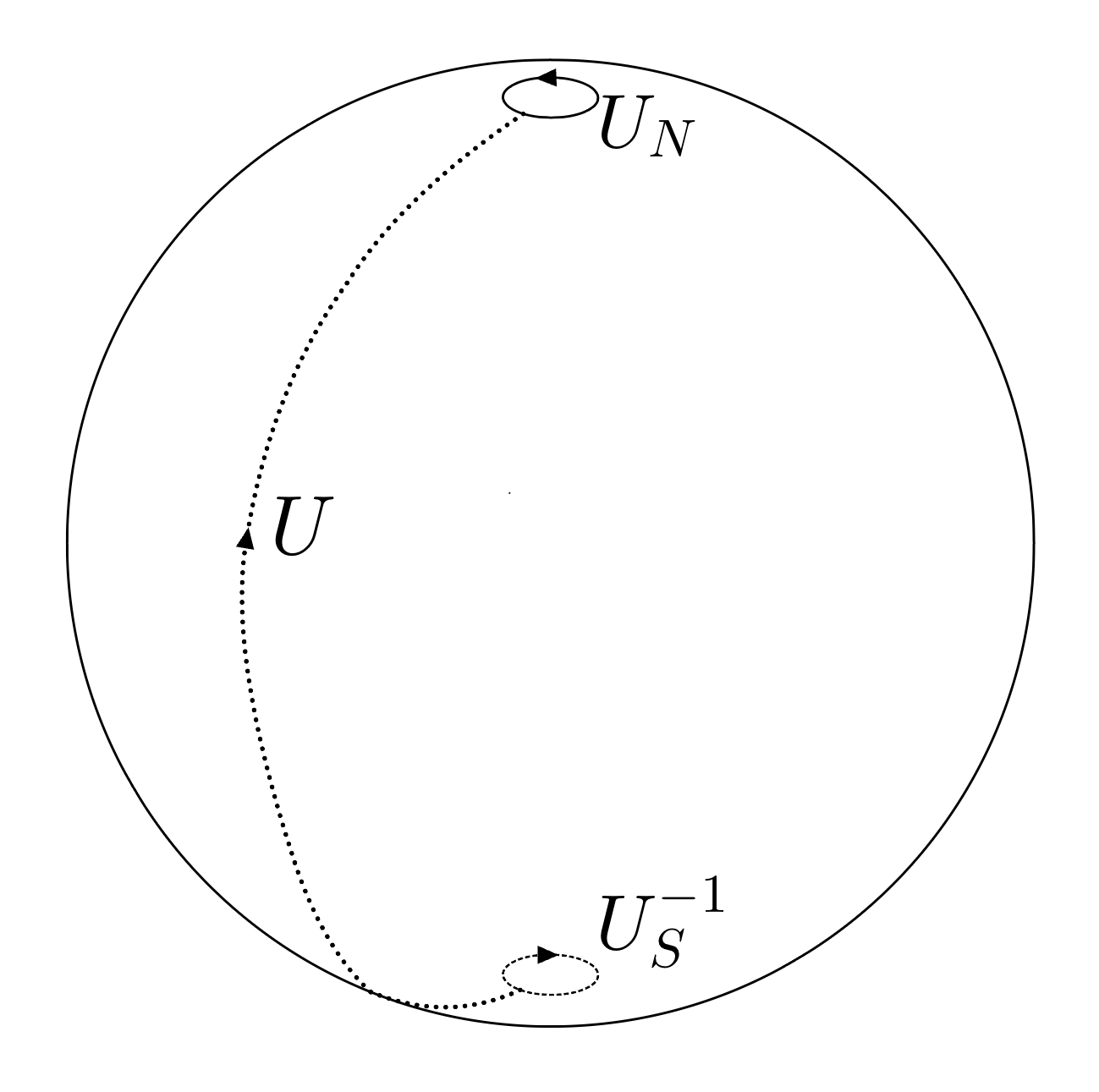}
\caption{Covering of punctured sphere by a polygon.} \label{figure:ZS2}
\end{figure}

In our case, we have punctures at the North and South poles with holonomies, therefore we can decompose the sphere into a single polygon according to figure \ref{figure:ZS2}. One subtle difference compared to the ordinary $2d$ Yang-Mills is that since the structure group at the boundary is $T_G$ (see discussion below \eqref{eq:2dbackground}), we would expect an additional overall factor related to the difference of the gauge group. We leave the precise determination of this factor for the future work, and merely call it as $V_{\al,\be}$ since it will generically depends on the holonomies at the boundary. In any case, it drops out in normalized correlators. This way we get

\beq \label{eq:2dYMvevfull}
Z_{\Sigma}^{YM_2}=V_{\al,\be}\,e^{ 8\pi^2 \be^2 \ov g_{2d}^2\CA}\sum_{\CR\in \text{Irrep} \,G } \chi_{\CR }(U_N) \chi_{\CR}(U_S^{-1}) e^{-{1\ov 4} g^2_{2d} \CA C_2(\CR)} ,
\eeq
where $\CA=4\pi$ is an area of $S^2$, $U_{N,S}$ are holonomies at north and south poles in (\ref{eq:2dbackground}), and the sum is over irreducible representations of $G=U(N)$, where in our case, it is labeled by N-tuples of integers $\CR=( \lam_1,\dots, \lam_N)$ which satisfy $\infty >\lam_1\geq \lam_2\geq \dots\geq \lam_N\geq -\infty$. The quadratic Casimir is given by 
\beq
C_2(\CR)=\sum_{i=1}^N \lam_i^2 +(N+1- 2i)  \lam_i,
\eeq
which can be simplified in terms of $l_i=\lam_i+N-i$ so that $\{l_i\}$ are strictly decreasing $\infty>l_1>l_2 >\dots >l_N>-\infty$ as
\beq
C_2(\CR)&=-{1\ov 12}N(N^2-1)+\sum_{i=1}^N \left(l_i-{N-1\ov 2} \right)^2.
\eeq
Now $U_{N,S}\in G_{\mathbb C}\simeq GL(N,\mathbb C)$, and since any $U(N)$ representation can be naturally extended to $GL(N,\mathbb C)$, we naturally take $\chi_{\CR}$ to be a character of $GL(N,\mathbb C)$. The explicit representation of the character for $e^y\in T_{\mathbb C}=(\mathbb C^*)^N$ is given by
\beq \label{eq:characterT}
\chi_{\CR}(e^y)={\det e^{y_a l_b}\ov \det e^{y_a(N-b)} },
\eeq
where $a$ and $b$ denote row and column of matrices with $y=\text{diag}(y_1,\dots,y_N)$. Note that in our case $y=\pm\al+i\be$, and hence the character formula \eqref{eq:characterT} only makes sense for $\mathbb L=U(1)^N$ and becomes singular for generic $\mathbb L=\prod_{j=1}^M U(N_j)$. In this case, we apply L'Hôpital's rule to obtain the character formula for singular element in $GL(N,\mathbb C)$ as
\beq
~&\chi_{\CR}(e^y)={1\ov D(y)\prod_{j=1}^M \prod_{s=1}^{N_j}s!}\sum_{w\in S_N}\ve(w)  e^{ \sum_{a=1}^N y_{w(a)} l_a } \prod_{j=1}^M\Delta(l_{w^{-1}(1+\sum_{k=1}^{j-1} N_k)},\dots, l_{w^{-1}(N_j)}) 
\\& D(y)\equiv e^{\sum_{j=1}^M {1\ov 2}N_j(N_j-1)y_{[j]}} \prod \limits_{\substack{a<b=1
\\ [a]\neq [b]}}^N
(e^{y_{a}}-e^{y_{b}}), \quad \Delta(l_1,\dots,l_N)\equiv\prod_{a<b=1}^N(l_a-l_b),
\eeq
where we ordered $y$ such that $y=\text{diag}(y_1,\dots, y_N)=\text{diag}(\overbrace{y_{[1]},\dots,y_{[1]}}^{N_1},\dots, \overbrace{y_{[M]},\dots y_{[M]}}^{N_M})$ so that $y_a$ belongs to $[a]$-th block.   

To extract the zero-instanton sector of $Z_\Sigma^{YM_2}$ \eqref{eq:2dYMvevfull} we use the Poisson summation formula \cite{Gross:1994mr,Bassetto:1998sr},
\beq
\sum_{-l_i=-\infty}^\infty f(l_1,\dots,l_N)&=\sum_{m_i=-\infty}^\infty \int_{-\infty}^\infty dz_1 \dots dz_N \,e^{2\pi i \sum_{i=1}^N m_i z_i} f(z_1,\dots, z_N)
\\&\equiv \sum_{m_i=-\infty}^\infty F(m_1,\dots,m_N),
\eeq
which identifies the zero-instanton sector as $m_1=\dots=m_N=0$ contribution. This way, we get
\beq
Z_{\Sigma}^{YM_2}|_{0-inst}
=& V_{\al,\be} e^{ 8\pi^2 \be^2 \ov g_{2d}^2\CA} c(\{N_i\},g,\al,\be)
\\&\left(\sum_{w\in S_N}\ve(w)  e^{2\pi(\al+i\be)_{w(a)} z_a } \prod_{j=1}^M\Delta(z_{w^{-1}(1+\sum_{i=1}^{j-1} N_i)},\dots, z_{w^{-1}(N_j)})
\right)
\\& 
\left(\sum_{\tilde w\in S_N}\ve(\tilde w)  e^{2\pi(-\al+i\be)_{\tilde w(a)}  z_a } \prod_{j=1}^M\Delta( z_{\tilde w^{-1}(1+\sum_{i=1}^{j-1} N_i)},\dots,  z_{\tilde w^{-1}(N_j)})
\right)
\\& \exp\left(-{1\ov 4}g^2_{2d} \mathcal A \sum_{i=1}^N(z_i-{N-1\ov 2})^2\right) ,
\eeq
where
\beq \label{eq:c}
c_W(\{N_i\},g_{2d},\al,\be)={e^{{1\ov 48}g^2_{2d}\mathcal A N(N^2-1)}\ov N! |D(\al+i\be)|^2 \prod_{j=1}^M \prod_{s=1}^{N_j} (s!)^2 } .
\eeq
The sums over $w$ and $\tilde w$ reflect the freedom of independent Weyl transformation on the holonomies around the North and South poles in the standard $2d$ Yang-Mills setup. This is natural, because the standard $2d$ Yang-Mills path integral localizes to   instantons \cite{Witten:1992xu} and the independent Weyl transformations on the North and South poles generate different saddle points (c.f. \cite{Hosomichi:2017dbc}).

However, our holonomies at the Sorth and South poles are fixed and hence cannot have an independent Weyl transformation. Therefore, the true $2d$ sector describing $4d$ sector is described by the pair $(w,\tilde w)\in W\times W$ satisfying $\pm\al+i\be=w\cdot(\pm \al+i\be)=\tilde w \cdot (\pm \al+i\be)$. There are    $(  \prod_{j=1}^M N_j!)^2$ such pairs. The remaining integration over $\{z_a\}$ is purely Gaussian and we get $4d$ vev of the surface operator from the $B^3\times S^1$ localization as
\beq \label{eq:ZYM2}
\ex{\CO_\Sigma}_{B^3\times S^1}= V_{\al,\be}   \left(\prod_{j=1}^M N_j! \right)^2  c_W(\{N_i\},g_{2d},\al,\be)
e^{2\pi i(N-1)\sum_i\be_i}
\left(\prod_{p=1}^M G_p\right),
\eeq
where
\beq \label{eq:Gp}
G_p&=\int \left(\prod_{a=1}^{N_p}d x_a \right) \Delta(x_1,\dots, x_{N_p})^2e^{-{1\ov 4}g_{2d}^2 \CA \sum_a x_a^2}
\\&=\left({g_{2d}^2 \CA/ 2 }\right)^{-N_p^2/2}(2\pi)^{N_p/2} \left(\prod_{n=1}^{N_p}n!\right).
\eeq
Therefore, we see that \eqref{eq:ZYM2}   is also a Gaussian matrix model with a one-loop measure $\Delta_{\mathbb L}$ as in \eqref{eq:2d4dvev}.

Therefore in terms of $4d$ gauge coupling $g^2$, we finally get
\beq \label{eq:vev2dYM}
\ex{\CO_\Sigma}_{B^3\times S^1}={V_{\al,\be}  e^{2\pi i(N-1)\sum_{i=1}^N\be_i} \ov |D(\al+i\be)|^2 }{  \left(\prod_{j=1}^M N_j!\right) (2\pi)^{N/2} (-g^2)^{-\text{dim}(\mathbb L)/2} e^{-{1\ov 24}g^2 N(N^2-1)}\ov  N! \prod_{j=1}^M \prod_{s=1}^{N_j-1} s! },
\eeq
where we note that the exponential term $e^{-{1\ov 24}g^2 N(N^2-1)}=e^{{1\ov 48}g_{2d}^2\CA N(N^2-1)}$ can be removed by a local counterterm in $2d$ proportional to $\int_{S^2}\sqrt{g_{S^2}}$, and the denominator $\prod_{j=1}^M \prod_{s=1}^{N_j-1} s! $ is proportional to the $\text{vol}(\mathbb L)$. However, there are still some qualitative differences compared to the $4d/2d$ localization computation of the vev of the $S^2$ surface operator in $S^4$ \eqref{eq:finalvevS2S4} and we comment on them.

First, we see that there is no radius dependence of the vev. This is consistent with our claim of the absence of the surface Weyl anomaly on $B^3\times S^1$ discussed at the end of section \ref{sec: localization}.

Secondly, a difference between overall powers of $g^2$, which   originated from the Gaussian matrix model with exponent proportional to $g^2$ compared to $1/g^2$ in section \ref{sec:4d2d}. Since this cannot be compensated by any local counterterm on $S^2$, it is natural to guess that the one-loop determinant from the $B^3\times S^1$ localization would produce such a factor difference (the localization determinant was not computed in \cite{Pestun:2009nn}). This factor is also needed for the vacuum partition function (c.f. \cite{Wang:2020seq}).

Thirdly, there are purely $\al,\be$ dependent factors appearing in the first factor in \eqref{eq:vev2dYM}. These factors have to be canceled since we expect the vev of the surface operators on $B^3\times S^1$ are trivial as the planar surface operators on $\mathbb R^4$. One observation is that 
\beq
|D(\al+i\be)|^2=e^{\sum_{j=1}^M iN_j(N_j-1)\be_{[j]}} \prod \limits_{\substack{a<b=1
\\ [a]\neq [b]}}^N
|e^{(\al+i\be)_{a}}-e^{(\al+i\be)_{b}}|^2\,,
\eeq
where the 2nd product factor is proportional to the (analytically continued) volume of the conjugacy class of $e^{\al+i\be}$ (or equivalently $e^{-\al+i\be}$) in $G=U(N)$. Therefore we anticipate that this volume factor is canceled by $V_{\al,\be}$. On the other hand, using $\ex{\bm F}=i\beta \text{vol}_{S^2}$, we might expect the $e^{\be_j}$ factors will be canceled by the ABJ anomaly of massless fermions at the level of constrained $2d$ YM (c.f. \eqref{eq:2dYMpathintegral}), which is responsible for the restriction to the perturbative sector.

In the end, it turns out that these subtle factors will similarly appear in the correlator $\ex{W_F \CO_\Sigma}$, and therefore the above subtleties are inconsequential when we compute the normalized correlator. Therefore we proceed and leave a careful determination of the vev on $B^3\times S^1$ from this  localization to the future. The computation in a different localization was already presented in section \ref{sec:4d2d}.

\begin{figure}
\centering    \includegraphics[width=0.4\textwidth]{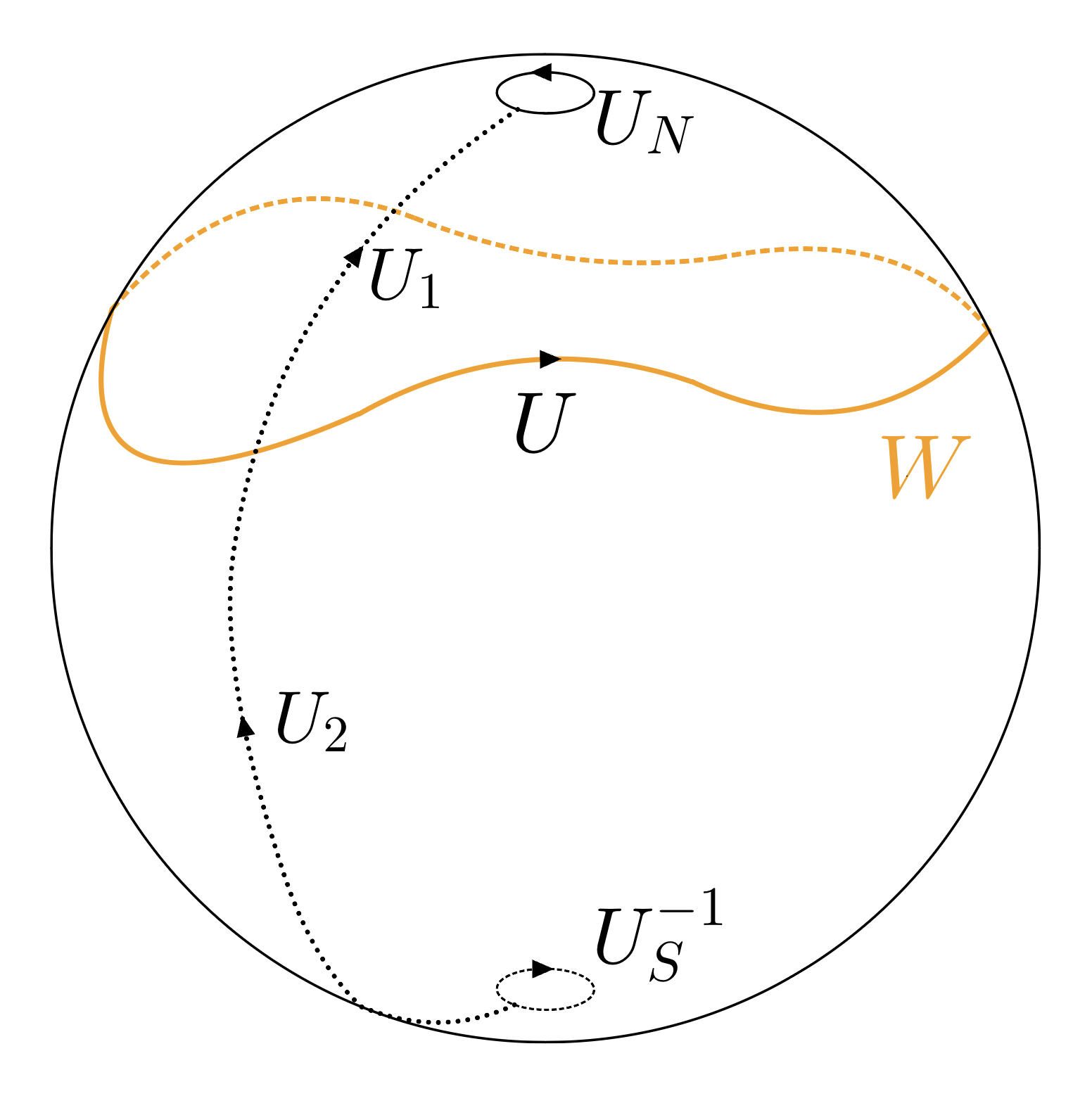}
\caption{Covering of punctured sphere with Wilson loops by polygons.} \label{figure:w}
\end{figure}

Next, we compute the correlation function of the fundamental  Wilson loop wrapping non-trivially the North and the South poles as in figure \ref{figure:wilson}. Using the lattice decomposition as in figure \ref{figure:w}, we get 
\beq
\ex{W_F \CO_\Sigma }^{YM_2}=&V_{\al,\be}\,e^{ 8\pi^2 \be^2 \ov g_{2d}^2\CA} \sum_{\CR_1,\CR_2} \chi_{\CR_1}(U_N)\chi_{\CR_2}(U_S^{-1})e^{-{1\ov 2}g_{2d}^2 [\CA_1 C_2(\CR_1)+\CA_2C_2(\CR_2)]} 
\\&\int dg \chi_F(g) \chi_{R_1}(g^{-1}) \chi_{R_2}(g).
\eeq

The fundamental representation is given by $F=(1,0,\dots,0)$. Then we can similarly extract the 0-instanton sector as
\beq
\ex{W_F \CO_\Sigma }^{YM_2}|_{0-inst}=&V_{\al,\be}\,e^{ 8\pi^2 \be^2 \ov g_{2d}^2\CA} c_W(\{N_i\},g_{2d},\al,\be)
\\&\int   dz_a\sum_{k=1}^N 
\left(\sum_{w\in S_N}\ve(w)  e^{2\pi(\al+i\be)_{w(a)} z_a } \prod_{j=1}^m\Delta(z_{w^{-1}(1+\sum_{i=1}^{j-1} N_i)},\dots, z_{w^{-1}(N_j)})
\right)
\\& 
\left(\sum_{w\in S_N}\ve(w)  e^{2\pi(-\al+i\be)_{w(a)} \tilde z_a } \prod_{j=1}^m\Delta(\tilde z_{w^{-1}(1+\sum_{i=1}^{j-1} N_i)},\dots, \tilde z_{w^{-1}(N_j)})
\right)
\\& \exp\left(-{1\ov 4}g^2_{2d} \mathcal A_1 \sum_{i=1}^N(z_i-{N-1\ov 2})^2-{1\ov 4}g^2_{2d} \mathcal A_2 \sum_{i=1}^N(z_i-\del_{k,i}-{N-1\ov 2})^2 \right).
\eeq
Again, the true $0$-instanton sector is described by further extracting only $(w,\tilde w)\in W\times W$ satisfying $\pm \al+i\be=w\cdot(\pm \al+i\be)=\tilde w \cdot (\pm \al+i\be)$. Therefore we get the Wilson loop correlator in $4d$ on $B^3\times S^1 $ as
\beq 
\ex{W_F \CO_\Sigma }_{B^3\times S^1}
=& V_{\al,\be} \left(\prod_{j=1}^m N_j! \right)^2 c_W(\{N_i\},g_{2d},\al,\be)
e^{2\pi i(N-1)\sum_i\be_i}
e^{-{1\ov 4}g_{2d}^2 \CA_2}
 e^{{g_{2d}^2 \CA_2^2 \ov 2\CA}}
\\&\sum_{l=1}^m N_l e^{2\pi(\al_l-i\be_l{\CA_1-\CA_2\ov \CA})} F_l \prod_{\substack{p=1, \dots ,m\\p\neq l}} G_p,
\eeq
where $G_p$ is defined in \eqref{eq:Gp} and $F_l$ is given by
\beq 
 F_l&=\int \left(\prod_{a=1}^{N_l}d x_a \right) \Delta(x_a- \del_{a,1}\CA_1/ \CA) \Delta(x_a+\del_{a,1}\CA_2/\CA)e^{-{1\ov 4}g_{2d}^2 \CA \sum_a x_a^2}
 \\&=
(g_{2d}^2 \CA/2)^{-N_l^2/2} 
(N_l-1)!
(2\pi)^{N_l/2}\left(\prod_{n=1}^{N_l-1}n!\right) L^1_{N_l-1}\left(g_{2d}^2{\CA_1\CA_2\ov  2\CA} \right).
\eeq
As a consequence, we reach  at the exact normalized correlator
\beq \label{eq:locWilsonfinal}
{\ex{W_F\CO_\Sigma}_{B^3\times S^1}\ov \ex{\CO_\Sigma}_{B^3\times S^1}}=\sum_{l=1}^m {N_l\ov N} e^{2\pi(\al_l-i\be_l{\CA_1-\CA_2\ov \CA})} {1\ov N_l} L_{N_l-1}^1\left(g_{2d}^2{\CA_1\CA_2\ov 2\CA} \right)e^{-g_{2d}^2{\CA_1\CA_2\ov 4\CA}}.
\eeq
To compare with the supergravity result in \cite{Drukker:2008wr}, we first take the large $N$ limit $N_l\gg 1$ and $\lam=g^2 N$ fixed, which gives
\beq \label{eq:1/8 BPS final large N}
{\ex{W_F\CO_\Sigma}_{B^3\times S^1}\ov \ex{\CO_\Sigma}_{B^3\times S^1}}\Bigg|_{\text{large $N$}}&\simeq \sum_{l=1}^m {N_l\ov N} e^{2\pi(\al_l-i\be_l{\CA_1-\CA_2\ov \CA})} { J_{1}\left (2 \sqrt{N_l g^2_{2d} {\CA_1 \CA_2 \ov 2\CA}}\right) \ov  \sqrt{N_l g^2_{2d} {\CA_1 \CA_2 \ov 2\CA}}}
\\& = \sum_{l=1}^m {N_l\ov N} e^{2\pi(\al_l-i\be_l{\CA_1-\CA_2\ov \CA})} {2 I_{1}\left ( \sqrt{  {\lam N_l \ov N} {4\CA_1 \CA_2 \ov \CA^2 }}\right) \ov 
 \sqrt{  {\lam N_l \ov N} {4\CA_1 \CA_2 \ov \CA^2 }
 }}.
\eeq

If we further take the strong coupling limit $\lam\rightarrow \infty$, we get
\beq
{\ex{W_F\CO_\Sigma}_{B^3\times S^1}\ov \ex{\CO_\Sigma}_{B^3\times S^1}}\Bigg|_{\text{strong coupling}}
\simeq  \sum_{l=1}^m {N_l\ov N} e^{2\pi(\al_l-i\be_l{\CA_1-\CA_2\ov \CA})} \sqrt{2\ov \pi}{ \exp \left ( \sqrt{  {\lam N_l \ov N} {4\CA_1 \CA_2 \ov \CA^2 }}\right) \ov 
\left(  {\lam N_l \ov N} {4\CA_1 \CA_2 \ov \CA^2 }
 \right)^{3/4}}.
\eeq

This is beautifully consistent with the supergravity result \cite{Drukker:2008wr} upon converting parameters accordingly, where the computation was restricted to the 1/4 BPS Wilson loop so that it has a circular geometry with a fixed latitude on $S^2$. 
 
In Appendix \ref{appsec:Wilson}   we analyze  the effective      propagator of the  circular BPS Wilson loop holonomy   in the surface operator background (using background field method and Feynman gauge) and show that it is constant and   non-zero in the Levi blocks. 
This  implies that  summing   over the rainbow diagrams     in the surface operator background exactly reproduces 
\eqref{eq:locWilsonfinal} (c.f.\cite{Erickson:2000af,Drukker:2000rr}).

\section*{Acknowledgments}

The authors would like to thank C. Closset, L. Di Pietro, N. Drukker,  D. Gaiotto, K. Hosomichi, Z. Komargodski, B. Le Floch, K. Lee, S. Lee,  M. Nguyen, T. Okuda,   M. Ro\v{c}ek, Y. Tachikawa, L. A. Takhtajan, M. Ünsal, C. Vafa and Y. Wang   for useful discussions. Research at Perimeter Institute is supported in part by the Government of Canada through the Department of Innovation, Science and Economic Development Canada and by the Province of Ontario through the Ministry of Colleges and Universities.

\vfill\eject

\begin{appendix}

\section{Geometry of $B^3\times S^1$ } \label{appsec:B3S1}
The relation between $\mathbb R^4$ and $B^3\times S^1$ can be understood as   consecutive Weyl transformations, first from $\mathbb R^4$ to $S^4$ and then from $S^4$ to $B^3\times S^1$. From $\mathbb R^4$ coordinates $\{x_i\}$ with a standard Euclidean metric $ds^2=dx_i^2$, we have a standard stereographic projection to $S^4$ of radius $R/2$ 
\beq
g_{\mu\nu}(S^4)=e^{2w_1}g_{\mu\nu}(\mathbb R^4),\quad  e^{w_1}=\left(1+{x^2\ov R^2}\right)^{-1}.
\eeq
Next we write $S^4$ coordinate in terms of the embeeding coordinates $\{X_A\}$ with $A=1,\dots,5$ satisfying $\sum_A X_A^2=r^2$. We divide the embedding coordinates into $(X_1,X_5)$ and $(X_2,X_3,X_4)$ such that $SO(2)\times SO(3)$ isometry is explicit. Then relation to the $B^3\times S^1$ coordinates is given by 
\beq
~&X_1+i X_5=r\left( 1-{\tilde x^2\ov R^2}\right)\left( 1+{\tilde x^2\ov R^2}\right)^{-1}e^{i\tau}
\\& X_i={\tilde x_i}\left( 1+{\tilde x^2\ov R^2}\right)^{-1},\quad i=2,3,4,
\eeq
together with the Weyl transformation 
\beq
g_{\mu\nu}(B^3\times S^1)=e^{2w_2}g_{\mu\nu}(S^4),\quad e^{w_2}=\left( 1+{\tilde x^2\ov R^2}\right).
\eeq
Therefore the total Weyl transformation from $\mathbb R^4$ to $B^3\times S^1$ is given by
\beq
g_{\mu\nu}(B^3\times S^1)=e^{2w} g_{\mu\nu}(\mathbb R^4), \quad e^{w}={1+\tilde x^2+(1-\tilde x^2)\cos\tau\ov 2},
\eeq
which gives the metric \eqref{eq:B3S1 metric}

\section{$\mathcal N=4$ Super Yang-Mills Theory} \label{appsec:N=4}

The Euclidean $\CN=4$ SYM on Riemannian four-manifold $(M,h_{\mu\nu})$ is given by
\beq \label{eq:N=4action}
S=\int_M\mathcal L_{bulk} \sqrt{h}d^4 x= \int {1\ov g^2}\left({1\ov 2}F_{MN}F^{MN}-\lam \Gamma^M D_M \lam  +{\bm R\ov6 } \phi^I\phi_I\right)\sqrt{h}\,d^4 x,
\eeq
where $\bm R$ is the scalar curvature and we are using 10d notation following \cite{Pestun:2007rz} such that $M=1,2,\dots,9,0$ and $A_\mu$ with $\mu=1,\dots, 4$ is a four-dimensional gauge field, $A_I=\phi_I$ with $I=5,\dots,9,0$ denotes 6 adjoint scalars, and $\lam\in S^+$ is a dimensionally reduced $Spin(10)$ Weyl Spinor with positive chirality. Also, see footnote \ref{ft:bilinear} for our convention for the bilinear form on $\mathfrak g$.

Following \cite{Pestun:2007rz}, our choice of basis $\gamma_M\in\{\gamma_1,\dots,\gamma_9,\gamma_0\}$ of the Clifford algebra $Cl(10)$ is such that 
\beq
\{\gamma^M,\gamma^N\}=2\del^{MN}.
\eeq
In the chiral representation, we have
\beq
\gamma^M=\begin{pmatrix} 0 & \tilde \Gamma^M \\ \Gamma^M &0 \end{pmatrix},
\eeq
where $\tilde \Gamma^M=(\Gamma^1,\dots,\Gamma^9,-\Gamma^0)$, and we choose them explicitly as
\beq
 \Gamma_{1}=\begin{pmatrix} 1_{8\times 8} & 0 \\ 0&  i1_{8\times 8}\end{pmatrix}
,\quad \Gamma_{M=2,\dots,9}=\begin{pmatrix} 0 & E_M^T \\   E_M&0\end{pmatrix} ,\quad ~&\Gamma_{0}=\begin{pmatrix} i1_{8\times 8} & 0 \\ 0&  i1_{8\times 8}\end{pmatrix},
\eeq
where $E_{M=2,\dots,9}$ are matrices representing the left multiplication of the octonion and we take $E_9$ to be identity such that $E_9=1_{8\times 8}$. Explicitly, we have $(E_i)^k_{j}=c^k_{ij}$ where $c^k_{ij}$ is the structure constant of the octonion algebra $e_i\cdot e_j=c_{ij}^k e_k$. We choose a cylic quaternionic triple multiplication table as $(234),(256),(357),(458),(836),(647),(728)$ such that $e_9$ is an identity and $e_2e_3=e_4$, etc.

The chirality matrix is given by
\beq
\gamma_c=-i \gamma^1 \dots \gamma^9\gamma^0,
\eeq
under which, 32 components spinors in Dirac representation $\mathcal S$ decomposes in to two 16 components Weyl spinors of positive and negtaive chirality as $\mathcal S\simeq \mathcal S^+\oplus \mathcal S^-$.

Given the conformal-Killing spinor $\e$ on $M$ whose defining equation is
\beq
\nabla_\mu \e=\tilde \Gamma_\mu \tilde \e,
\eeq
there is an superconformal symmetry given by
\beq
\del_\e A_M=\e \Gamma_M \lam,\quad \del_\e \lam={1\ov 2}F_{MN}\Gamma^{MN}\e+{1\ov 2}\Gamma_{\mu A }\phi^A  \nabla^\mu \e.
\eeq
Note that superconformal algebra is only closed on-shell (for off-shell realization see \cite{Berkovits:1993hx,Pestun:2007rz}.)

The variation of the action under $\del_\e$ is given by
\beq  \label{eq:susybd}
\del_\e S={1\ov g^2}\int_M  \nabla_\mu\left(\e\Gamma_M\lam F^{\mu M}+{1\ov 2}F_{MN}\e \Gamma^{\mu MN}\lam+2\phi_I \tilde\e\Gamma^{I\mu} \lam \right)\sqrt{h} \,d^4 x.
\eeq
On the other hand, the Euler-Lagrange variation is given by
\beq \label{eq:ELbd}
\del S={1\ov g^2}\int_M  \left[\nabla_\mu\left(2 \del A_M F^{\mu M}-\lam \Gamma^\mu \del \lam\right) +\del A_M \cdot (eom)_{A_M} +\lam \cdot (eom)_{\lam} \right ]\sqrt{h} \,d^4 x,
\eeq
where $eom_X$ denotes the Euler-Lagrange equations of motion come from the variation $\del_X$.

\section{$\langle {\cal O}_{S^2}\rangle_{S^4}$ From Disorder Definition}
\label{app:disorder}

In this Appendix we compute the vev of a surface operator wrapping a maximal $S^2$ in $S^4$  using the  disorder definition of ${\cal O}_\Sigma$ given in 
section \ref{sec:surface operator}. This yields the same answer (\ref{vevanswer}) found in section \ref{sec:surfacevev} using the coupled $4d/2d$ system.

We choose the localizing supercharge used by Pestun in \cite{Pestun:2007rz} to compute the $S^4$ partition function and Wilson loops of $4d$ ${\cal N}=2$ SCFTs.  In \cite{Gomis:2011pf} the computation of the 't Hooft loop disorder operator was performed,  and here we compute  $\langle {\cal O}_{S^2}\rangle_{S^4}$.  

The first step is to determine the space of supersymmetric saddle points in the presence  of the   singularity produced by the surface operator ${\cal O}_{S^2}$. The saddle point is      the field configuration  in Pestun's original computation with the additional   surface operator background turned on \cite{Hosomichi:2016flq}. 
 The ${\cal N}=4$ SYM action evaluated on the saddle points yields \cite{Hosomichi:2016flq}
\beq
S=-{8\pi^2\over g^2} \left(a^2+ 2ia \alpha\right)\,.
\eeq

Then there are the perturbative and non-perturbative corrections. The nonperturbative corrections are captured by the the ramified instanton partition function at the North pole and the ramified anti-instanton partition function at the South pole (see e.g. \cite{Kanno:2011fw} are references within for details on ramified instanton partition functions).

If we consider ${\cal N}=4$ SYM with the  insertion of ${\cal O}_{S^2}$
 then the expectation is that $Z_{\text{ramified}}$=1 due to fermion zero modes. These are lifted for  ${\cal N}=2$ theories, including for ${\cal N}=2^*$ (for a related discussion in the absence of surface operators see \cite{Okuda:2010ke}).

 There remains the one-loop contribution around the localization locus. This factorizes into the product from the North pole and the South pole. Therefore, $\langle{\cal O}_{S^2}\rangle_{S^4}$ in ${\cal N}=4$ SYM is given by
\beq
\langle{\cal O}_{S^2}\rangle_{S^4}=\int da e^{-{8\pi^2\over g^2} \left(a^2+ 2ia \alpha\right)}
Z_{\text{1-loop}}(a(N))
\bar Z_{\text{1-loop}}(a(S))\,.
\eeq
$a(N)$ and $a(S)$ are the values of the scalar field in ${\cal N}=2$ vector multiplet in the localizing locus in the presence of the background produced by ${\cal O}_{S^2}$. They are  given by  \cite{Hosomichi:2016flq}
\beq
a(N)=a(S)=a+i (\alpha-\ceil{\alpha})\,,
\eeq
 where $\ceil{\alpha}$ is the ceiling function.  This combination ensures periodicity of the path integral under $\alpha\rightarrow \alpha+1$. Physically, it encapsulates that in the presence of surface operator there are normalizable, singular modes that contribute to the one-loop determinant. The importance of these singular modes  has been found in other dimensions for vortex loops \cite{Drukker:2008jm,Kapustin:2012iw,Drukker:2012sr,Hosomichi:2021gxe} in $3d$
 and vortex operators in $2d$ \cite{Hosomichi:2015pia,Hosomichi:2017dbc}.

 Both in the classical action and one-loop determinants we can shift the integration variable 
\beq
a\rightarrow a+ i \alpha
\eeq
   and then the one-loop determinant in ${\cal N}=4$ SYM in presence of ${\cal O}_{S^2}$ are (see e.g. \cite{Pestun:2007rz,Gomis:2011pf})
   \beq
Z_{\text{1-loop}}=\prod_{\bm\alpha}{{\Upsilon(i \bm\alpha\cdot a+ \ceil{\bm\alpha\cdot \alpha})}\over   \Upsilon(1+ i\bm\alpha\cdot a+  \floor{\bm\alpha\cdot \alpha})}\,,
   \eeq
where $\bm\alpha$ are the roots of the Lie algebra of $G$, the gauge group of ${\cal }=4$ SYM. $\floor{\alpha}$ is the floor function.

The formula can be greatly simplified by separating the product over roots into those that have $\bm\alpha\cdot \alpha>0, \bm\alpha\cdot \alpha<0$ and $\bm\alpha\cdot \alpha=0$. Remarkably, only the roots with $\bm\alpha\cdot \alpha=0$ contribute, the rest cancel between the vector multiplet and hypermultiplet, and yield
\beq
Z_{\text{1-loop}}=\prod_{\bm\alpha\cdot \alpha=0}{\Upsilon(i \bm\alpha\cdot a )\over \Upsilon(  i\bm\alpha\cdot a+1)}=\prod_{\bm\alpha\cdot \alpha=0, \bm\alpha}{1\over \gamma(i\bm\alpha\cdot \alpha)
\gamma(-i\bm\alpha\cdot \alpha)}=\Delta_{\mathbb L}(a)\,,
\eeq
where we have used the shift properties of $\Upsilon(x+1)=\gamma(x)\Upsilon(x)$.

Putting everything together, and reintroducing the sphere radius,  we recover what we found from the computation  in section \ref{sec:surfacevev} using the $4d/2d$ definition
\beq
\langle{\cal O}_{S^2}\rangle_{S^4}= \int da\,  \Delta_{\mathbb L}(a) e^{-{8\pi^2\over g^2}  a^2 } ={r}^{- \text{dim} \,\mathbb L } \left({g^2\over 4\pi}\right)^{\text{dim} \,\mathbb L/2}e^{-8\pi^2 \alpha^2/g^2}\,. 
\eeq

\section{Supersymmetric Boundary Conditions and Boundary Term of Surface Operator in $\mathcal N=4$ Super Yang-Mills Theory}  \label{appsec:bdry}

\subsection{Planar Surface Operator in $\mathbb R^4$} \label{sec:R2R4}
In this section, we determine the supersymmetric boundary conditions and boundary terms for the planar surface operator in $\mathbb R^4$ using the guiding principles explained in section \ref{sec: localization}. We set the planar surface operator's profile to be $x^3=x^4=0$. Introducing $\phi_w={1\ov \sqrt{2}}(\phi^7-i\phi^8)$ and $z=x^3+i x^4=re^{i\psi}$, the background produced by the surface operator is\footnote{Here we choose the orientation opposite compared to \eqref{eq:scalarbackground} to be directly comparable with $B^3\times S^1$ analysis, see footnote \eqref{ft:orientation}.}
\beq \label{eq:R2background}
\phi_w={\be+i\gamma\ov \sqrt{2} z},\quad A=\al d\psi.
\eeq
The conformal Killing spinors in $\mathbb R^4$ are given by $\e=\e_s+x^\mu \tilde \Gamma_\mu \e_c$, and the surface defect background is preserved by the half of supercharges characterized by the following half-BPS condition
\beq \label{eq:R4BPS}
\Gamma^{wz}\e_s=0,\quad \Gamma^{wz}\e_c=0.
\eeq
Next, we regularize the surface operator where we excise the infinitesimal neighborhood around $x_3=x_4=0$. The most natural cutoff is to have a boundary $\pa M\simeq S^1\times  \mathbb R^2$, such that  $M:x_3^2+x_4^2>\rho^2$ has a $\mathbb R^2\times S^1$ with an infinitesimal boundary $\pa M:x_3^2+x_4^2=\rho$.

Our task is to find boundary conditions and possible boundary term on $\pa M$ which makes the theory with and has consistent Euler-Lagrange variational principle and manifest supersymmetry.

We first analyze the possible boundary conditions on fermions. Since it is a first-order system, the boundary condition should be a projection that removes half degrees of freedom at the boundary. If we look at \ref{eq:ELbd}, there is the following fermionic part of boundary contribution from the EL principle 
\beq \label{eq:ELfer}
\int_{\pa M} n_\mu \lam \Gamma^\mu \del \lam  \sqrt{h|_{\pa M}}\,d^3x ,
\eeq
where $n_\mu$ is a (inward) normal vector at the boundary. Note that there is no possible boundary term to compensate it purely algebraically, therefore \eqref{eq:ELfer} has to be purely vanish by the fermionic boundary condition. The fermionic boundary condition invariant under $SO(4)\times U(1)\times SO(2)_{\parallel}$ of surface operator and makes  \eqref{eq:ELfer} identically vanish is given by 
\beq \label{eq:fbc}
\Gamma^{w\bar z}\lam=\Gamma^{\bar wz}\lam=0\quad\text{or}\quad \Gamma^{3478}\lam=-\lam\quad \text{(fermionic boundary condition)}.
\eeq
From this, we can determine the bosonic boundary conditions by looking at the superpartner of \eqref{eq:fbc}. Explicitly, supersymmetry invariance of the boundary condition requires ($P_+={1+\Gamma^{3478}\ov 2}$)
\beq
0&=\del_\e {P_+\lam}
\\&=F_{12}\Gamma^{12}\e+F_{i\hat I}\Gamma^{i\hat I}
\e
+{1\ov 2}F_{\hat I\hat J}\Gamma^{\hat I\hat J}\e
+F_{z\bar z}\Gamma^{z\bar z}\e
+F_{w  \bar w}\Gamma^{w \bar w}\e
+F_{z\bar  w}\Gamma^{z  \bar w}\e
+F_{\bar z  w}\Gamma^{\bar z w}\e
 -2\phi_{\hat I} \tilde\Gamma^{\hat I} \tilde \e,
\eeq
where we introduced a label $i,j,\dots=1,2$,~ $a,b,\dots=3,4$, ~$\hat I,\hat J,\dots=5,6,9,0$,~ $\hat A,\hat B,\dots=7,8$. Now using \eqref{eq:R4BPS}, we see that the bosonic boundary conditions are determined as 
\beq \label{eq:bbc}
~& A_i=\phi_{\hat I}=0,\quad D_z \phi_{\bar w}=D_{\bar z}\phi_w=F_{z\bar z}-{1\ov 2}F_{w\bar w}=0 \quad  \text{(bosonic boundary condition)},
\eeq
therefore classically unexcited fields $(A_i,\phi_{\hat I})$ obey Dirichlet boundary conditions and excited fields $(A_a,\phi_{\hat A})$ obey Neumann boundary conditions.

Given the boundary conditions, the remaining task is to determine the boundary term. First, we look at the Euler-Lagrange variation. Imposing the boundary conditions on \eqref{eq:ELbd}, we have an on-shell bulk contribution to the boundary as 
\beq 
g^2 \del S|_{OS}=&\int_{\pa M} \left[{i}\left(\del A_a F^{\bar z a} +\del \phi_{\hat A} D^{\bar z} \phi^{\hat A}\right)d z-{i}\left(\del A_a F^{za} +\del \phi_{\hat A} D^z \phi^{\hat A}\right)d\bar z \right] d^2 x_i,
\eeq
where we choose the integration contour for $dz$, $d\bar z$ to be  counter-clockwise. This naturally determines that the following boundary term\footnote{We note that this boundary term is natural since it can also be obtained from the Bogomolnyi trick as in \cite{Giombi:2009ek}
}
\beq \label{eq:R4bdryterm}
g^2 S_{bdry}=\int_{\pa M} -2i\phi_{\bar w}( D_z \phi_w dz+ D_{\bar z}\phi_w d\bar z) d^2 x_i,
\eeq
which makes the action principle well-defined for the total bulk-boundary action $S_{tot}=S+S_{bdry}$, where one can check that $\del S_{tot}|_{OS}=0$ under the boundary conditions \eqref{eq:fbc}, \eqref{eq:bbc}. 

At the same time, one can check that the boundary term \eqref{eq:R4bdryterm} automatically makes the total action supersymmetric as well, i.e. $\del_\e S_{tot}=0$ under the boundary conditions and the BPS supercharges \eqref{eq:R4BPS}.

We can additionally convince ourselves of the consistency of the above procedures by applying them to the case of straight line 1/2 BPS `t Hooft operator in $\mathbb R^4$. The regulator introduces a boundary $S^2_\rho\times R\subset R^3\times R$ and the consistent boundary term turns out to be $ S_{bdry}=-{2\ov g^2}\int_{S^2_\rho\times R} \Phi F\wedge d\tau$ ($d\tau$ is a canonical one-form on $R$) which matches with the known result \cite{Giombi:2009ek,Gomis:2011pf}. On the other hand, this sheds new light on the boundary conditions for `t Hooft operator, which naturally imposes a Dirichlet for unexcited fields and Neumann-like for excited fields which has the  form of the Bogomolnyi equations $*_{R^3}F-d^{R^3}_A\Phi=0$.

\subsubsection{$\langle {\cal O}_{\mathbb R^2}\rangle$ From Semiclassics } \label{appsec:R2onshell}

We can evaluate $\langle {\cal O}_{\mathbb R^2}\rangle$ semiclassically by computing the total bulk+boundary on-shell action using the supersymmetric boundary term given in \eqref{eq:R4bdryterm}. Given the profile of background fields \eqref{eq:R2background}, it is straightforward compute the semiclassical contribution to the bulk and boundary term to get
\beq
S_{tot}=S+S_{bdry}=0.
\eeq
Semiclassically, there are $1/\rho^2$ divergence from the bulk action and it is exactly cancelled by the boundary term without any non-trivial finite piece. Therefore, we see that the semiclassical expectation value of the planar surface operator in $\mathbb R^4$ is trivial, and it is natural to expect this to be true at the quantum level.

\subsection{Spherical Surface Operator in $\mathbb R^4$} \label{sec:S2R4}

We can also determine the supersymmetric boundary conditions and boundary terms of the spherical surface operator in $\mathbb R^4$ with radius $R$. In terms of the Cartesian coordinates $(x_1,\dots,x_4)$, we choose the location of the surface operator to be $\sqrt{x_1^2+x_2^2+x_3^2}=R$ and $x_4=0$. The planar and spherical defect are related by the conformal transformation, and hence it has a following singular background
\beq \label{eq:S2R4 defect profile}
\phi_w&=-{\be+i\gamma \ov\sqrt{2} }{2R \ov [x_1^2+x_2^2+x_3^2+(x_4-iR)^2 ]}={\be+i\gamma\ov \sqrt{2}\tilde r e^{i\psi}}, \quad A=\al d\psi.
\eeq
where $\psi$ parametrizes $S^1$ in $AdS_3\times S^1$ coordinates $ds^2=\tilde r^2 (d\rho^2+\sinh^2 \rho (d\vartheta^2+\sin^2\vartheta  d\varphi^2)+d\psi^2)$ and $\tilde r$ is a conformally invariant distance from the sphere
\beq
\tilde r={\sqrt{(x_1^2+x_2^2+x_3^2+x_4^2-a^2)^2+4R^2 x_4^2}\ov 2R}={R\ov \cosh\rho+\cos\psi}.
\eeq
Therefore, it is natural to regulate the spherical surface operator so that the cutoff is located at fixed $\tilde r=\tilde r_0$ and we take it to be infinitesimal so that $\tilde r_0\ll a$.

Let's introduce a spherical coordinate which respects a manifest $SO(3)$ symmetry of the spherical defect
\beq
ds^2=dr^2+r^2(d\vartheta^2+\sin^2\vartheta  d\varphi^2)+dx_4^2
\eeq
where $r=\sqrt{x_1^2+x_2^2+x_3^2}$. To determine the boundary terms and boundary conditions, it is convenient to use a complex coordinate $(\zeta,\bar \zeta, \vartheta,\varphi)$ where $\zeta=x_4+ir$ parametrizes the upper half-plane. Then the half-BPS sector of the bulk conformal Killing spinors $\e=\e_s+x^\mu \tilde \Gamma_\mu \e_c$ preserved by the background are characterized by
\beq \label{eq:S2R4 projector spinor}
\Gamma^{4w}\e_s=ia \tilde \Gamma^{w} \e_c, \quad \Gamma^{4\bar w}\e_s=-ia \tilde \Gamma^{\bar w} \e_c.
\eeq

At the boundary, the projection equation \eqref{eq:S2R4 projector spinor} can be simplified up to $o(\tilde r_0)$ as
\beq 
\Gamma^{w\zeta}\e=0 ~~\text{or}~~ \Gamma^{4r78}\e=\e,
\eeq
which suggests that natural fermionic boundary condition is (up to $o(\tilde r_0)$)
\beq \label{eq:S2R4fbc}
\Gamma^{w\bar \zeta}\lam=0  ~~\text{or}~~ \Gamma^{4r78}\lam=-\lam.
\eeq

We can determine the bosonic boundary conditions from the superpartner of the fermionic boundary condition \eqref{eq:S2R4fbc}. We label the coordinates so that $i,j,\dots=\vartheta,\vp$, $a,b,\dots=\zeta,\bar\zeta$, $I,J,\dots=5,6,9,0$, $X,Y,\dots=w,\bar w$, then we have (up to $o(\tilde r_0)$)
\beq
0=&\del_\e {(1+P)\ov 2}\lam
\\=&F_{\vartheta\vp}\Gamma^{\vartheta\vp}\e+{1\ov 2}F_{IJ}\Gamma^{IJ}\e+F_{iI}\Gamma^{iI}\e
+(F_{\zeta \bar\zeta}-{1\ov 2}F_{w\bar w})\Gamma^{\zeta\bar\zeta}\e+F_{w\bar \zeta}\Gamma^{w\bar \zeta}\e+F_{\bar w \zeta}\Gamma^{\bar w \zeta}\e
\\&-2\phi_I \tilde \Gamma^I {1+P\ov 2} \e_c-2\phi_X \tilde \Gamma^X{1-P\ov 2} \e_c.
\eeq
From this, we again obtain Dirichlet boundary conditions for unexcited fields and Neumann-like boundary conditions on the excited fields as
\beq \label{eq:S2R4bbc}
A_i=\phi_I=0,\quad F_{\zeta\bar\zeta}-{1\ov 2}F_{w\bar w}=0
,\quad F_{\bar\zeta w}+{i\ov 2R} \phi_w=0
,\quad F_{ \zeta \bar w}-{i\ov 2R} \phi_{\bar w}=0,
\eeq
again which should be understood with an $o(\tilde r_0)$ accuracy.

Now armed with boundary conditions \eqref{eq:S2R4fbc} and \eqref{eq:S2R4bbc}, we can find a boundary term which has a well-defined action principle and desired supersymmetry. After trial and error, we can find following boundary term
\beq \label{eq:S2R4bdryterm}
~& S_{bdry}=S_{bdry,1}+S_{bdry,2}
\\& g^2S_{bdry,1}=\int_{\pa M} 
-{i}\phi_{
\bar w}(D_{\zeta}  \phi_w d\zeta +D_{\bar \zeta} \phi_w d\bar \zeta)\text{Im}(\zeta)^2 d\Omega
 +{i}\phi_w (D_{\zeta}\phi_{\bar w}d\zeta+D_{\bar \zeta}\phi_{\bar w}d\bar\zeta) \text{Im}(\zeta)^2d\Omega_2
\\&
g^2 S_{bdry,2}=\int_{\pa M}{1\ov a}\phi_w \phi_{\bar w}(d\zeta+d\bar\zeta) \text{Im}(\zeta)^2 d\Omega_2,
\eeq
where again we take the integration contour $d\zeta,d\bar\zeta$ to be counter-clockwise. Note that $S_{bdry,1}$ is an analog of the planar surface operator boundary term \eqref{eq:R4bdryterm} while $S_{bdry,1}$ is genuienly a new contribution compared to the planar case.

\subsubsection{$\langle {\cal O}_{S^2}\rangle$ From Semiclassics }
\label{appsec:S2onshell}
We can semiclassically evaluate $\langle {\cal O}_{S^2}\rangle$ on $\mathbb R^4$ by computing the total bulk+boundary on-shell action with a boundary term \eqref{eq:S2R4bdryterm}. Using the background profile \eqref{eq:S2R4 defect profile}, we obtain
\beq \label{eq:on-shellS2}
S_{tot}=S+S_{bdry,1}+S_{bdry,2}={2\pi^2 (\be^2+\gamma^2)\ov g^2},
\eeq
where the $o(1/\rho^2)$ divergences appear only in $S$ and $S_{bdry,1}$ in \eqref{eq:S2R4bdryterm} which cancel each other, and each terms $S$, $S_{bdry,1}$ and $S_{bdry,2}$ contributes to the non-zero finite piece to the total on-shell action \eqref{eq:on-shellS2}.

We note that the semiclassical $\ex{\CO_{S^2}}$ in \eqref{eq:on-shellS2} is different from the exact result separately obtained by the $4d/2d$ point of view in section \ref{sec:surfacevev} and the disorder definition \ref{app:disorder}. This is not a contradiction but merely suggests that the description of the surface operators with a geometric UV cutoff preserves a different massive subalgebra other than $SU(2|1)_A$. Furthermore, \eqref{eq:on-shellS2} is also consistent with S-duality since it is invariant under \eqref{actionSS} and \eqref{eq:s-dual}.

\section{Perturbative Correlation Function with Chiral Primary Operators} \label{appsec:CPO}

In this Appendix we perform the perturbative analysis of the normalized correlation function of the surface operator with the chiral primary operators (CPO) $\Oo_{\Delta,k}$. Using   conformal invariance, we consider a planar surface operator in $\mathbb R^4$. Surprisingly, we will identify finite number of Feynman diagrams which exactly capture the results obtained using the supersymmetric localization in section \ref{sec:CPOloc}.

As discussed in section \ref{sec:surface operator}, the chiral primary operators with scaling dimension $\Delta$ are given by \cite{Drukker:2008wr}: 
\begin{equation}
    \label{cpo}
    \mathcal{O}_{\Delta,k}^I= \frac{(8 \pi^2)^{\Delta/2}}{\lambda^{\Delta/2} \sqrt{\Delta}}C^{\Delta,k}_{i_1 \ldots i_\Delta} \Tr \left( \phi^{\{i_1}\ldots \phi^{i_\Delta \}} \right),
\end{equation}
where 
$$
Y^{\Delta,k}= C^{\Delta,k}_{i_1 \ldots i_\Delta} x^{i_1}\ldots x^{i_\Delta}
$$
are the scalar $SO(4)$ invariant $SO(6)$ spherical harmonics. $C^{\Delta,k}_{i_1 \ldots i_\Delta}$ is a symmetric traceless tensor. These operators have $U(1)_R$ charges $k=-\Delta, -\Delta+2, \ldots, \Delta-2, \Delta$. The $SO(6)$ spherical harmonics are solutions for Laplace's equation on $S^5$, where we consider the metric: 
\begin{equation}
    \label{metricS5g}
    ds^2= \cos^2\theta d\Omega_3 + d\theta^2+ \sin^2 \theta d\phi^2, \quad \theta \in [0,\pi/2] \text{ and } \phi \in [0,2 \pi].
\end{equation}
The $SO(4)$ invariant spherical harmonics can be written as 
\begin{equation}
\label{spherical}
    Y^{\Delta,k}= c^{\Delta, k} y^{\Delta, k}(\theta) e^{i k \phi},
\end{equation}
where $\Delta \in \mathbb{N}$ and $k \in \{-\Delta, -\Delta+2, \ldots, \Delta-2,\Delta\}$. The constant $c^{\Delta,k}$ is defined so that the CPO are unit normalized, and the function $y^{\Delta,k}$ is given by  
$$
y^{\Delta, k} = \sin ^{| k| }(\theta ) \, _2F_1\left(\frac{| k| -\Delta }{2},2+\frac{\Delta +| k| }{2},| k| +1;\sin ^2(\theta )\right).
$$
It will also be useful to introduce the following notation: 
\begin{equation}
\label{constantspherical}
     Y^{\Delta,k}(\theta=\pi/2,\phi)=C_{\Delta,k}\; e^{i k \phi}\,.
\end{equation}
Also, we impose the following normalization condition in the spherical harmonics
\begin{equation}
    \label{normalization}
    \int Y^{\Delta_1,j_1} Y^{\Delta_2,j_2}= \pi^3 z(\Delta) \delta_{\Delta_1 \Delta_2} \delta_{j_1 j_2},
\end{equation}
where 
$$
z(\Delta)=\frac{2^{-\Delta+1}}{\Delta+1}.
$$

From this, we can find that $\Delta=2,\ldots,6$ CPO are given by (In this Appendix, we follow the convention in section \ref{sec:surface operator}  so that the six scalars are labeled by $\phi_{1,\dots, 6}$ and the complex scalar is chosen to be $\Phi={1\ov \sqrt{2}}({\phi_5+i\phi_6})$)
\begin{equation}
\small
    \label{operators}
    \begin{aligned}
        \mathcal{O}_{2,0} &= \frac{4 \pi^2}{\sqrt{6}\lambda}\Tr\left( 4 \Phi \Bar{\Phi} - \sum_{i=1}^4 \phi^i \phi^i \right), \\
        \mathcal{O}_{2,2} &= \frac{8 \pi^2}{\sqrt{2}\lambda} \; \Tr\left( \Phi \Phi \right), \\
        \mathcal{O}_{3,1} &= \frac{8 \pi^3}{\lambda^{3/2}} \; \Tr\left( 2 \Phi^2 \Bar{\Phi} - \Phi \sum_{i=1}^4 \phi^i \phi^i \right), \\
        \mathcal{O}_{3,3} &= \frac{32 \pi^3}{\sqrt{6}\lambda^{3/2}} \; \Tr\left( \Phi^3 \right),\\
        \mathcal{O}_{4,0} &= \frac{8 \pi^4}{\sqrt{5}\lambda^2} \Tr\left(12 \frac{1}{6}\left( 2 (\Phi \Bar{\Phi})^2 + 4 \Phi^2 \Bar{\Phi}^2 \right)-12 \frac{1}{12} \left( 8 \Phi \Bar{\Phi} \phi_i \phi_i + 4 \Phi \phi_i \Bar{\Phi} \phi_i \right) + \frac{1}{6} \left(4 \phi_i \phi_i \phi_j \phi_j + 2 \phi_i \phi_j \phi_i \phi_j \right) \right),\\
        \mathcal{O}_{4,2} &= \frac{32 \pi^4}{\sqrt{10}\lambda^2} \Tr\left( 4 \Phi^3 \Bar{\Phi} - 3 \frac{1}{6} \left( 4 \Phi^2 \phi_i \phi_i + 2 \Phi \phi_i \Phi \phi_i \right) \right),\\
        \mathcal{O}_{5,1}&= \frac{64 \pi^5}{\sqrt{5}\lambda^{5/2}} \Tr\left( 4 \frac{1}{2}\left(\Phi^3 \Bar{\Phi}^2 + \Phi (\Phi \Bar{\Phi})^2  \right) \right.
        \\&\left.\quad-6 \frac{1}{6} \left( \Phi^2 \Bar{\Phi} \phi_i \phi_i + \Phi \Bar{\Phi} \Phi \phi_i \phi_i + \Phi^2 \phi_i \phi_i \Bar{\Phi}  + \Phi \phi_i \Phi \Bar{\Phi} \phi_i + \Phi \phi_i \Bar{\Phi} \Phi \phi_i + \Phi^2 \phi_i \Bar{\Phi} \phi_i \right) \right. \\
        & \left. \quad +  \frac{1}{3} \left( \Phi \phi_i \phi_i \phi_j \phi_j +   \Phi \phi_i \phi_j \phi_i \phi_j + \Phi \phi_i \phi_j \phi_j \phi_i  \right) \right),\\
        \mathcal{O}_{5,3}&= \frac{256 \pi^5}{\sqrt{10}\lambda^{5/2}} \Tr\left( \Phi^4 \Bar{\Phi} - \frac{1}{2} \left( \Phi^3 \phi_i \phi_i + \Phi^2 \phi_i \Phi \phi_i\right) \right),\\
    \mathcal{O}_{5,5}&= \frac{256 \pi^5}{\sqrt{10}\lambda^{5/2}} \Tr\left( \Phi^5 \right),\\
    \mathcal{O}_{6,0}&= \frac{64 \pi^6}{\sqrt{42}\lambda^{3}} \Tr\left( 32 \frac{3}{10} \left( \Phi^3 \Bar{\Phi}^3 + \Phi^2 \Bar{\Phi} \Phi \Bar{\Phi}^2 + \Phi^2 \Bar{\Phi}^2 \Phi \Bar{\Phi}  + \frac{1}{3} \left(\Phi \Bar{\Phi} \right)^3 \right) + \ldots \right),\\
\mathcal{O}_{6,2}&= \frac{1280 \pi^6}{\sqrt{70}\lambda^{3}} \Tr\left( 2 \frac{1}{10} \left( 6 \Phi^4 \Bar{\Phi}^2 +6 \Phi (\Phi \Bar{\Phi})^2 + 3(\Phi^2 \Bar{\Phi})^2\right) + \ldots \right) ,\\
\mathcal{O}_{6,4}&= \frac{256 \pi^6}{\sqrt{14}\lambda^{3}} \Tr\left( 4 \Phi^5 \Bar{\Phi}  + \ldots  \right),\\
\mathcal{O}_{6,6}&= \frac{512 \pi^6}{\sqrt{6}\lambda^{3}} \Tr\left( \Phi^6 \right),\\
    \end{aligned}
\end{equation}
where $\ldots$ includes the permutations needed to make the operator $SO(4)$ invariant.

The normalized correlation function between the surface operator $\Oo_\Sigma$ and these local operators is computed as the expectation value of these CPO in the theory with the defect. The surface operator is defined by the singularity profiles of the fields near the surface as given in section \ref{sec:surface operator}. At the quantum level, one should perform the path integral over fluctuations of the fields around the classical singularities, i.e. decomposes fields into classical and the quantum fluctuations. For scalars, we decompose as $\phi^I=\phi_0^I+\vp^I$, and accordingly $\Phi=\Phi_0+\vp$.

The tree level result is the semiclassical computation presented in \cite{Drukker:2008wr}, obtained by substituting the classical singular profile for $\Phi, \bar \Phi$ into the operators \eqref{operators}. Now we will compute the quantum correction given by sum of all possible Wick contractions of fields inside the operators \eqref{operators} by using the tree level propagator of quantum fluctuations. In order to do this we need the near coincident limit of the propagators among all the quantum fields that appear in the CPO: $\varphi, \, \bar{\varphi}, \, \varphi^i$. 
We will shortly perform this computation and prove that the only nonzero propagator that includes $\varphi$ or $\bar{\varphi}$   is  $\langle \varphi_{ij} \bar{\varphi}_{kl} \rangle$. Together with the vanishing one-point function of $\varphi$, $\bar{\varphi}$\footnote{Any contraction with an odd number of $\varphi$ or $\bar{\varphi}$ will vanish.}, we can see the structure of quantum corrections from Wick contractions follows.

The simplest case is when $\Delta=|k|$, where we see that there is no quantum corrections from the Wick contractions, which agrees with the dual supergravity computations using both the probe brane aproximation and bubbling geometry configuration in \cite{Drukker:2008wr}. 

For $\Delta<|k|$ case, there are non-trivial quantum corrections from the Wick contractions. For example, let's consider an operator $\Tr \Phi^4 \bar{\Phi}^2$ which is a part of $\CO_{6,2}$. In \ref{fig:enter-label}, we show the possible Wick contractions. At each cross there is one operator and the dashed circle in which they are inserted represents the trace. The external lines correspond to insertions of the background value $\Phi_0$ and the lines joining two nodes represent the propagator $\langle \varphi \bar{\varphi} \rangle$. Since the propagator is of order $g^2$, (d) and (e) in Figure \ref{fig:enter-label} correspond to the higher order corrections coming from Wick contractions in $\Tr \Phi^4 \bar{\Phi}^2$. In this example there are enough fields so that different planarity degree contributions appear: (d) is planar and will be the leading term in the large $N$ limit whereas (e) is nonplanar and will be subleading. 
 Indeed, with the propagator we will introduce (equation \eqref{eq:final propagator}) the diagram in (e) vanishes. 
% Later we will show that the result obtained this way exactly coincides with the exact result proven using localization, so it makes sense to say that all interacting diagrams cancel at each order in $g^2$, so that the full answer arises from Wick contractions. 

A striking fact about the CPO $\CO_{\Delta,k}$ is that the quantum corrections from the Wick contractions solely depends on the  classical fields $\Phi_0$ and the near coincident limit of the difference of the propagators\footnote{where we used the $SO(4)$ invariance for the equality.} 
$\langle \varphi \Bar{\varphi} -{1\ov 4} \sum_{I=1}^4\varphi_{ I} \varphi_{ I} \rangle=\langle \varphi \Bar{\varphi} -\varphi_{1} \varphi_{1} \rangle$,
as in the following examples:
\begin{equation}
    \label{ops2}
    \begin{aligned}
         \langle \mathcal{O}_{2,0} \rangle |_{\Oo_\Sigma}&= \frac{16 \pi^2}{\sqrt{6}\lambda}  \;\Tr\left( \;   \left(\Phi_0 \Bar{\Phi}_0\right)  +\langle \varphi \Bar{\varphi} - \varphi_1 \varphi_1 \rangle  \right) , \\
         \langle \mathcal{O}_{2,2} \rangle |_{\Oo_\Sigma}&= \frac{8 \pi^2}{\sqrt{2}\lambda} \; \Tr\left( \Phi_0 \Phi_0 \right), \\
         \langle \mathcal{O}_{3,1} \rangle |_{\Oo_\Sigma}&=  \frac{16 \pi^3}{\lambda^{3/2}} \; \Tr\left(   \left( \Phi_0^2 \Bar{\Phi}_0 \right) + 2 \Phi_0 \langle \varphi \Bar{\varphi} - \varphi_1 \varphi_1 \rangle  \right) , \\
         \langle \mathcal{O}_{3,3} \rangle |_{\Oo_\Sigma}&= \frac{32 \pi^3}{\sqrt{6}\lambda^{3/2}} \; \Tr\left( \Phi_0^3 \right) , \\
        \langle \mathcal{O}_{4,0} \rangle|_{\Oo_\Sigma} &=  \frac{8 \pi^4}{\sqrt{5}\lambda^2} \left( 12 (\Phi_0^{ij} \Bar{\Phi}_0^{jk}\Phi_0^{kl} \Bar{\Phi}_0^{li}) + 32 \Phi_0^{ij} \Bar{\Phi}_0^{jk} \langle \varphi^{kl} \bar{\varphi}^{li} - \phi_1^{kl} \phi_1^{li} \rangle + \right. \\
        &\left. +16  \Phi_0^{ij} \Bar{\Phi}_0^{kl} \langle \varphi^{jk} \bar{\varphi}^{li} - \phi_1^{jk} \phi_1^{li} \rangle  +  16 \langle \varphi^{ij} \bar{\varphi}^{jk} - \phi_1^{ij} \phi_1^{jk} \rangle \langle \varphi^{kl} \bar{\varphi}^{li} - \phi_1^{kl} \phi_1^{li} \rangle   + \right. \\
        &+ \left. 8 \langle \varphi^{ij} \Bar{\varphi}^{kl} - \phi_1^{ij} \phi_1^{kl}\rangle  \langle \varphi^{jk} \bar{\varphi}^{li} - \phi_1^{jk} \phi_1^{li} \rangle \right),\\
        \langle \mathcal{O}_{4,2} \rangle|_{\mathcal{O}_\Sigma} &= \frac{32 \pi^4}{\sqrt{10}\lambda^2} \left( 12 \Phi_0^{ij} \Phi_0^{jk}\Phi_0^{kl} \Bar{\Phi}_0^{li} + 8 \Phi_0^{ij} \Phi_0^{jk} \langle \varphi^{kl} \bar{\varphi}^{li} - \phi_1^{kl} \phi_1^{li} \rangle  4  \Phi_0^{ij} \Phi_0^{kl} \langle \varphi^{jk} \bar{\varphi}^{li} - \phi_1^{jk} \phi_1^{li} \rangle \right), \\
         \langle \mathcal{O}_{4,4}\rangle|_{\mathcal{O}_\Sigma} &= \frac{32 \pi^4}{\lambda^2} \Tr\left( \Phi_0^4 \right) ,\\
         \langle \mathcal{O}_{5,1} \rangle|_{\mathcal{O}_\Sigma} &= \frac{128 \pi^5}{\sqrt{5}\lambda^{3/2}}  \left( 2 \Phi_0^{ij} \Phi_0^{jk} \Phi_0^{kl} \Bar{\Phi}_0^{lm}\Bar{\Phi}_0^{mi} + 6 \Phi_0^{ij} \Phi_0^{jk} \bar{\Phi}_0^{kl}  \langle \varphi^{lm} \bar{\varphi}^{mi} - \phi_1^{lm} \phi_1^{mi} \rangle   + \right. \\
     &+ \left. 4 \Phi_0^{ij} \Phi_0^{lm} \bar{\Phi}_0^{jk}  \langle \varphi^{kl} \bar{\varphi}^{mi} - \phi_1^{kl} \phi_1^{mi} \rangle  + 2 \Phi_0^{ij} \Phi_0^{jk} \bar{\Phi}_0^{lm}  \langle \varphi^{kl} \bar{\varphi}^{mi} - \phi_1^{kl} \phi_1^{mi} \rangle  +   \right. \\
     &+ 4 \Phi_0^{ij} \langle \varphi^{jk} \bar{\varphi}^{kl} - \phi_1^{jk} \phi_1^{kl} \rangle \langle \varphi^{lm} \bar{\varphi}^{mi} - \phi_1^{lm} \phi_1^{mi} \rangle +  \\
     &+  4 \Phi_0^{ij} \langle \varphi^{jk} \bar{\varphi}^{lm} - \phi_1^{jk} \phi_1^{lm} \rangle \langle \varphi^{mi} \bar{\varphi}^{kl} - \phi_1^{mi} \phi_1^{kl} \rangle + \\
     &+ \left. \Phi_0^{ij}\langle \varphi^{jk} \Bar{\varphi}^{lm} - \phi_1^{jk} \phi_1^{lm}\rangle  \langle \varphi^{kl} \bar{\varphi}^{mi} - \phi_1^{kl} \phi_1^{mi} \rangle   \right), \\
     \langle \mathcal{O}_{5,3} \rangle|_{\mathcal{O}_\Sigma} &= \frac{256 \pi^5}{\sqrt{10}\lambda^{5/2}} \Tr\left( \Phi_0^{ij} \Phi_0^{jk} \Phi_0^{kl} \Phi_0^{lm}\Bar{\Phi}_0^{mi}+ 2 \Phi_0^{ij} \Phi_0^{jk} \Phi_0^{kl}  \langle \varphi^{lm} \bar{\varphi}^{mi} - \phi_1^{lm} \phi_1^{mi} \rangle+ \right. \\
     +& \left. 2 \Phi_0^{ij} \Phi_0^{lm} \Phi_0^{jk}  \langle \varphi^{kl} \bar{\varphi}^{mi} - \phi_1^{kl} \phi_1^{mi} \rangle \right),\\
     \langle \mathcal{O}_{5,5} \rangle|_{\mathcal{O}_\Sigma} &= \frac{256 \pi^5}{\sqrt{10}\lambda^{5/2}} \Tr\left( \Phi_0^5 \right).
        \eeq
        
%         \beq   
%      \langle \mathcal{O}_{5,3} \rangle|_{\mathcal{O}_\Sigma} &= \frac{256 \pi^5}{\sqrt{10}\lambda^{5/2}} \Tr\left( \Phi_0^{ij} \Phi_0^{jk} \Phi_0^{kl} \Phi_0^{lm}\Bar{\Phi}_0^{mi}+ 2 \Phi_0^{ij} \Phi_0^{jk} \Phi_0^{kl}  \langle \varphi^{lm} \bar{\varphi}^{mi} - \phi_1^{lm} \phi_1^{mi} \rangle+ \right. \\
%      +& \left. 2 \Phi_0^{ij} \Phi_0^{lm} \Phi_0^{jk}  \langle \varphi^{kl} \bar{\varphi}^{mi} - \phi_1^{kl} \phi_1^{mi} \rangle \right),\\
%      \langle \mathcal{O}_{5,5} \rangle|_{\mathcal{O}_\Sigma} &= \frac{256 \pi^5}{\sqrt{10}\lambda^{5/2}} \Tr\left( \Phi_0^5 \right).
% \end{aligned}
% \end{equation}

Note that we have used that in the coincident limit the difference difference of propagators is symmetric under the exchange of $\varphi$ and $\bar \varphi$\footnote{This can be seen in equation \eqref{eq:final propagator}.}:  
\begin{equation}
    \lim_{x' \to x} \langle \varphi_{ij}(x) \bar{\varphi}_{kl}(x') - \phi^1_{ij}(x) \phi^1_{kl}(x') \rangle = \lim_{x' \to x} \langle \bar \varphi_{ij}(x) {\varphi}_{kl}(x')  - \phi^1_{ij}(x) \phi^1_{kl}(x')  \rangle.
\end{equation}

\begin{figure}
    \centering
\input{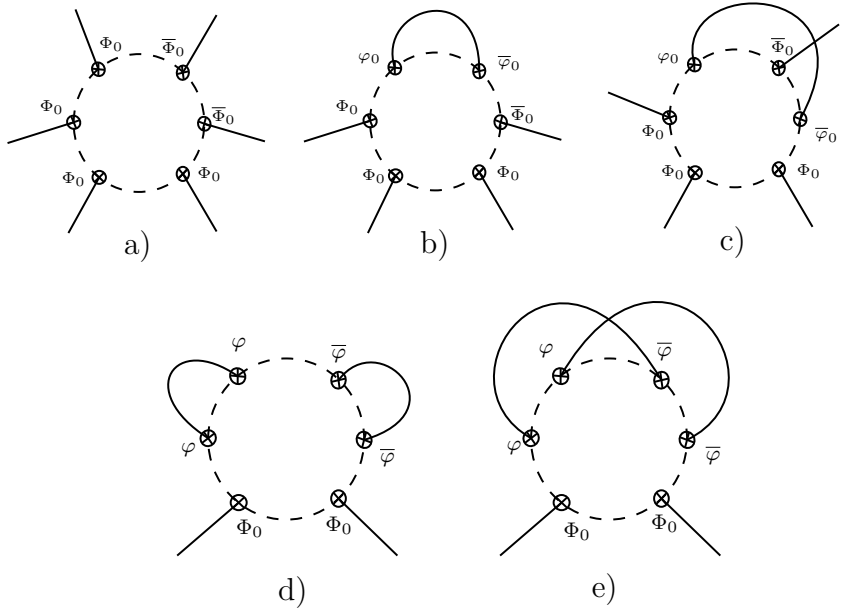}
    \caption{Feynman diagrams corresponding to Wick contractions in $\Tr \Phi^4\bar{\Phi}^2$.}
    \label{fig:enter-label}
\end{figure}
 
 In the following analaysis we will compute the propagators of all the scalars fields of the theory in the background of the planar surface operator, and this difference of propagators. 

\subsection{Background Field Method}

The goal of this section is to compute the propagators of the scalar fields of $\mathcal{N}=4$ super Yang-Mills in the presence of the planar surface operator $\mathcal{O}_{\mathbb R^2}$. To do that, we will use the background field method, since picking this gauge will simplify the perturbative calculations of the correlation functions. This gauge fixing condition can be written in a covariant way in $10$ dimensions \cite{Gomis:2009ir}.

We construct the action of $\mathcal{N}=4$ super Yang-Mills in $\mathbb{R}^4$ by dimensional reduction of $\mathcal{N}=1$ super Yang-Mills in 10 dimensions as in \eqref{eq:N=4action}: 
\begin{equation}
    \label{10dN1}
    S = \frac{1}{g^2} \int d^{10}x \; \sqrt{g}\; \left( \frac{1}{2} F^{MN}F_{MN} - {\lam}\Gamma^M D_M \lam\right),
\end{equation}
and we take the gauge group to be $G=U(N)$.

Since the surface operator is a disorder operator, any correlation function will involve a path integral in which the fields of $\mathcal{N}=4$ SYM are expanded around the singular configurations fixed by $\mathcal{O}_{\Sigma}$: 
\begin{equation}
    A^M = (A_0)^M + a^M, 
\end{equation}
or, in the $4d$ language: 
\begin{align}
\label{background4dexpansion}
  A^\mu &= A_0^\mu + a^\mu\\
\phi^I &= \phi_0^I + \varphi^I.  
\end{align}

The background field gauge fixing condition is\footnote{The same gauge fixing that was used in \cite{Gomis:2009ir,Kristjansen:2023ysz} to study 't Hooft loops.}
\begin{equation}
    G(a_M)= D_0^M a_M = \nabla^M a_M + [A_0^{ M}, a_M] =0,
\end{equation}
that in terms of the $4d$ fields is given by 
\begin{equation}
\label{gaugefixing4d}
  \begin{aligned}
D_0^\mu a_\mu + [(\phi_0^{I}), (\varphi^{I})] =0.
  \end{aligned}
\end{equation}
Note that $D_0$ denotes covariant derivatives with respect to the background field $(A_0)_M$ and the $10d$ metric $g$.

In order to compute the propagator of the scalar fields we need to recast the quadratic terms in $a_M$ in the gauge fixed action: 
\begin{equation}
    \label{gaugefixed}
    S_{gf} = \frac{1}{g^2} \int d^{10}x \; \sqrt{g}\; \left( \frac{1}{2} F^{MN}F_{MN} + (D_0^M a_M) (D_0^N a_N) - \Bar{\Psi}\Gamma^M D_M \Psi  -  \Bar{c}D_0^M D_M c \right).
\end{equation}
If now we expand with respect to the background, the quadratic terms in the quantum fluctuation $a_M$ are given --up to total derivatives-- by 
\begin{equation}
    S^2_{10d}=\frac{1}{g^2} \int d^{10}x \; \sqrt{g}\;  \left( a^{ N} (-D_0^2)a_N + a^{ N} R_{MN} a^{ M} +2 \,( F_0) _{MN} [a^{ M},a^{ N}] \right), 
\end{equation}
where we have used that 
\begin{equation}
[D_{0\,M}, D_{0\,N}] a^{L} = R^L_{O M N} a^{O} + [  F_0^{ MN},a^{  L}],
\end{equation}
with $R^L_{O M N}$ the $10$d Riemann tensor, $R^L_{M L N}=R_{MN}$ the Ricci tensor, both computed from the 10$d$ metric $g$. We find the $4d$ effective action by dimensional reduction over the directions $0,5,\ldots,9$, so that $A_\mu$ with $\mu=1,\ldots4$ is the $4d$ gauge field and $A_I=\phi_I$ with $I=0,5,\ldots,9$ are the six real scalar fields. The effective action is given by
\begin{equation}  
\label{Seff4d}
    \begin{aligned}
        S_2 &= \frac{1}{g^2}  \int d^4x \sqrt{h} \bigg( a^{ \mu}  \left( -D_0^2 h_{\mu \nu} + R_{\mu \nu }    \right) a^{\nu} +2  F_0^{\mu \nu} \; [a_\mu,a_\nu] + [a_\mu, \phi_0^I][a^\mu,\phi^I_0] +  \\
        &+ 4  \left( (D_0)_\mu \phi_0^I \right) \; [a^\mu,\varphi^I] +  \varphi^{b i}  \left( -D_0^2  + \xi R \,  \right)\varphi^{b i }  + 2 [\phi_0^I,\phi_0^J] [\varphi^I,\varphi^J]  + [\varphi^J, \phi_0^I][\varphi^J,\phi^I_0]\bigg),  
    \end{aligned} 
\end{equation}
where $h_{\mu \nu}$ is the $4d$ metric, $R_{\mu \nu}$ is the Ricci tensor and $R$ the curvature of the $4$ dimensional metric. We have added by hand the term $\xi R \phi^2$, the conformal coupling of the scalar fields with $\xi=\frac{1}{6}$ in 4 dimensions. Here $D_0$ denotes a covariant derivative with respect to both the metric $h_{\mu \nu}$ and the background gauge field $(A_0)_\mu$.

The insertion of $\mathcal{O}_{\Sigma}$ breaks the gauge group $U(N)$ to a Levi subgroup $\mathbb L= \prod_{I=1}^M U(N_I)\subset U(N)$ near the surface $\Sigma$. According to this symmetry breaking pattern, we can distinguish between the color modes of the fields that are affected by the insertion of the surface operator:
\begin{equation}
    \left(\begin{array}{c:c:c:c}
         b_{N_1, N_1} \otimes \mathbb{I}_{N_1}& \tilde{b}_{N_1, N_2} &\cdots & \tilde{b}_{N_1, N_M} \\
         \hdashline
         \tilde{b}_{N_2, N_1}& b_{N_2} \otimes \mathbb{I}_{N_2,N_2}&\cdots & \tilde{b}_{N_2,N_M} \\
         \hdashline
         \vdots &\vdots &\ddots&\vdots \\
         \hdashline
         \tilde{b}_{N_M,N_1}&\tilde{b}_{N_M,N_2}&\cdots& b_{N_M,N_M}\otimes \mathbb{I}_{N_M}
    \end{array} \right) \in \mathfrak{u}(N),
\end{equation}
the modes without the tilde, inside of the diagonal blocks dictated by the symmetry breaking, are invariant under $\mathbb L$. They are massless modes with the usual $\mathcal N = 4$ propagator in the Lorenz gauge. On the other hand, the modes with tilde have a different propagator, since new terms appear in their quadratic action because of the insertion of $\mathcal{O}_\Sigma$. 

In addition, there is a nontrivial flavour mixing between the scalar fields and the gauge field in \eqref{Seff4d}. It is induced by the singular background, that is nontrivial for the gauge field $(A_0)_\mu$ and two of the six real scalar fields (which we packaged into the complex field $\Phi_0, \bar \Phi_0$). These are the fields that we call ``complicated fields'', since to compute their propagators we will need to solve a more complicated eigenvalue problem. Note that the remaining four scalar fields do not have any mixing with the gauge field $a_\mu$ according to \eqref{Seff4d}, and that is why we call them ``easy fields''. 

We will consider the case in which the surface operator is inserted on a plane $\Sigma= \mathbb R^2 = \{ x_3 = x_4=0 \}$. We consider the case where the background for the gauge field is set to zero and leave the general case for future work. The singular background for the complex scalar field is given by 
\begin{equation}
\label{backgroundaphi}
    \begin{aligned}
        % A_0 &= \alpha \, d\theta, \\
        \Phi_0 &= \frac{\beta + i \gamma}{\sqrt{2}z},
    \end{aligned}
\end{equation}
where $z = x_3 + i x_4 = r \, e^{i \theta}$ is an holomorphic coordinate in the normal plane to $\Sigma$. 

Substituting the background in \eqref{Seff4d} we find the quadratic action of the fields:
\begin{equation} 
    \label{Seff-plane}
    \begin{aligned}
        S_2 &= -\frac{1}{ g^2}\int d^4x \sqrt{h} \bigg(  2 \sum_{i<j}  h_{\mu \nu}   a^{ \mu}_{ij}  \left( -\nabla^2 + \frac{z_{ij} \bar z_{ij}}{r^2}   \right) a^{ \nu}_{ji} + \sum_{i} a^{ \mu}_{ii}  \left( -\nabla^2 h_{\mu \nu}    \right) a^{ \nu}_{ii}+  \\
        &+2 \sum_{a=1}^4 \sum_{i<j}    \varphi^a_{ij}  \left( -\nabla^2 + \frac{z_{ij} \bar z_{ij}}{r^2}   \right) \varphi^a_{ji} + \sum_{i} \varphi^a_{ii}  \left( -\nabla^2   \right) \varphi^a_{ii}+  \\
        &  + 2 \sum_{i<j} \varphi_{ij}  \left( -\nabla^2 + \frac{z_{ij} \bar z_{ij}}{r^2} \right)\bar \varphi_{ji} + i \leftrightarrow j +2\sum_{i} \varphi_{ii}  \left( -\nabla^2  \right) \bar \varphi_{ii} +\\
        & \left. - 4 i  \left(  \sum_{i<j}  \nabla_\mu \left( \frac{1}{\sqrt{2}re^{i\theta}} \right)   \left( a^\mu_{ij} \bar \varphi_{ji} - a^\mu_{ji} \bar \varphi_{ij} \right) z_{ij}   +    \nabla_\mu \left( \frac{1}{\sqrt{2}re^{-i\theta}} \right)  \left( a^\mu_{ij} \varphi_{ji}- a^\mu_{ji} \varphi_{ij} \right) \bar z_{ij} \right) \right), \\
    \end{aligned} 
\end{equation}
where $z_{ij}= i (\beta_i + i \gamma_i)-(\beta_j + i \gamma_j)$,\footnote{And $\bar z_{ij}= i (\beta_i - i \gamma_i)-(\beta_j - i \gamma_j)$, since $\beta, \gamma$ are anti-hermitian.} and $\nabla^2$  denotes the scalar or vector laplacian in polar coordinates.

\subsection{Propagator of the Easy Scalar Fields}

In order to compute $\langle \phi^1_{ij} \phi^1_{kl}\rangle$ we want the eigenfunctions of the quadratic action of $\phi^1$, so that we can build the propagator using the spectral theorem. 

The general solution of 
\begin{equation}
    -\left( \partial_{rr} +  \frac{1}{r^2} \partial_{\theta \theta} + \frac{1}{r} \partial_r  +\partial_{xx} +\partial_{yy} +\frac{ z_{ij} \bar z_{ij} }{r^2}  \right)\phi^1_{ji}= \lambda^2 \phi^1_{ji}.
\end{equation}
is given by 
\begin{equation}
    \phi_{ji}^1 = e^{i n \theta} e^{i \vec p \cdot \vec x} ( c_1 J_\nu ( a r) + c_2 Y_\nu ( a r) ), 
\end{equation}
where $\lambda^2 = a^2 + p^2$, and $J_\nu, Y_\nu$ are Bessel functions of the first and second kind with order $\nu$, that depends on the Fourier mode $n$ and the surface operator parameters $\beta, \gamma$ through  
\begin{equation}
\label{eq:nueasy}
    \nu= \sqrt{n^2 + z_{ij} \bar z_{ij}}. 
\end{equation}

Since we use the description of $\mathcal{O}_\Sigma$ as a disorder operator, the best way to deal with UV divergences is to define the operator in a tubular neighborhood around the surface $\Sigma$. Here we consider a plane $\Sigma = \mathbb R^2$, and this regulating procedure consists in inserting the surface operator at $r = \epsilon$ and taking later the $\epsilon \to 0$ limit. We will consider eigenfunctions that are regular at $\epsilon \to 0$, which fixes $c_1=1, c_2=0$.

The color structure of the propagator can be read from \eqref{Seff-plane}: 
\begin{equation}
    \langle \phi^1_{ij}(r,\theta,\Vec{x}) \, \phi^1_{kl}(r',\theta',\Vec{x}') \rangle = \delta_{il} \delta_{jk} \; G(r,\theta,\Vec{x};r',\theta',\Vec{x}'),
\end{equation}
where the Green's function $G$ obeys
\begin{equation}
    -\left( \partial_{rr} + \frac{1}{r^2} \partial_{\theta \theta} + \frac{1}{r} \partial_r  + \nabla_{\vec x}^2 -\frac{ z_{ij} \bar z_{ij} }{r^2} \right)G(r,\theta,\Vec{x};r',\theta',\Vec{x}') = -\frac{g^2}{2} \frac{\delta(r-r') }{r} \, \delta(\theta - \theta') \, \delta(\vec x - \vec x') .
\end{equation}
It is not surprising that because of the symmetries of the set up -- a surface operator inserted on a plane, so preserving the maximal amount of conformal symmetry for a codimension 2 operator -- the propagator can be written as a sum of $AdS_3$ propagators: 
\begin{equation}
\label{easypropfinite}
    \langle \phi^1_{ij}(r,\theta,\Vec{x}) \, \phi^1_{kl}(r',\theta',\Vec{x}') \rangle = -\delta_{il} \delta_{jk} \frac{g^2}{4 \pi }  \frac{1}{r r'}  \sum_n e^{i n (\theta-\theta')} G_{\nu}(r,\Vec{x}; r',\Vec{x}'),
\end{equation}
where 
\begin{equation}
\label{ads3prop}
    G_\nu(r,\Vec{x};r',\Vec{x}') =  \; (r \, r') \; \int_{R^2} \frac{d^2k}{(2 \pi)^2} \; e^{i\Vec{k}(\Vec{x}-\Vec{x}') } \; \int_0^\infty \, da \, a \; \frac{1}{\left(   a^2 + \Vec{k}^2 \right)}  \; J_\nu( a r)  \;  J_\nu( a r')
\end{equation}
is the $AdS_3$ propagator in Poincar\'e coordinates and $\nu$ is defined in \eqref{eq:nueasy}.

\subsection{Propagator of the Complicated Scalar Fields}

In order to compute the $\langle \varphi \bar \varphi \rangle$ propagator, we need to solve the flavour mixing problem between the gauge field and the scalar fields. It is more illuminating to write the equations of motion derived from \eqref{Seff-plane} in ``matrix form'': 
\begin{equation} 
\label{matrixkinetic}
        \left( \begin{array}{ccc}
            h_{\mu \nu}  \left( -\nabla^2 + \frac{||z_{ij}||^2}{r^2}   \right) & -2 i \nabla_\mu \left( \frac{1}{\sqrt{2}re^{-i\theta}} \right)     \left(  \bar z_{ij} \right) & -2 i \nabla_\mu \left( \frac{1}{\sqrt{2}re^{i\theta}} \right)     \left(  z_{ij} \right)   \\
            2 i \nabla_\nu \left( \frac{1}{\sqrt{2}re^{i\theta}} \right)     \left(  z_{ij} \right)  &\left( -\partial^2 +\frac{||z_{ij}||^2}{r^2} \right) & 0\\
             2 i \nabla_\nu \left( \frac{1}{\sqrt{2}re^{-i\theta}} \right)     \left(  \bar z_{ij} \right)&0 &\left( -\partial^2 +\frac{||z_{ij}||^2}{r^2} \right) \\
        \end{array}\right) \cdot
        \left(\begin{array}{c}
              a^\nu_{ji}\\
             \varphi_{ji}\\
             \bar \varphi_{ji}
        \end{array}\right)=0 ,
\end{equation}
using the vector and scalar laplacian in polar coordinates we find the system of equations we need to solve to diagonalize this quadratic action:
\begin{equation} 
    \begin{aligned} 
     &-\left( \partial_{rr} + \frac{1}{r^2}\partial_{\theta \theta} + \frac{1}{r} \partial_r  + \nabla_{\vec x}^2 - \frac{ z_{ij} \bar z_{ij} }{r^2} \right)\varphi -i \sqrt{2} \frac{ z_{ij} }{r^2 e^{i \theta}} \, a^r +\sqrt{2} \frac{ z_{ij} }{r^2 e^{i \theta}} \, a^\theta= \lambda^2  \varphi_{ji} \\
     &-\left( \partial_{rr} + \frac{1}{r^2}\partial_{\theta \theta} + \frac{1}{r} \partial_r  + \nabla_{\vec x}^2 - \frac{ z_{ij} \bar z_{ij} }{r^2} \right)\bar \varphi -i \sqrt{2} \frac{ \bar z_{ij} }{r^2 e^{-i \theta}} \, a^r - \sqrt{2} \frac{ \bar z_{ij} }{r^2 e^{-i \theta}} \, a^\theta= \lambda^2  \bar \varphi_{ji} \\ 
     &-\left( \partial_{rr} + \frac{1}{r^2}\partial_{\theta \theta} + \frac{1}{r} \partial_r  + \nabla_{\vec x}^2 - \frac{1}{r^2}   - \frac{ z_{ij} \bar z_{ij} }{r^2} \right) a^r + \frac{2}{r^2} \partial_\theta a^\theta +i \sqrt{2} \frac{ \bar z_{ij} }{r^2 e^{-i \theta}} \, \varphi + i \sqrt{2} \frac{  z_{ij} }{r^2 e^{i \theta}} \, \bar \varphi= \lambda^2  a^r_{ji} \\
     &-\left( \partial_{rr} + \frac{1}{r^2}\partial_{\theta \theta} + \frac{1}{r} \partial_r  + \nabla_{\vec x}^2 - \frac{1}{r^2} - \frac{ z_{ij} \bar z_{ij} }{r^2} \right) a^\theta - \frac{2}{r^2} \partial_\theta a^r +\sqrt{2} \frac{ \bar z_{ij} }{r^2 e^{-i \theta}} \, \varphi -  \sqrt{2} \frac{  z_{ij} }{r^2 e^{i \theta}} \, \bar \varphi= \lambda^2  a^\theta_{ji}
    \end{aligned}
\end{equation}

As we did before, we expand the fields in eigenfunctions of the laplacian in polar coordinates, with the sole difference of a shift in the Fourier modes for the fields $\varphi$ and $\bar \varphi$, so that we can translate from a system of partial differential equations to a linear system of algebraic equations: 
\begin{equation}
\label{expansionalphanonzero }
    \begin{aligned}
        a^r_{ji} &= (a^r_{ji})_{n,m}\, J_m(a r) \, e^{i \, n \, \theta  }, \\
        a^\theta_{ji} &=  (a^\theta_{ji})_{n,m}\, J_m(a r ) \, e^{i \, n \, \theta }, \\
        \varphi_{ji} &= (\varphi_{ji})_{n,m}\, J_m(a r) \,  e^{i \,( n -1)\, \theta }, \\
        \bar \varphi_{ji} &= (\bar \varphi_{ji})_{n,m}\, J_m(a r) \,  e^{i \,(n +1)\, \theta },
    \end{aligned}
\end{equation}
where $\lambda^2 = a^2 + p^2$, and we find a linear system of equations for $\{\varphi_{n,m} ,\bar{\varphi}_{n,m} ,a^r_{n,m}, a^\theta_{n,m}\}$: 
\begin{equation}
(M-m^2 Id) 
\left(\begin{array}{c}
     \varphi_{n,m}  \\
     \bar{\varphi}_{n,m} \\
     a^r_{n,m}\\
     a^\theta_{n,m}
\end{array}\right) = 0,
\end{equation}
where
\begin{equation}
    M = \left(
\begin{array}{cccc}
 z_{ij} \bar z_{ij}+(n-1)^2   & 0 & -i \sqrt{2} \,  z_{ij} & \sqrt{2} \, z_{ij} \\
 0 & z_{ij} \bar z_{ij} + (n+1)^2  &- i \sqrt{2} \, \bar z_{ij}&  -\sqrt{2} \bar z_{ij}\\
 i \sqrt{2} \bar z_{ij}& i \sqrt{2} z_{ij} & z_{ij} \bar z_{ij}+n^2+1 & 2 i n \\
 \sqrt{2} \bar z_{ij} &- \sqrt{2} z_{ij} &- 2 i n & z_{ij} \bar z_{ij}+n^2+1 \\
\end{array}
\right).
\end{equation}

The eigenfunctions of the quadratic action in \eqref{matrixkinetic} are \eqref{expansionalphanonzero }, where the coefficients are the components of the eigenvectors of $M$ and the order of the Bessel function is given by the square root of the corresponding eigenvalue. The second step corresponds to decoupling the flavour mixing between $\varphi$ and $a^z$ in \eqref{matrixkinetic}.

$M$ has two degenerate eigenspaces, with eigenvalues 
\begin{equation}
    \label{eigenvaluesMatrix}
        m_\pm^2=\left( \sqrt{n^2 + z_{ij} \bar z_{ij} } \pm 1\right)^2
\end{equation}
and a set of orthonormal eigenvectors is given by 
\begin{equation}
\label{ortheig}
    \begin{aligned}
        u_-&=\left(-\frac{\sqrt{z_{ij} \bar z_{ij}} \left(\sqrt{z_{ij} \bar z_{ij}+n^2}+n\right)}{2  \bar z_{ij} \sqrt{z_{ij} \bar z_{ij}+n^2}},-\frac{\sqrt{z_{ij} \bar z_{ij}} \left(n-\sqrt{z_{ij} \bar z_{ij}+n^2}\right)}{2 z_{ij} \sqrt{z_{ij} \bar z_{ij}+n^2}},0,\frac{\sqrt{z_{ij} \bar z_{ij}}}{\sqrt{2} \sqrt{z_{ij} \bar z_{ij}+n^2}}\right),\\
        \tilde u_-&= \left( i\frac{z_{ij}}{2 \sqrt{z_{ij} \bar z_{ij}+n^2}},\frac{i  \bar z_{ij}}{2 \sqrt{z_{ij} \bar z_{ij}+n^2}},\frac{1}{\sqrt{2}},\frac{i n}{\sqrt{2} \sqrt{z_{ij} \bar z_{ij}+n^2}} \right),\\
        u_+&=\left(\frac{\sqrt{z_{ij} \bar z_{ij}} \left(\sqrt{z_{ij} \bar z_{ij}+n^2}-n\right)}{2 \bar z_{ij} \sqrt{z_{ij} \bar z_{ij}+n^2}},-\frac{\sqrt{z_{ij} \bar z_{ij}} \left(\sqrt{z_{ij} \bar z_{ij}+n^2}+n\right)}{2 z_{ij} \sqrt{z_{ij} \bar z_{ij}+n^2}},0,\frac{\sqrt{z_{ij} \bar z_{ij}}}{\sqrt{2} \sqrt{z_{ij} \bar z_{ij}+n^2}}\right),\\
        \tilde u_+&=\left(- \frac{i z_{ij}}{2 \sqrt{z_{ij} \bar z_{ij}+n^2}},-i\frac{ \bar z_{ij}}{2 \sqrt{z_{ij} \bar z_{ij}+n^2}},\frac{1}{\sqrt{2}},-\frac{i n}{\sqrt{2} \sqrt{z_{ij} \bar z_{ij}+n^2}}  \right)
    \end{aligned}
\end{equation}

Using the eigenvectors of $M$ to find the full eigenfunctions of the quadratic action, we can write the propagator of the complicated field as
\begin{equation}
\label{phiphibarprop}
\begin{aligned}
     &\langle \bar \varphi_{ij} (r,\theta, \vec x) \varphi_{kl} (r',\theta', \vec x') \rangle = -\delta_{il} \delta_{jk} \frac{g^2}{4 \pi }  \frac{1}{r r'}   \sum_n e^{i(n-1)(\theta-\theta')} \times \\
     & \quad \quad \quad \quad \quad \quad \quad \quad  \times \Big( (|(u_-)_1|^2 + |(\tilde{u}_-)_1|^2 ) \, G_{\nu_-}  + (|(u_+)_1|^2 + |(\tilde{u}_+)_1|^2 ) \, G_{\nu_+}   \Big)=\\
     &=-\delta_{il} \delta_{jk} \frac{g^2}{4 \pi }  \frac{1}{r r'}   \sum_n e^{i(n-1)(\theta-\theta')} \left( \frac{1}{2} \left( G_{\nu_-}+\; G_{\nu_+} \right) - \frac{n}{2 \sqrt{n^2 + z_{ij} \bar z_{ij} }} \left( G_{\nu_-}-\; G_{\nu_+} \right) \right)
     \end{aligned}
\end{equation} 
where $G_\nu$ is the $AdS_3$ propagator defined on \eqref{ads3prop} and 
\begin{equation}
    \nu_\pm=\left| \sqrt{n^2+ z_{ij} \bar z_{ij}} \pm 1\right|. 
\end{equation}

One can also show that 
\begin{equation}
    ((u_+)_1)^* (u_+)_2 + ((\tilde u_+)_1)^* (\tilde u_+)_2 = ((u_-)_1)^* (u_-)_2 + ((\tilde u_-)_1)^* (\tilde u_-)_2 = 0,
\end{equation}
which proves that $\left< \varphi \varphi \right>$ and $\left< \bar \varphi \bar \varphi \right>$ vanish.

\subsection{Coincident Points Limit of the Difference of Propagators}

The quantity that plays the important role in the computation of the correlation functions with the CPO is 
\begin{equation}
    \lim_{x' \to x} \langle \varphi_{ij}(x) \bar{\varphi}_{kl}(x') \rangle - \langle \phi^1_{ij}(x) \phi^1_{kl}(x') \rangle .
\end{equation}
We need to choose a regularization scheme to evaluate the propagators at coincident points. Since we have written all the propagators as an infinite sum of $AdS_3$ propagators, we will use the same strategy as in \cite{Kristjansen:2023ysz} in the context of the insertion of a 't Hooft loop. We will dimensionally regularize the $AdS_3$ propagators. In their case, they set $\theta = \theta'$ and even if each propagator diverges, the difference is a convergent series. That is not true here, so we will need to take the limit $\vec x' \to \vec x, r' \to r$ and use dimensional regularization on $AdS_3$, but keeping $\Delta \theta $ finite and later we will use a different regularization procedure to take care of the divergent infinite sums. 

We start with the $AdS_3$ propagator in \eqref{ads3prop} and compute the integral over $a$: 
\begin{equation}
    \begin{aligned}
        G_\nu(r,\Vec{x};r',\Vec{x}') &= (r \, r') \; \int_{R^2} \frac{d^2k}{(2 \pi)^2} \; e^{i\Vec{k}\cdot (\Vec{x}-\Vec{x}') } \; \int_0^\infty \, da \, a \; \frac{1}{\left(   a^2 + \Vec{k}^2 \right)}  \; J_\nu( a r)  \;  J_\nu( a r')= \\
        &= (r \, r') \int \frac{d^2\Vec{k} }{(2\pi)^2} \, e^{i\Vec{p} \, (\vec{x}-\vec{x}') } I_\nu ( k r^<) \, K_\nu( k  r^>),
    \end{aligned}
\end{equation}
where $k=\sqrt{\Vec{k}^2} $ and  $r^>$ (resp. $r^<$) is the largest (smallest) between $r,r'$. Now we can set $r=r', \vec x = \vec x'$ and do the momentum integral in $d=2-2\varepsilon$ dimensions: 
\begin{equation}
\label{ads3propcoincident}
    \begin{aligned}
          G_\nu(r,\Vec{x};r,\Vec{x}) &=r^2  \int \frac{d^{2-2\varepsilon} \Vec{k} }{(2\pi)^2}  I_\nu ( k r ) \, K_\nu( k  r )=\\
         &=\frac{r^2 }{(2\pi)^2} \frac{2 \pi^{1-\varepsilon}}{\Gamma(1-\varepsilon)} \int_0^\infty dp \, p^{1-2 \varepsilon}   I_\nu ( p r) \, K_\nu( p r) = \\
         &= \frac{r^2 }{(2\pi)^2} \left( \frac{\pi ^{\frac{1}{2}-\varepsilon } r^{2 \varepsilon -2} \Gamma \left(\varepsilon -\frac{1}{2}\right) \Gamma (-\varepsilon +\nu +1)}{2 \Gamma (\varepsilon +\nu )} \right) = -\frac{ \nu }{4 \pi} 
    \end{aligned}
\end{equation}

We will consider $\alpha=0$. Inserting \eqref{ads3propcoincident} into \eqref{easypropfinite} we can find the  propagator of the easy field at different angular coordinates $\theta, \theta'$: 
\begin{equation}
\begin{aligned}
     \langle \phi^1_{ij}(r,\theta,\Vec{x}) \, \phi^1_{kl}(r,\theta',\Vec{x}) \rangle &= \delta_{il} \delta_{jk} \frac{g^2}{16 \pi^2 }  \frac{1}{r^2}  \sum_{n=-\infty}^\infty \sqrt{n^2 + z_{ij} \bar z_{ij}} \; e^{i n (\theta-\theta')}.
\end{aligned}
\end{equation}

We repeat this calculation for the $\langle \bar \varphi \, \varphi \rangle$ propagator, which is written in \eqref{phiphibarprop} as a sum of $AdS_3$ propagators with two different arguments $\nu_{\pm}$. If now we plug in the near coincident points limit for the $AdS_3$ propagator\footnote{ We are assuming that $ z_{ij} \bar z_{ij} > 1$, where large $|z_{ij}|$ corresponds to the semiclassical regime. Note that this is not true for the Levi block modes, but those have the massless free propagator and the difference of propagators vanishes exactly} we find the following expression for the propagator of the complex field $\varphi$ at finite $\Delta \theta$: 
\begin{equation} 
    \begin{aligned}
    \langle \bar \varphi_{ij} (r,\theta, \vec x) &\varphi_{kl} (r,\theta', \vec x) \rangle = \delta_{il} \delta_{jk} \frac{g^2}{16\pi^2 }  \frac{1}{r^2}   \sum_{n = - \infty}^\infty \left(\sqrt{n^2 + z_{ij} \bar z_{ij}} - \frac{n}{\sqrt{n^2 + z_{ij} \bar z_{ij}}} \right) e^{i (n-1) \Delta  \theta }\\ 
\end{aligned} 
\end{equation}

We want to compute the leading order term in $\Delta \theta$ in the difference of propagators: 
\begin{equation} 
    \begin{aligned}
    &\langle \bar \varphi_{ij} (r,\theta, \vec x) \varphi_{kl} (r,\theta', \vec x) \rangle - \langle \phi^1_{ij} (r,\theta, \vec x) \phi^1_{kl} (r,\theta', \vec x) \rangle = \delta_{il} \delta_{jk} \frac{g^2}{16\pi^2 }  \frac{1}{r^2}  \times \\
    &\left( (1-e^{-i \Delta \theta})\sum_{n = - \infty}^\infty  \sqrt{n^2 + z_{ij} \bar z_{ij}} \; e^{i n \Delta  \theta } + e^{-i \Delta \theta} \sum_{n = - \infty}^\infty  \frac{n}{\sqrt{n^2 + z_{ij} \bar z_{ij}}} \; e^{i n \Delta  \theta } \right) \\ 
\end{aligned} 
\end{equation}
If we set $\Delta \theta=0$ this two series are divergent. However, we can Fourier transform and find a series that is convergent even at $\Delta \theta=0$ except for a single term that diverges: 
\begin{equation}
\label{differencepropsseries}
\begin{aligned}
    \sum_{n = - \infty}^\infty  &\sqrt{n^2 + z_{ij} \bar z_{ij}} \; e^{i n \Delta  \theta } =    \left( - 2 \sqrt{-z_{ij} \bar z_{ij}} \right)  \sum_{m=-\infty}^\infty \ \frac{K_1(| \sqrt{z_{ij} \bar z_{ij}}  (\Delta \theta  + 2\pi m) |)}{|\Delta \theta  + 2\pi m|} , \\
    \sum_{n = - \infty}^\infty  &\frac{n}{\sqrt{n^2 + z_{ij} \bar z_{ij}}} \; e^{i n \Delta  \theta }  =  \left( 2 i  \sqrt{z_{ij} \bar z_{ij}} \right) \sum_{m=-\infty}^\infty \text{sgn} (\Delta \theta  + 2\pi m) K_1\left( \sqrt{z_{ij} \bar z_{ij}}  |\Delta \theta  + 2\pi m| \right) ,
\end{aligned}
\end{equation}
where we have used that
\begin{equation}
\begin{aligned}
    &\int_{-\infty}^\infty \sqrt{a^2+x^2 } e^{i x k} dx = - 2 \left|\frac{a}{k}\right| K_{1}(|a \, k | ) ,\\
    &\int_{-\infty}^\infty \frac{x}{\sqrt{a^2+x^2 }} e^{i x k} dx = 2 i  \text{sgn} (k) |a| K_1(|a \, k |) .
\end{aligned}
\end{equation}

We see that at finite $\Delta \theta$ both sums are absolutely convergent, and that at $\Delta \theta = 0$ the only divergent term is $m=0$. In particular, the second series reduces to the $m=0$ term when $\Delta \theta=0$, because all other terms cancel. Since in \eqref{differencepropsseries} the first term is multiplied by  an $1-e^{-i \Delta \theta} \sim O(\Delta \theta)$ factor, only $m=0$ will contribute to the result up to order $O(\Delta \theta)$. Using the asymptotics of the modified Bessel function we find that there is a $\frac{1}{\Delta \theta}$ divergence that cancels and we find the finite result: 
\begin{equation}
    \langle \bar \varphi_{ij} (r,\theta, \vec x) \varphi_{kl} (r,\theta', \vec x) \rangle-\langle \phi^1_{ij} (r,\theta, \vec x) \phi^1_{kl} (r,\theta', \vec x) \rangle = -\delta_{il} \delta_{jk} \frac{g^2}{16 \pi^2 }  \frac{1}{r^2}  \left( 1+O\left(\Delta \theta ^1\right) \right),
\end{equation}
that does not depend on the surface operator parameters $\beta,\gamma$. 

As we said before, the modes inside of the Levi blocks do not see the surface operator: their quadratic action is that of a free massless scalar -- the same as in $\mathcal{N}=4$ SYM before the insertion of the surface defect -- and the propagator of the easy and complicated scalar field cancel exactly: 
\begin{equation} \label{eq:final propagator}
    \langle \bar \varphi_{(ai)(bj)}  \varphi_{(ck)(dl)}  \rangle-\langle \phi^1_{(ai)(bj)} \phi^1_{(ck)(dl)}  \rangle (r,\theta, \vec x)= -\delta_{ad} \delta_{bc} \delta_{il} \delta_{jk} (1-\delta_{ab}) \; \frac{g^2}{16 \pi^2 }  \frac{1}{r^2}  
\end{equation}
where the indices $a,b,c,d=1,\ldots,M$ run over the blocks of the background and $i,j,k,l$ over indices inside each block (there is no sum over repeated indices).

\subsection{Correlation Function with CPO}

Finally, we are ready to write down the quantum corrections from the Wick contractions to the correlators with CPO using the difference of the propagators \eqref{eq:final propagator}. We present the result for $\Delta=2,\dots,5$ and see that this is consistent with the exact $2d$ YM computation for $\Delta=2,3$ in \eqref{eq:2dYMO23} and $\Delta=4,5$ in Appendix \ref{appsec:higherCPO} using the dictionary explained in section \ref{sec:CPOloc}.
\begin{equation}\label{gaugeresults1}
    \begin{aligned}
         \langle \mathcal{O}_{2,0} \rangle |_{\Oo_\Sigma}&= \frac{16 \pi^2}{\sqrt{6}\lambda}  \;   \sum_l N_l \left(\left(\frac{\beta_l^2+\gamma_l^2}{2} \right)  -\frac{\lambda }{8 \pi ^2}  \frac{N-N_l}{2N} \right)  + \mathcal{O}(\lambda), \\ 
         \langle \mathcal{O}_{2,2} \rangle |_{\Oo_\Sigma}&= \frac{4 \pi^2}{\sqrt{2}\lambda}  \;   \sum_l N_l (\beta_l  + i \gamma_l)^2 , \\ 
         \langle \mathcal{O}_{3,1} \rangle |_{\Oo_\Sigma}&=  \frac{16 \pi^3}{\lambda^{3/2}} \;  \sum_l N_l \frac{(\beta_l  + i \gamma_l)}{\sqrt{2}} \left(  \left(\frac{\beta_l^2+\gamma_l^2}{2} \right)  - 2 \frac{\lambda }{8 \pi ^2}  \frac{N-N_l}{2N} \right), \\
         \langle \mathcal{O}_{3,3} \rangle |_{\Oo_\Sigma}&=  \frac{8 \pi^3}{\sqrt{3}\lambda^{3/2}} \;  \sum_l N_l  (\beta_l  + i \gamma_l)^3, \\
    \end{aligned}
\end{equation}

\begin{equation*}
    \begin{aligned}
    \langle \mathcal{O}_{4,0} \rangle|_{\Oo_\Sigma} &=  \frac{8 \pi^4}{\sqrt{5}\lambda^2} \left(  \sum_l N_l \left( 3  \left(\beta_l ^2+\gamma_l ^2\right)^2 -\frac{\lambda }{2 \pi ^2 N} \left(\beta_l ^2+\gamma_l ^2\right) (2 N- 3N_l) \right. \right. \\
         &\quad\quad-  \left. \left.  \frac{\lambda ^2 }{16 \pi ^4 N^2}(N-N_l)^2  \right) +  \frac{\lambda }{ \pi ^2 N} \left( \Tr(\Phi_0)\Tr(\bar{\Phi}_0)   \right) \right), \\
         \langle \mathcal{O}_{4,2} \rangle|_{\Oo_\Sigma}&= \frac{32 \pi^4}{\sqrt{10}\lambda^2} \left( \sum_l  N_l \left( (\beta_l +i \gamma_l )^2  \left( \left(\beta_l ^2+\gamma_l ^2\right) -\frac{\lambda  }{8 \pi ^2 N}  (2 N-3 N_l) \right) \right)  \right.\\
          &\quad\quad- \left.  \frac{\lambda }{4 \pi ^2 N} \left( \Tr(\Phi_0)^2   \right) \right), \\
          \langle \mathcal{O}_{4,4} \rangle|_{\Oo_\Sigma} &= \frac{8 \pi ^4 (\beta +i \gamma )^4}{\lambda ^2} ,\\
        \langle \mathcal{O}_{5,1} \rangle|_{\Oo_\Sigma} &=  \frac{128 \pi^5}{\sqrt{5}\lambda^{3/2}}  \left( \frac{1}{2 \sqrt{2}}\sum_l   N_l \left(\beta_l^2+\gamma_l^2\right)^2 \left(\beta_l+ i \gamma_l\right)  \right. +\\
        & \quad\quad- \frac{3\lambda}{8 \pi^2N}   \sum_l N_l (N-2N_l) \frac{ \left(\beta_l^2+\gamma_l^2\right) \left(\beta_l+ i \gamma_l\right)}{2 \sqrt{2}}  + \\
        &\quad\quad+  \left(\frac{\lambda^2}{64 \pi^4N^2}  \sum_l N_l (N-N_l) (N-2N_l)  \frac{ \left(\beta_l+ i \gamma_l\right)}{\sqrt{2}}\right) + \\
        &\quad\quad- \frac{\lambda}{8 \pi^2N}  \left( \left( \Tr\bar \Phi_0 \right)\left( \Tr( \Phi_0^2) \right) \right) - \frac{\lambda}{4 \pi^2N}  \left( \left( \Tr \Phi_0 \right)\left( \Tr( \Phi_0 \bar \Phi_0) \right)  \right) + \\
       &\quad\quad+4\left( \frac{\lambda}{8 \pi^2} \frac{1}{2 N} \right)^2 \left( \Tr ( \Phi_0) \sum_l N_l (N-N_l) \right) \Bigg), \\
     \langle \mathcal{O}_{5,3} \rangle|_{\Oo_\Sigma} &=  \frac{256 \pi^5}{\sqrt{10}\lambda^{5/2}} \Tr\left( \sum_l N_l \frac{  \left(\beta_l+ i \gamma_l\right)^3}{2 \sqrt{2}} \left(\frac{\left(\beta_l^2+\gamma_l^2\right)}{2} -  \frac{\lambda}{8 \pi^2 N} (N-2N_l) \right) \right.
     \\&\left. \quad\quad- \frac{\lambda}{8 \pi^2 N} \Tr \Phi_0\Tr( \Phi_0^2)  \right)\\
     \langle \mathcal{O}_{5,5} \rangle|_{\Oo_\Sigma} &=  \frac{32 \pi ^5 }{\sqrt{5}} \sum_l  N_l (\beta_l +i \gamma_l)^5 .
\end{aligned}
\end{equation*}

Surprisingly, we can explicitly see that this exactly matches with the localization result in \ref{sec:CPOloc} (see \ref{appsec:higherCPO} for $\Delta=4,5$). Therefore we see that only a finite number of simply Wick contracted Feynman diagrams as in Figure \label{fig:enter-label} contribute to the correlator with CPO.

\section{Perturbative Correlation Function with Wilson Loop} \label{appsec:Wilson}

Now we want to use the propagators in Appendix \ref{appsec:CPO} to compute the one-loop correlation function between the surface operator $\Oo_\Sigma$ and a BPS Wilson loop linking with $\Sigma$. For   simplicity, we restrict our Wilson loop to be 1/4-BPS (which is a special case of 1/8-BPS Wilson loop) as in \cite{Drukker:2008wr}, which becomes a fixed latitude circular Wilson loop supported on $S^2$ with a unit radius. We will see that the result agrees with the correlation function computed using localization in section \ref{sec:Wilsonloc}.

We consider the surface operator $\Oo_\Sigma$ inserted on the plane $r=0$ in the polar coordinate and a Wilson loop that links with this surface: it is inserted on a circle in the transverse plane to $\Sigma$, so it can be parametrized as $\{x_1,x_2,r,\psi\}$. The standard parmeterization of the 1/4 BPS Wilson loop is given by \cite{Drukker:2006ga,Drukker:2008wr}
\begin{equation}
    W_{\theta_0, \psi_0} = \frac{1}{N} \; \Tr \exp \int d \psi \left( i A_{\psi} + |z| \cos \theta_0 \, \phi^1 + \sqrt{2} \sin \theta_0 \, \Re \left( z \, \Phi e^{-i \psi_0} \right)   \right) ,
\end{equation}
where $z=r e^{i \psi}$ is the holomorphic coordinate in the normal plane to $\Sigma$, and $\theta_0$ is related to the latitude $\theta$ of $S^2$ as $\theta_0=\theta-\pi/2$.

For the one-loop computation, we expand the fields around the background \eqref{backgroundaphi} and expand the correlation function to next order:
\begin{equation}
\label{wilsonloopexpansion}
\footnotesize 
\begin{aligned}
   & \frac{\langle W_{\theta_0 , \psi_0} \; \mathcal{O}_\sigma \rangle }{\langle  \mathcal{O}_\sigma \rangle} = W_{\theta_0 , \psi_0}|_{\Phi_0, A_0} \left( 1 + \frac{1}{2} \Tr \int d\psi \int d\psi'  \left< \left( i a_{\psi} + |z| \cos \theta_0 \, \phi^1 + \sqrt{2} \sin \theta_0 \, \Re \left( z \varphi e^{-i \psi_0} \right)   \right) (r,\psi,\vec x)  \times \right. \right. \\
 &   \left. \left. \times  \left( i a_{\psi} + |z'| \cos \theta_0 \, \phi^1 + \sqrt{2} \sin \theta_0 \, \Re \left( z' \varphi e^{-i \psi_0} \right)   \right) (r,\psi',\vec x)  \right> + \ldots  \right) 
\end{aligned}
\end{equation}
The semiclassical value $W_{\theta_0, \psi_0}|_{\Phi_0,A_0}$ was computed in \cite{Drukker:2008wr}.

The conjecture is that this correlation function only depends on the surface operator parameters $\alpha, \beta, \gamma$ through the semiclassical factor that was computed in \cite{Drukker:2008wr} (which here we denoted by $ W_{\theta_0 , \psi_0}|_{\Phi_0, A_0} $) and that the quantum corrections to this result are computable via a matrix model that is independent of the surface operator parameters. To compute this observable we will need to take the trace of a certain linear combination of the propagators of the theory. We can split this trace in two terms: a sum that only contains modes inside of the Levi blocks defined by $\mathbb L$, and another sum that only contains ``off-block diagonal'' modes. We will show that the second sum vanishes, and the only contribution that survives comes from the modes whose propagators correspond to massless free fields, that do not depend in the surface operator parameters. 

We expand the integrand in \eqref{wilsonloopexpansion} using that $\langle \varphi \varphi \rangle$ and $\langle \bar \varphi \bar \varphi \rangle$ vanish\footnote{This can be seen using the eigenvectors in \eqref{ortheig}.}:
\begin{equation}
    \begin{aligned}
      &\Tr \Big(  - \langle a_\psi (z) a_\psi (z') \rangle + \frac{i \sin \theta_0 }{\sqrt{2}e^{-i \psi_0}} \left(  z ' \langle a_\psi (z) \varphi(z') \rangle + z \langle  \varphi(z) a_\psi (z) \rangle \right) + \\
      &+ \frac{i \sin \theta_0 }{\sqrt{2}e^{i \psi_0}} \left( \bar z ' \langle a_\psi (z) \bar \varphi(z') \rangle +\bar z \langle  \bar \varphi(z) a_\psi (z) \rangle \right) +\\
      &+\frac{\sin \theta_0^2}{2} \left( z \bar z ' \langle \varphi(z) \bar \varphi(z') \rangle + \bar z \, z' \langle \bar  \varphi(z) \varphi(z') \rangle \right) + |z| |z'| \cos^2 \theta_0^2 \langle   \phi^1(z) \phi^1(z') \rangle\Big) ,
    \end{aligned}
\end{equation}
note that because of the geometry of the Wilson loop, all points have the same $\vec x$ and $r$ coordinates ($z= r\, e^{i \psi},z'= r \, e^{i \psi'}$), and consequently we will need to use the coincident points expression for the $AdS_3$ propagators \eqref{ads3propcoincident}, but keeping the two angular coordiantes $\psi, \psi'$ separate.

Now we compute each propagator using the corresponding entries of the eigenvectors \eqref{ortheig}. First, we see that the ``mixed terms'' vanish:
\begin{equation}
\begin{aligned}
    \frac{i \sin \theta_0 }{\sqrt{2} e^{- i \psi_0}} \left(  |z| z ' \langle (a_\psi)_{ij} (z) \varphi_{ji}(z') \rangle + z |z'| \langle  \varphi_{ij}(z) (a_\psi)_{ji} (z) \rangle \right) = 0\\
    \frac{i \sin \theta_0 }{\sqrt{2}e^{i \psi_0}} \left( \bar z ' |z| \langle (a_\psi)_{ij} (z) \bar \varphi_{ji}(z') \rangle +\bar z  |z'| \langle  \bar \varphi_{ij}(z) (a_\psi)_{ji} (z) \rangle \right) = 0.
\end{aligned}
\end{equation}
So all that is left is the sum of the three terms containing the $a_\psi$ propagator, and the propagators of the scalar field:
\begin{equation}
\begin{aligned}
    &\sum_{i,j} \; -|z| |z'| \langle (a_\psi)_{ij} (r,\psi', \vec x)  (a_\psi)_{ji} (r,\psi, \vec x) \rangle + \frac{\sin \theta_0^2}{2} \left( z \bar z ' \langle \varphi_{ij}(z) \bar \varphi_{ji}(z') \rangle + \bar z \, z' \langle \bar  \varphi_{ij}(z) \varphi_{ji}(z') \rangle \right)  + \\
    &+|z| |z'| \cos^2 \theta_0^2 \langle   \phi^1_{ij}(z) \phi^1_{ji}(z') \rangle
    \\& = \sum_{i,j} \; \sum_n e^{in(\psi-\psi')} \left( -\frac{1}{2} (G_{\nu_-} + G_{\nu_+}) +  \frac{\sin \theta_0^2}{2}(G_{\nu_-} + G_{\nu_+}) + \cos \theta_0 ^2 G_{\nu_0} \right)  \\
    &= \sum_{i,j} \; \sum_{n=-\infty}^\infty e^{in(\psi-\psi')}  \cos \theta_0 ^2 \left( -\frac{1}{2} (G_{\nu_-} + G_{\nu_+}) + G_{\nu_+} \right),
\end{aligned}
\end{equation}
So everything is reduced to computing
\begin{equation}
    -\frac{1}{2} (G_{\nu_-} + G_{\nu_+}) + G_{\nu_0},
\end{equation}
where $G_\nu$ is the $AdS_3$ coincident point propagator \eqref{ads3propcoincident}:
\begin{equation}
\begin{aligned}
    -\frac{1}{2} \left( \left| \sqrt{-z_{ij} \bar z_{ij}+(n-i \alpha_{ij})^2} + 1\right| + \left| \sqrt{-z_{ij} \bar z_{ij}+(n-i\alpha_{ij})^2} - 1\right| \right) + \sqrt{-z_{ij} \bar z_{ij}+(n-i\alpha_{ij})^2},
\end{aligned}
\end{equation}
and this vanishes for all $n$ if $z_{ij} \bar z_{ij} > 1$. However, for the modes $_{ij}$ inside of the Levi blocks this reads
\begin{equation}
    -\frac{1}{2} \left( |n+1| + |n-1| \right) -|n|,
\end{equation}
which vanishes for all $n \neq 0$. Then, only  $n=0$ of the Levi block modes contribute: 

\begin{equation}
    \frac{\langle W_{\theta_0 , \psi_0} \; \mathcal{O}_\Sigma \rangle }{\langle  \mathcal{O}_\Sigma \rangle} = W_{\theta_0 , \psi_0}|_{\Phi_0, A_0} \left( 1 + \frac{ \sum_{l=1}^M N_l^2 }{N} \frac{g^2}{8}  \,  \cos^2 \theta_0 \right).
\end{equation}
This clearly captures the $O(g^2)$ contribution to the exact answer \eqref{eq:locWilsonfinal} upon correctly identifying the parameters as $\CA_1=1+\sin\theta_0$, $\CA_2=1-\sin\theta_0$. This  implies that  summing   over the rainbow diagrams     in the surface operator background exactly reproduces 
\eqref{eq:locWilsonfinal} (c.f.\cite{Erickson:2000af,Drukker:2000rr}).

\section{More on $O_\Delta$ and $\mathcal O_{\Delta,k}$} \label{appsec:higherCPO}

In this Appendix, we further work on the relation between CPOs $O_\Delta$ and $\CO_{\Delta,k}$ by explicitly working on $\Delta=4,5$. Unlike the $\Delta=2,3$ cases, the $SO(4)$ averaged $O_\Delta$ has a non-trivial remainder $\CR[O_\Delta]$ in the sense of equation \eqref{eq:remainder}.

\subsection{$\Delta=4$}
First, we consider $\Delta=4$. The $2d$ YM localization picture determines the normalized correlator $\ex{O_4}$ as
\beq \label{eq:O42d}
\ex{O_4}=\sum_l N_l \be_l^4-{\lam \ov 16 \pi^2 N}[2(\sum_l N_l \be_l)^2+ \sum_l 2(2N-3N_l)N_l \be_l^2]+(N-N_l)^2N_l {\lam^2\ov 128 \pi^2 N}
\eeq

Using the non-commutative binomial expansion, we have ($\phi_B=-\phi_9$ in the convention of section \eqref{sec:CPOloc})
\beq
O_4(x)=&\Tr\left({ z\Phi +\bar z\bar\Phi \ov \sqrt{2}} +x_2\phi_6+i\phi_B\right)^4 
\\=&\Tr\left({z\Phi +\bar z\bar\Phi \ov \sqrt{2}}\right )^4 
+4\Tr\left( {z\Phi +\bar z\bar\Phi \ov \sqrt{2}}\right)^2 (x_2\phi_6+i\phi_B)^2 
\\&+2\Tr \left({z\Phi +\bar z\bar\Phi \ov \sqrt{2}}\right) (x_2\phi_6+i\phi_B)\left({z\Phi +\bar z\bar\Phi\ov \sqrt{2}}\right) (x_2\phi_6+i\phi_B) 
+\Tr(x_2\phi_6+i\phi_B)^4 .
\eeq 
After straightforwardly performing an $SO(4)$ averaging, we obtain
\beq
~&\CP[O_4]= {\lam^2 \ov 128\pi^4}(z^4\CO_{4,4}+\bar z^4\CO_{4,-4})+{\sqrt{10}\lam^2\ov 128\pi^4}(z^3 \bar z\CO_{4,2}+z \bar z^3\CO_{4,-2})+{\sqrt{5}\lam^2\ov 64\pi^4}z^2 \bar z^2\CO_{4,0}+\CR[O_4]
\\&\CR[O_4]={1\ov 24}(1+x_2^2)^2 \sum_{i,j=1,\dots,4}(6\phi_i^4-2\phi_i^2 \phi_j^2-\phi_i\phi_j\phi_i\phi_j).
\eeq

It is remarkable that we can have an alternative understanding of the vanishing of $\ex{\CR[O_4] \CO_\Sigma}$ as follows. First we using the $SO(4)$ invariance, we rewrite as
\beq
\ex{\CR[O_4]}={1\ov 2}(1+x_2^2)^2 \ex{(\phi_1^4-2\phi_1^2 \phi_2^2-\phi_1\phi_2\phi_1\phi_2)},
\eeq
where $1,2$ are any pair of components in $SO(4)$ indices. Now consider an $SO(4)$ generator $\del_{12}$ which acts on the vector representation $\phi_i$ as $\del_{12} \phi_1=-\phi_2$, $\del _{12}\phi_2=\phi_1$ and $\del_{12} \phi_{3,4}=0$. Then we remarkably see that
\beq \label{eq:so4exact}
\del_{12}(\phi_1^3\phi_2)=\phi_1^4-2\phi_1^2 \phi_2^2-\phi_1\phi_2\phi_1\phi_2.
\eeq
Therefore $\CR[O_4]$ is an $\del_{12}$ exact operator and because of the $SO(4)$ global symmetry, related Ward identity directly gives $\ex{\CR[O_4]}=0$.

As a result, we can substitute the perturbative result \eqref{gaugeresults1} to $\ex{\CP[O_4]}=\ex{O_4}$ and confirm that the result is in precise agreement with \eqref{eq:O42d}.

\subsection{$\Delta=5$}

The $2d$ YM localization picture gives the normalized $\Delta=5$ correlator s
\beq \label{eq:O52d}
\ex{O_5}=&\sum_l N_l \be_l^5
-{5\lam \ov 16 \pi^2 N}\left[(\sum_a N_a \be_a  )(\sum_b N_b \be_b^2) +(N-2N_l)N_l \be_l^3)\right]
\\&+{5\lam^2 \ov 256\pi^4 N^2} N_l(N-N_l)\left[(\sum_a N_a\be_a) +(N-2N_l)\be_l\right].
\eeq

To compare with the results of $\ex{O_{\Delta,k}}$, we similarly find
\beq \label{eq:PO_5}
\CP[O_5]=&{\sqrt{10}\lam^{5/2}\ov 256 \pi^5}{1\ov 2^{5/2}}(z^5\CO_{5,5} +\bar z^5 \CO_{5,-5}+5z^4 \bar zO_{5,3}+5 z\bar z^4O_{5,-3})+{\sqrt{5}\lam^{5/2}\ov 128 \pi^5}{5\ov 2^{5/2}}(z^3 \bar z^2 O_{5,1}+z^2 \bar z^3O_{5,-1})
\\&+{5\ov 24\sqrt{2}}(1+x_2^2)^2 ( z\Phi+\bar z\bar \Phi)\sum_{i,j=1,\dots,4}(6\phi_i^4-2\phi_i^2\phi_j^2-\phi_i\phi_j\phi_i\phi_j).
\eeq
Therefore, using again \eqref{eq:so4exact}, we explicitly see that $\ex{\CR[O_5]}=0$. And if we further plugging back the perturbative results \eqref{gaugeresults1} to \eqref{eq:PO_5}, we see an exact agreement with \eqref{eq:O52d}.

\end{appendix}

\newpage

% BIBLIOGRAPHY

\bigskip

\bigskip

\renewcommand\refname{\bfseries\large\centering References\\ \vspace{-0.4cm}
\addcontentsline{toc}{section}{References}}

\bibliographystyle{utphys.bst}
\bibliography{Refs.bib}
	
\end{document}